\documentclass[twocolumn,english,aps,pra,10pt,superscriptaddress,floatfix]{revtex4-2}
\usepackage{times}
\usepackage{graphicx}
\usepackage{amssymb}
\usepackage{bm}% bold math
\usepackage{amssymb,amsfonts,amsmath,amsbsy,bm,t1enc,latexsym}
\usepackage{float}
\usepackage[colorlinks=true,citecolor=blue,linkcolor=magenta]{hyperref}
\usepackage[markup=blue, authormarkupposition=left]{changes}
\usepackage[english]{babel}
\usepackage{url}
\usepackage{siunitx}
\DeclareSIUnit \dbc {dBc}
\usepackage{soul}
\usepackage{changes}
\usepackage{cancel}
\usepackage{times}

\newcommand{\SiN}[0]{Si$_3$N$_4$~}
\newcommand{\HzsHz}[0]{$\rm{Hz}^{2}/\rm{Hz}$~}
\definechangesauthor[name={TJK}, color=red]{TJK}

\begin{document}
	
	\title{Monolithic piezoelectrically tunable hybrid integrated laser with sub-fiber laser coherence}
	
	\author{Andrey Voloshin}
	\email[]{authors have contributed equally; andreysvoloshin@icloud.com}
	\affiliation{Institute of Physics, Swiss Federal Institute of Technology Lausanne (EPFL), CH-1015 Lausanne, Switzerland}
	\affiliation{Center of Quantum Science and Engineering, EPFL, CH-1015 Lausanne, Switzerland}
	\affiliation{Institute of Electrical and Micro-Engineering, EPFL, CH-1015 Lausanne, Switzerland}
	\affiliation{Deeplight SA, St Sulpice CH-1025, Switzerland}
	
	\author{Anat Siddharth}
	\email[]{authors have contributed equally}
	\affiliation{Institute of Physics, Swiss Federal Institute of Technology Lausanne (EPFL), CH-1015 Lausanne, Switzerland}
	\affiliation{Center of Quantum Science and Engineering, EPFL, CH-1015 Lausanne, Switzerland}
	\affiliation{Institute of Electrical and Micro-Engineering, EPFL, CH-1015 Lausanne, Switzerland}
	
	\author{Simone Bianconi}
	\email[]{authors have contributed equally}
	\affiliation{Institute of Physics, Swiss Federal Institute of Technology Lausanne (EPFL), CH-1015 Lausanne, Switzerland}
	\affiliation{Center of Quantum Science and Engineering, EPFL, CH-1015 Lausanne, Switzerland}
	\affiliation{Institute of Electrical and Micro-Engineering, EPFL, CH-1015 Lausanne, Switzerland}
	
	\author{Alaina Attanasio}
	\email[]{authors have contributed equally}
	\affiliation{OxideMEMS Lab, Purdue University, 47907 West Lafayette, IN, USA}
	
	\author{Andrea Bancora}
	\email[]{authors have contributed equally}
	\affiliation{Institute of Physics, Swiss Federal Institute of Technology Lausanne (EPFL), CH-1015 Lausanne, Switzerland}
	\affiliation{Center of Quantum Science and Engineering, EPFL, CH-1015 Lausanne, Switzerland}
	\affiliation{Institute of Electrical and Micro-Engineering, EPFL, CH-1015 Lausanne, Switzerland}
	\affiliation{Deeplight SA, St Sulpice CH-1025, Switzerland}
	
	\author{Vladimir Shadymov}
	\email[]{authors have contributed equally}
	\affiliation{Institute of Physics, Swiss Federal Institute of Technology Lausanne (EPFL), CH-1015 Lausanne, Switzerland}
	\affiliation{Center of Quantum Science and Engineering, EPFL, CH-1015 Lausanne, Switzerland}
	\affiliation{Institute of Electrical and Micro-Engineering, EPFL, CH-1015 Lausanne, Switzerland}
	\affiliation{Deeplight SA, St Sulpice CH-1025, Switzerland}
	
	\author{Sebastien Leni}
	\affiliation{Deeplight SA, St Sulpice CH-1025, Switzerland}
	
	\author{Rui Ning Wang}
	\affiliation{Institute of Physics, Swiss Federal Institute of Technology Lausanne (EPFL), CH-1015 Lausanne, Switzerland}
	\affiliation{Center of Quantum Science and Engineering, EPFL, CH-1015 Lausanne, Switzerland}
	\affiliation{Institute of Electrical and Micro-Engineering, EPFL, CH-1015 Lausanne, Switzerland}
	
	\author{Johann Riemensberger}
	\affiliation{Institute of Physics, Swiss Federal Institute of Technology Lausanne (EPFL), CH-1015 Lausanne, Switzerland}
	\affiliation{Center of Quantum Science and Engineering, EPFL, CH-1015 Lausanne, Switzerland}
	\affiliation{Institute of Electrical and Micro-Engineering, EPFL, CH-1015 Lausanne, Switzerland}
	
	\author{Sunil A. Bhave}
	\email[]{bhave@purdue.edu}
	\affiliation{OxideMEMS Lab, Purdue University, 47907 West Lafayette, IN, USA}
	
	\author{Tobias J. Kippenberg}
	\email[]{tobias.kippenberg@epfl.ch}
	\affiliation{Institute of Physics, Swiss Federal Institute of Technology Lausanne (EPFL), CH-1015 Lausanne, Switzerland}
	\affiliation{Center of Quantum Science and Engineering, EPFL, CH-1015 Lausanne, Switzerland}
	\affiliation{Institute of Electrical and Micro-Engineering, EPFL, CH-1015 Lausanne, Switzerland}
	
	\medskip
	
	\maketitle
	
	%%%%%%%%%%%%%%%%%%%%%%%%%%%%%%%%%%%%%%%%%%%%%%%%%%%%%%%%%%%%%%%%%%%%%%%%%%%%%%%%%%%%%%%%%%%%%%%%%%%%%%%%%%%%%%%
	%%%%%%%%%%%%%%%%%%%%%%%%%%%%%%%%%%%%%%%%%%%%%%%%% Abstract %%%%%%%%%%%%%%%%%%%%%%%%%%%%%%%%%%%%%%%%%%%%%%%%%%%%
	%%%%%%%%%%%%%%%%%%%%%%%%%%%%%%%%%%%%%%%%%%%%%%%%%%%%%%%%%%%%%%%%%%%%%%%%%%%%%%%%%%%%%%%%%%%%%%%%%%%%%%%%%%%%%%%
	
	\noindent\textbf{Ultra-low noise lasers are essential tools in a wide variety of applications, including data communication, light detection and ranging (LiDAR), quantum computing and sensing, and optical metrology. Recent advances in integrated photonics, specifically the development of ultra-low loss silicon nitride (\SiN) platform, have allowed attaining performance that exceeds conventional legacy laser systems, including the phase noise of fiber lasers. This platform can moreover be combined with monolithic integration of piezoelectrical materials, enabling frequency agile low noise lasers.
	However, this approach has to date not surpassed the trade-off between ultra-low frequency noise and frequency agility.
	Here we overcome this challenge and demonstrate a fully integrated laser based on the \SiN platform with frequency noise lower than that of a fiber laser, while maintaining the capability for high-speed modulation of the laser frequency. 
	The laser achieves an output power of 30 mW with an integrated linewidth of 4.3 kHz and an intrinsic linewidth of 3 Hz, demonstrating phase noise performance that is on par with or lower than commercial fiber lasers. Frequency agility is accomplished via a monolithically integrated piezoelectric aluminum nitride (AlN) micro-electro-mechanical system (MEMS) actuator, which enables a flat frequency actuation bandwidth extending up to 400 kHz.
	Such a MEMS device is one of the largest fabricated structures, featuring MHz-level bandwidth, which is significantly higher than the typical kHz-level bandwidth of similarly sized mm-scale MEMS devices. The non-linearity of the frequency-modulated linearly chirped output reaches 0.08\% without any linearization or pre-distortion, making complaint with the requirement for long-range FMCW LiDAR. This combination of ultra-low noise and frequency agility is a useful feature enabling tight laser locking for frequency metrology, fiber sensing, and coherent sensing applications. Our results demonstrate the ability of 'next generation' integrated photonic circuits (beyond silicon) to exceed the performance of legacy laser systems in terms of coherence and frequency actuation.}
	%%%%%%%%%%%%%%%%%%%%%%%%%%%%%%%%%%%%%%%%%%%%%%%%%%%%%%%%%%%%%%%%%%%%%%%%%%%%%%%%%%%%%%%%%%%%%%%%%%%%%%%%%%%%%%%
	%%%%%%%%%%%%%%%%%%%%%%%%%%%%%%%%%%%%%%%%%%%%%%%% Introduction %%%%%%%%%%%%%%%%%%%%%%%%%%%%%%%%%%%%%%%%%%%%%%%%%
	%%%%%%%%%%%%%%%%%%%%%%%%%%%%%%%%%%%%%%%%%%%%%%%%%%%%%%%%%%%%%%%%%%%%%%%%%%%%%%%%%%%%%%%%%%%%%%%%%%%%%%%%%%%%%%%
	
	\section{Introduction}
	Low-noise lasers are critical for a variety of applications, including coherent data communication protocols \cite{Kikuchi_fundametals_communications, OFDM, FDOR_1981, Carrier_Recovery_2007}, optical metrology, quantum sensing \cite{Cappellaro_quantum_sensing}, and atomic clocks \cite{Ludlow_atomic_clocks}. They are also valuable for long-distance coherent light detection and ranging (LiDAR) \cite{laser_radar} and distributed optical fiber sensing \cite{rogers1999distributed}. %Lasers have become a unique tool to study the universe with accuracy going beyond the order of one part in $10^{18}$ (for instance, in the imaging spectroscopy of the optical clock transition of lattice-trapped degenerate fermionic Strontium \cite{lattice_clock}) using table-top laser systems and complicated optical setups. The observation of gravitational waves became feasible due to laser systems with the frequency noise (FN) 0.01-1 \HzsHz in the range of Fourier frequency offsets 1-1000 Hz.
%	Ultra-low noise laser systems with the frequency noise below 1 \HzsHz at 100 Hz paved the way to the observation of gravitational waves.
	%Currently, laser-based systems are used not only for fundamental studies but also for various real-world applications, 
%	Other applications of low noise lasers are long-distance coherent light detection and ranging (LiDAR) \cite{laser_radar}, and distributed optical fiber sensing \cite{rogers1999distributed}.
		%Lasers that have the radiation with well-defined frequencies are usually called ultra-low noise lasers and allow us to bridge optical and radio frequencies, in other words, perform down- and up-conversion.
	%The principle “never measure anything but frequency” \cite{hansch_schawlow} highly relies on the FN of a laser and allows to measure the frequency ratio between the fundamental and its second harmonic of a 1064 nm laser with a fractional uncertainty of $10^{-22}$ \cite{10_minus_22_ratio_measurement}.
	%Speaking about the limitation of sensing systems based on coherent light detection, the frequency noise becomes an important figure of merit of a laser system. In most cases, it sets the noise floor of the detected signal.
	In many applications, both low noise and frequency agility are essential. Examples include carrier recovery in coherent communications, triangular chirping in frequency-modulated continuous-wave (FMCW) LiDAR, locking to frequency references, and frequency modulation in fiber sensing.
	Currently, most commercial lasers rely on technologies that require manual assembly using discrete components. Frequency agility in these systems is typically achieved through external frequency-shifting devices, such as acousto-optic or electro-optic single-sideband modulators. However, these components and systems tend to be bulky and expensive, making them impractical for widespread industrial and mass adoption of low-noise laser-based optical technologies, such as those used in optical gyroscopes \cite{optical_gyro_review}. For instance, long-range FMCW LiDAR requires the laser linewidth to be significantly lower (kHz-level) than that of a conventional semiconductor distributed-feedback (DFB) laser \cite{zhang2023long} and requires a complicated scheme for waveform compensation or calibration, in other words, linearization \cite{Martin18}. One of the first demonstrations of the silicon photonic FMCW LiDAR \cite{rogers2021universal} utilizes a <100 Hz fiber laser.
		
	\begin{figure*}[htb!]
		\centering\includegraphics[width=0.95\linewidth]{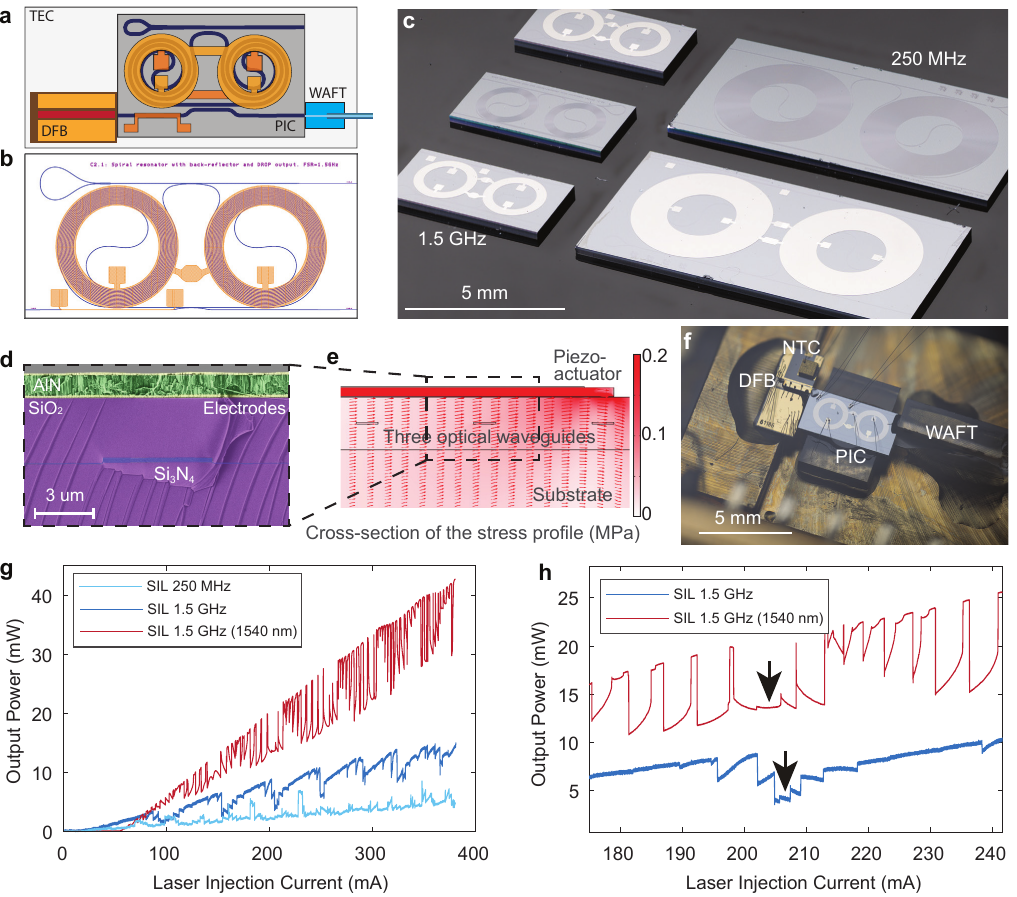}
		\caption{
			\footnotesize
			\label{fig:concept}\textbf{Hybrid integrated laser based on low-loss \SiN photonic chip monolithically integrated with AlN piezoelectrical actuators.}
			(a) The laser concept is based on self-injection locking a distributed feedback laser (DFB) to a \SiN photonic integrated circuit based microresonator. A photonic chip has a bus waveguide, a dual spiral cavity, piezo-actuators, a drop-port with a loop mirror and an output taper. The output component is a waveguide array to fiber transposer.
			(b) Layout of a chip with 1.5 GHz spiral cavity. A chip size of 5 $\times$ 2.5 mm allowed us to use standard butterfly packages.
			(c) Photograph of chips with and without MEMS actuators (FSR 1.5 GHz: \texttt{D134\_03\_F4\_C1.3\_4} and \texttt{D134\_03\_F3\_C1.3\_7}; FSR 250 MHz: \texttt{D134\_01\_F6\_C3.1\_0} and \texttt{D134\_03\_F2\_C3.1\_3}).
			(d) SEM image of the MEMS-photonic chip showing its vertical cross-section.
			(e) Stress profile at the chip cross-section on applying 1 V across the AlN piezoactuator.
			(f) Image of the hybrid integrated laser assembly.
			(g) SIL laser output power as function of DFB drive current.. The 1540 nm 1.5 GHz SIL laser reaches a power level of 15-30 mW without any degradation in the frequency noise.
			(h) Self-injection locked operation points of the 1.5 GHz SIL laser. Operating points for the 250 MHz SIL laser are not shown since they exist almost at all points of the driving current (due to the dense mode spectrum of the 250 MHz microresonator).
		}
	\end{figure*}
	
	Recent advances in low-loss \SiN have established a platform for photonic integrated circuits (PICs) with propagation losses below 3 dB/m \cite{pfeiffer2016damascene, liu2021highyield, blumenthal2011lowloss}, driving significant advancements in the development of photonic integrated lasers to make them complient with strong requirement for photonic sensing systems, especially, long-range LiDAR.
	% and with the performance reaching the level of well-established optical systems \cite{bowers2021reaching, martin2018lidar, bowers2019tutorial, boller2020hybridlasers, fan2020hybrid}. Most importantly, photonic integration has allowed not only the reproduction of optical systems on a photonic chip but also the demonstration of new capabilities (for instance, fully integrated electrically pumped soliton microcombs with spacing from 10 GHz to 1 THz \cite{raja2019electrically}).
	Various integrated laser designs on \SiN chips with active semiconductor media now cover a wide range of wavelengths \cite{anat2022bluelaser, winkler2024redlaser, isichenko2023chip, 780Li23, 780Prokoshin24, Franken:21, frentrop2023800}. However, current state-of-the-art technologies offer either ultra-low frequency noise, achieving coherence comparable to fiber lasers \cite{bower_hertz_integrated, bowers2021reaching, siddharth2023hertz}, or high-speed tunability \cite{snigirev2023lnodlaser, li2022pockels, lihachev2022low}, but not both. Achieving both ultra-low frequency noise and fast laser frequency chirping often requires electro-optic materials in heterogeneously integrated or monolithically integrated photonic chips \cite{snigirev2023lnodlaser, wang2024tantalate, li2023lnoi, siddharth2024ultrafast, xue2024pockels, Shams-Ansari:22, franken2024high}, though charge noise limits frequency noise reduction at low frequency offsets \cite{zhang2023fundamental}.

	Here, we demonstrate a hybrid integrated laser by employing self-injection locking (SIL) of a DFB laser to an ultra-low-loss photonic integrated microresonator based on a \SiN platform. This approach achieves a noise reduction of five orders of magnitude, resulting in phase noise lower than that of a fiber laser across most offset frequencies. Additionally, the inclusion of a monolithically integrated AlN piezoactuator enables MHz-level tuning bandwidth, significantly surpassing the tuning speed of conventional fiber lasers (Fig. \ref{fig:concept}(f)). The hybrid integrated laser exhibits high output power with an integrated linewidth of 1.38 kHz (up to 30 mW for 4.30 kHz), comparable to a SIL laser based on a crystalline whispering-gallery-mode microresonator \cite{liang2015ultralow}.
	%that achieves phase noise lower than that of a fiber laser across most offset frequencies, while also being tunable with an MHz-level actuation bandwidth—faster than conventional fiber lasers (Fig. \ref{fig:concept}(f)). This is enabled by a MEMS actuator monolithically fabricated on top of a \SiN chip. The aluminum nitride (AlN) piezoelectric MEMS actuator provides high actuation bandwidth, reaching several GHz \cite{tian2020bulk}. The laser concept leverages a self-injection locking (SIL) technique, creating a simple yet efficient system through the optical feedback of a laser diode. We demonstrate a high-power laser (up to 30 mW) with an integrated linewidth of 1.38 kHz and 4.30 kHz, comparable to an SIL laser based on a crystalline whispering-gallery-mode microresonator \cite{liang2015ultralow}.
	
	We showcase one of the largest MEMS structures fabricated to date. Typically, MEMS devices range in size from 20 $\mu$m to 1 mm \cite{choudhary2017mems}. For comparison, surface acoustic wave AlN MEMS filters \cite{gao2020aln} and AlN bulk acoustic wave MEMS filters \cite{zou2022aluminum}, widely used in consumer communication devices, measure on the order of hundreds of micrometers. Larger structures, however, have been fabricated for specialized applications, such as millimeter-scale flapping PZT-based wings for micro air vehicles (2.5 mm in length) \cite{PZT_wings1, PZT_wings2, jimbo2020flight}, a device for focal length tuning with MEMS-integrated meta-optics (1 mm) \cite{han2022millimeter}, and acoustic microspeakers \cite{seo10speakers}. These larger structures typically operate at kHz-level frequencies. Notably, flapping mechanical modes have been widely researched and commercialized by the MEMS community \cite{870061, Demirci03IEEE, clark2013temperature}, including in high-volume manufacturing \cite{clark2017microchip}, and are now employed in advanced resonant systems.
	
	%Unlike these examples in which the piezo-material is thinned to achieve large displacement, our goal is efficient and linear transfer of piezo-electric stress from the transducer to the underlying photonic cavity. In our case, we fabricated MEMS actuators in two sizes, 8.6 $\times$ 4 mm and 4.1 $\times$ 1.8 mm. The smaller actuator covers 20\% of the chip surface and supports flat actuation up to 400 kHz, with MHz actuation achievable at certain frequencies, which significantly exceeds the actuation bandwidth of most large MEMS structures. A laser with such a MEMS actuator becomes a new tool for research and industrial applications eliminating the need for high-speed external optical frequency shifters (acousto-optic or electro-optic) providing flat frequency actuation bandwidth up to 400 kHz which is 10-fold faster than conventional modulation bandwidth of ultra-low noise lasers with bulk piezo-actuators.
	%Such a large bandwidth allows highly linear actuation of a ultra-low noise laser, which is critical for a long-range coherent LiDAR. The laser performance is ideal for optical quantum metrology, microwave photonics and varoius types of coherent sensing (LiDAR, long-range gas sensing, optical metrology), especially, distributed fiber optic sensing which often utilizes a pair of ultra-low noise laser and frequency shifters based on acouto-optic or electro-optic effects.
	In these examples, where the piezoelectric material is thinned to achieve large displacement, our objective is to attain an efficient and linear transfer of piezoelectric stress from the transducer to the underlying photonic cavity. Here, we demonstrate MEMS actuators in two sizes: 8.6 $\times$ 4 mm and 4.1 $\times$ 1.8 mm. The smaller actuator, which covers 20\% of the chip surface, provides a flat actuation bandwidth up to 400 kHz, with MHz-level actuation achievable at specific resonant frequencies. This performance significantly surpasses the typical actuation bandwidth of most large MEMS structures. A laser integrated with such a MEMS actuator is a valuable asset for both research and industrial applications, eliminating the need for high-speed external optical frequency shifters (such as acousto-optic or electro-optic modulators). The actuator's flat frequency actuation bandwidth of up to 400 kHz is ten times faster than the conventional modulation bandwidth of ultra-low noise lasers equipped with bulk piezo-actuators. We also use the fully hybrid integrated laser to do a FMCW LiDAR experiment.
	
	\section{Concept and noise performance of the hybrid integrated laser}
	
	\begin{figure*}[htb!]
		\centering\includegraphics[width=0.95\linewidth]{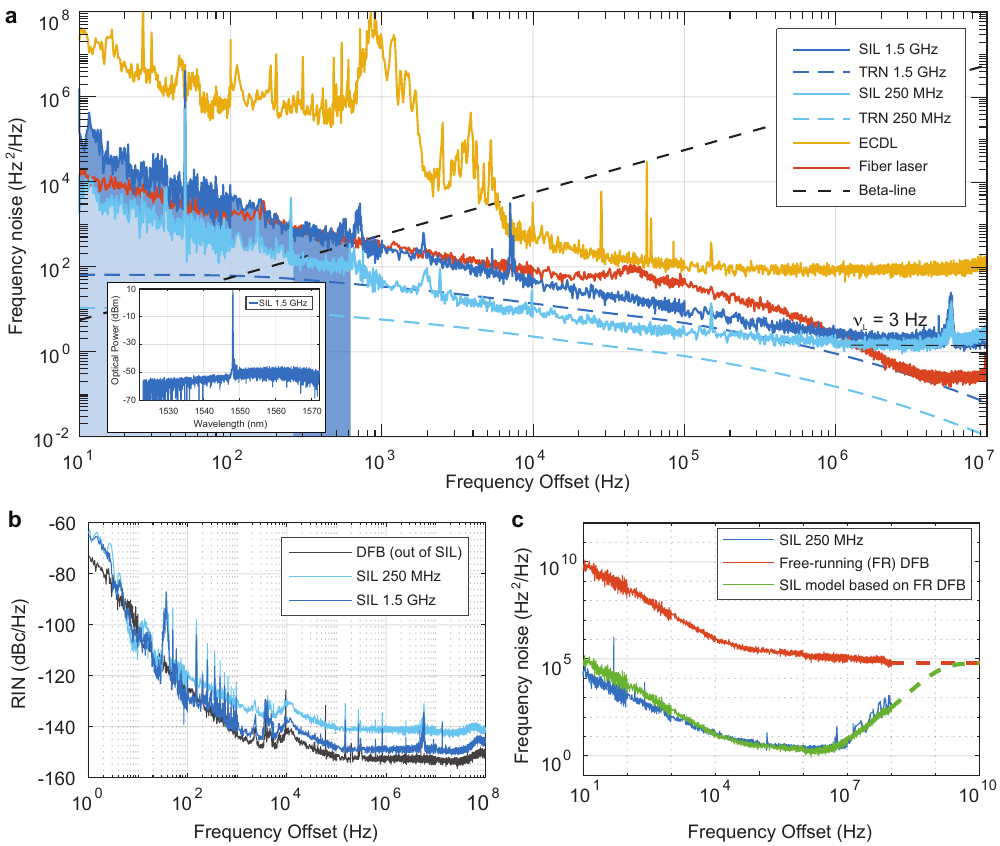}
		\caption{
			\footnotesize
			\label{fig:FN}\textbf{Comparison of the frequency noise of different laser systems.}
			(a) Light blue and dark lines correspond to FSR 250 MHz (based on \texttt{D134\_03\_F2\_C3.1\_3} PIC) and FSR 1.5 GHz (based on \texttt{D134\_03\_F4\_C1.3\_4} PIC) SIL lasers with integrated linewidths ($\tau_0 = $ 0.1 s) of 1.38 kHz and 4.30 kHz correspondingly. The Lorentzian linewidth is at 3 Hz for both lasers. The noise of the 250 MHz laser is lower than a fiber laser at all frequency offsets below 1 MHz. The fiber laser $\beta$-linewidth with an integration time of $\tau_0 =$ 0.1 s is 2.17 kHz. The noise of a commercial external cavity diode laser is shown for comparison. Inset: example of an optical spectrum exhibiting side-mode suppression ratio above 55 dB.
			(b) RIN of the self-injection locked lasers, which is higher than a free-running diode due to the amplitude noise-frequency noise transduction according to the light-intensity curve in Fig. \ref{fig:concept}(h).
			(c) Simulation of the self-injection locked frequency noise. The blue line is the experimental 250 MHz SIL laser noise. The red curve represents the measured and extrapolated noise of the laser diode. The green curve is the analytically estimated self-injection locked noise using the noise of a free-running laser ($\rho$ = 0.05, $Q_{\text{r}} = 10^{7}$, $Q_{\text{laser}} = 10^{4}$). The rise in noise can be eliminated by using a drop-port as the laser output, which filters out high-offset frequency noise \cite{jin2021hertz}.}
	\end{figure*}
	
	The self-injection locking (SIL) of a semiconductor laser diode to an optical cavity is a technique that relies on high Q-factors of optical microresonator and the light reflected back from the microresonator. Figure \ref{fig:concept}(a) illustrates the concept of a frequency-agile, hybrid-integrated SIL laser that optimizes both of these parameters. The laser comprises a DFB laser diode butt-coupled to a photonic chip fabricated on a 200 nm-thick, low-loss \SiN platform. This thickness provides an optimal balance between ultra-low propagation loss, achievable in weakly confining waveguides, and the tighter confinement needed to enable compact bends, which minimizes die size. Additionally, this configuration enhances piezoactuator tuning efficiency under a 3 $\mu$m top SiO$_2$ cladding layer. The material stack is shown in Fig. \ref{fig:concept}(d), which presents a cross-section of the photonic waveguide and piezoelectric actuator. Both sections of the spiral microresonator are covered with donut-shaped actuators. Figure \ref{fig:concept}(c) shows structures with free spectral ranges (FSRs) of 1.5 GHz and 250 MHz, with and without actuators. The intrinsic loss rates for both structures are $\kappa_0/2\pi = 30 \pm 10$~MHz.

	The size of the 250 MHz spiral structure is 8.6 $\times$ 4.0 mm, with a waveguide width of 5 $\mu$m and a waveguide pitch of 20~$\mu$m. The full chip size, including all auxiliary structures (i.e., mode converters, tapers, etc.), is 10 $\times$ 5 mm (eight devices per a reticle). The photonic structure of the SIL 135 MHz in \cite{bowers2021reaching} has dimensions of 9.2 $\times$ 7.2 mm (four devices per a reticle) with a waveguide pitch of 40 $\mu$m (the top cladding of 2.2 $\mu$m complicates the usage of any metal structures for controlling the spiral, as they introduce significant ohmic losses for the weakly confined optical mod in 100 nm \SiN). In contrast, the smaller pitch, reduced bending radius (0.5 mm), and top cladding of 3 $\mu$m in the \SiN platform used in this work allow us to achieve a compact footprint for photonic structures with integrated metal structures.
	%It is important to mention, that the authors of \cite{bower_hertz_integrated} reported much higher FN of the same structure, SIL 135 MHz.
	%However, the photonic structure of such SIL 135 MHz in \cite{bowers2021reaching} has dimensions of 9.2 $\times$ 7.2 mm, and the full PIC has bigger dimensions which limit the benefits of wafer-scale fabrication.
	%Also, the chip has a top cladding of 2.2 $\mu$m which makes not feasible usage of any metal structures for any kind of control of a spiral since they will introduce significant ohmic losses for such low confinement optical mode (\SiN thickness is 100 $\mu$m).	
	We investigate two different types of lasers, SIL 1.5 GHz and SIL 250 MHz, which correspond to the different spiral microresonator. The layout also includes a bus waveguide phase shifter to control the feedback phase and a loop reflector to improve the SIL linewidth reduction factor \cite{Kondratiev17, ulanov2024synthetic, ulanov2024laser}.	
	Different radii of the spiral microresonator experience varying levels of stress, leading to a reduction in actuation efficiency, as illustrated in Figure \ref{fig:concept}(e). The stress distribution across the waveguide cross-section reveals that the stress profile is not uniform, resulting in efficiency variations across the microresonator. Optimizing the actuation bandwidth involves numerous trade-offs. Key factors include the minimum bending radius of the photonic waveguide, the thickness of the top cladding, the mechanical modes of the actuator and the entire chip, the actuator's capacitance, and other considerations. In our case, we focus on optimizing the laser's optical noise performance in combination with a broad, flat actuation bandwidth for laser frequency, resulting in a reduced tuning efficiency of more than 4 times, yielding 2.5 MHz/V. Previous demonstrations \cite{lihachev2022low,liu2020monolithic} achieved tuning efficiencies of 10–20 MHz/V, depending on the radii of the microrings.
		
	Figure \ref{fig:concept}(g,h) presents key characteristics of these laser systems. The output power of the SIL 250 MHz laser reaches 1–5 mW at its operating point. Notably, due to the dense mode spectrum of the 250 MHz spiral cavity, the laser can achieve locking at almost any point, with the standard start-up procedure requiring no specific operating current. The average output power of the 1550 nm SIL 1.5 GHz laser is 5–10 mW, while a fully optimized system (incorporating optimized mode converters and a high-power DFB with high wall-plug efficiency) achieves output levels of 15–30 mW (low noise fiber lasers usually exhibit 30-40 mW output power).
	%These power levels, for the first time, enable photonic chip-based SIL lasers to meet the performance requirements for industrial applications.
	Our approach enables the development of an ultra-low noise laser in a compact footprint. Figure \ref{fig:FN}(a) presents the frequency noise characteristics of different laser systems. The SIL 1.5 GHz laser, with an integrated linewidth ($\tau_0 = 0.1$ s) of 4.30 kHz, maintains a consistent noise level across frequency offsets from 100 Hz to 1 MHz. The SIL 250 MHz laser achieves lower noise levels than a fiber laser across all frequency offsets from 10 Hz to 1 MHz, with an integrated linewidth ($\tau_0 = 0.1$ s) as low as 1.38 kHz. Additionally, we benchmark our results against other state-of-the-art SIL lasers based on crystalline resonators \cite{oewaves2023, liang2015ultralow} and photonic chip-based microresonators \cite{bowers2021reaching}, as detailed in the Supplementary Information.
	
	Fig. \ref{fig:FN}(c) compares the noise of the SIL 250 MHz laser with the analytically estimated noise using the experimentally measured noise of the free-running (FR) DFB laser. We use the model presented in~\cite{ousaid2024SIL} and described in Supplementary Information. The experimental data is in good agreement with the calculated noise, confirming that the noise is limited by the performance of the SIL, and not the thermo-refractive noise (TRN) limit.
	%Further improvement of Q-factors of \SiN PIC-based cavities monolithically integrated with MEMS actuators will significantly improve the SIL FN.
	It is also noteworthy that the increase in SIL noise at high frequency offsets is attributed to the dynamics of the locking mechanism, rather than shot noise.
	%It is possible to measure the SIL FN at frequencies above 1 GHz and confirm that the SIL FN is equal to the free-running DFB laser FN.
	
	\begin{figure*}[htb!]
		\centering\includegraphics[width=0.95\linewidth]{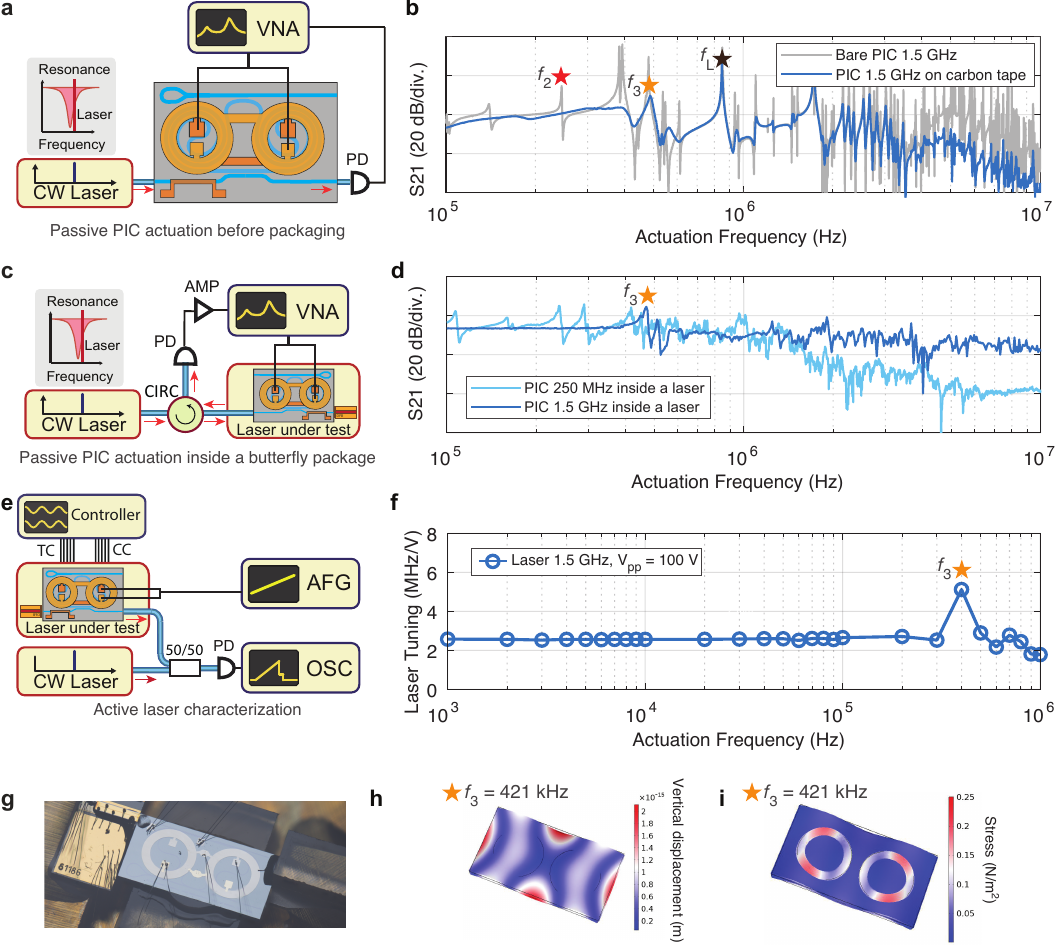}
		\caption{
			\footnotesize
			\label{fig:actuation}\textbf{Characterization of the monolithically integrated AlN piezoactuator.}
			(a) Experimental setup to measure the optomechanical $\rm{S}\textsubscript{21}$ response of the bare MEMS-photonic chips with two lensed fibers.
			(b) $\rm{S}\textsubscript{21}$ response with the flapping modes $f\textsubscript{2}$ = 220 kHz, $f\textsubscript{3}$ = 421 kHz and the length extension mode $f\textsubscript{L}$ = 865 kHz.
			(c) Experimental setup to measure the optomechanical $\rm{S}\textsubscript{21}$ response of the packaged MEMS-photonic chip inside a butterfly package. 
			(d) $\rm{S}\textsubscript{21}$ response of the 250 MHz and 1.5 GHz FSR chips inside the butterfly package. In this case, the actuator of the 1.5 GHz SIL laser shows flat actuation bandwidth up to 400 kHz.
			(e) The experimental setup to study the frequency-modulated continuous wave operation of the SIL lasers.
			(f) The actuator of a running laser is driven by a harmonic signal with different frequencies from 1 kHz to 1 MHz. For each frequency, 					we retrieve the laser frequency and the frequency excursion range using a heterodyne technique. The laser frequency has a flat 						actuation bandwidth up to 400 kHz.
			(g, h, i) demonstrate the image of a chip inside a butterfly package and the stress and the geometry distortion of a chip at the 					$f\textsubscript{3}$ mode.
		}
	\end{figure*}
	
	\section{Frequency agility of the hybrid integrated laser}
	
	%Next, we investigate the ability to chirp the laser frequency, i.e., continuously tune the laser frequency over a narrow range.
	%The \SiN PIC-based SIL lasers with piezo-electric actuators demonstrated both, low noise and MHz-level actuation bandwidth \cite{lihachev2022low}.

We conducted both active and passive characterization of the photonic chip-based laser to qualitatively and quantitatively assess its frequency agility. First, we carried out passive characterization of the piezo-actuators by measuring their actuation response $\rm{S}_{21}$, using an external continuous wave (CW) laser, as outlined in the experimental setup shown in Fig. \ref{fig:actuation}(a). The data presented in Fig. \ref{fig:actuation}(b) displays the $\rm{S}_{21}$ response of a 1.5 GHz chip, tested on both a flat metal surface and a carbon tape. The carbon tape serves to suppress certain mechanical modes; however, due to the high number of mechanical modes excited by millimeter-scale MEMS actuators, the $\rm{S}_{21}$ response is not entirely flat.	
	Without proper mechanical anchoring of the chip, even the smallest perturbations due to piezoelectric transducer excite flapping modes of the chip itself \cite{Demirci03IEEE}.
	In Fig. \ref{fig:actuation} we identify the first three flapping modes $f_1$, $f_2$, $f_3$ and the length extension mode $f_L$, which have significant impact on the chip geometry and the stress profile (cf. Supplementary Information for more details).
	
	Next, we measure the actuation response $\rm{S}\textsubscript{21}$ of chips inside photonic packages passively using an external CW laser and a circulator (Fig. \ref{fig:actuation}(d)).
	All photonic components (laser diode, chip, fiber array) are packaged inside a butterfly package (custom and standard Type 1) using active alignment and high-precision epoxies.
	In all instances, photonic packaging greatly enhances the laser's resilience to acoustic and vibrational disturbances and provides effective electrical shielding. This combination of improvements results in a stable frequency actuation response, ensuring that the frequency noise remains unaffected by external environmental factors. The dark blue curve in Fig. \ref{fig:actuation}(d) demonstrates a flat $\rm{S}_{21}$ response for the SIL 1.5 GHz chip at frequencies up to 400 kHz. Conversely, the $\rm{S}_{21}$ response of the SIL 250 GHz chip, shown in light blue in the same figure, exhibits additional resonances due to the larger dimensions of the chip and actuator.
	
	To directly assess the laser's frequency actuation bandwidth and tuning capabilities, we use an alternative characterization method by applying harmonic signals at varying frequencies to the actuator of the SIL 1.5 GHz laser. This induces chirping in the laser frequency, which is detected via heterodyne measurement with an external reference laser (cf. Fig. \ref{fig:actuation}(e)). It is important to note that this method specifically measures the laser's frequency actuation response under conditions of large amplitude signals.
	The flat tuning response of the laser up to 400 kHz validates its operation in line with the previously measured $\rm{S}_{21}$, which was obtained through small-signal characterization. While photonic packaging effectively suppresses most mechanical modes, it does not entirely eliminate them (as shown in Fig. \ref{fig:actuation}(b) and Fig. \ref{fig:actuation}(d)). The intentional overhang of the chip (cf. Fig. \ref{fig:actuation}(g)), required for photonic packaging, creates a free mechanical boundary condition. Consequently, the piezoelectric actuator excites the third harmonic mode, $f_3$, which is experimentally observed within the 400-500 kHz range and significantly enhances tuning efficiency through resonance. This third harmonic mode ultimately defines the upper limit of the actuation bandwidth. The same resonance, $f_3$, is also observed in the passive $\rm{S}_{21}$ characterization (cf. Fig. \ref{fig:actuation}(f)).
	
	\begin{figure*}[htb!]
		\centering\includegraphics[width=0.95\linewidth]{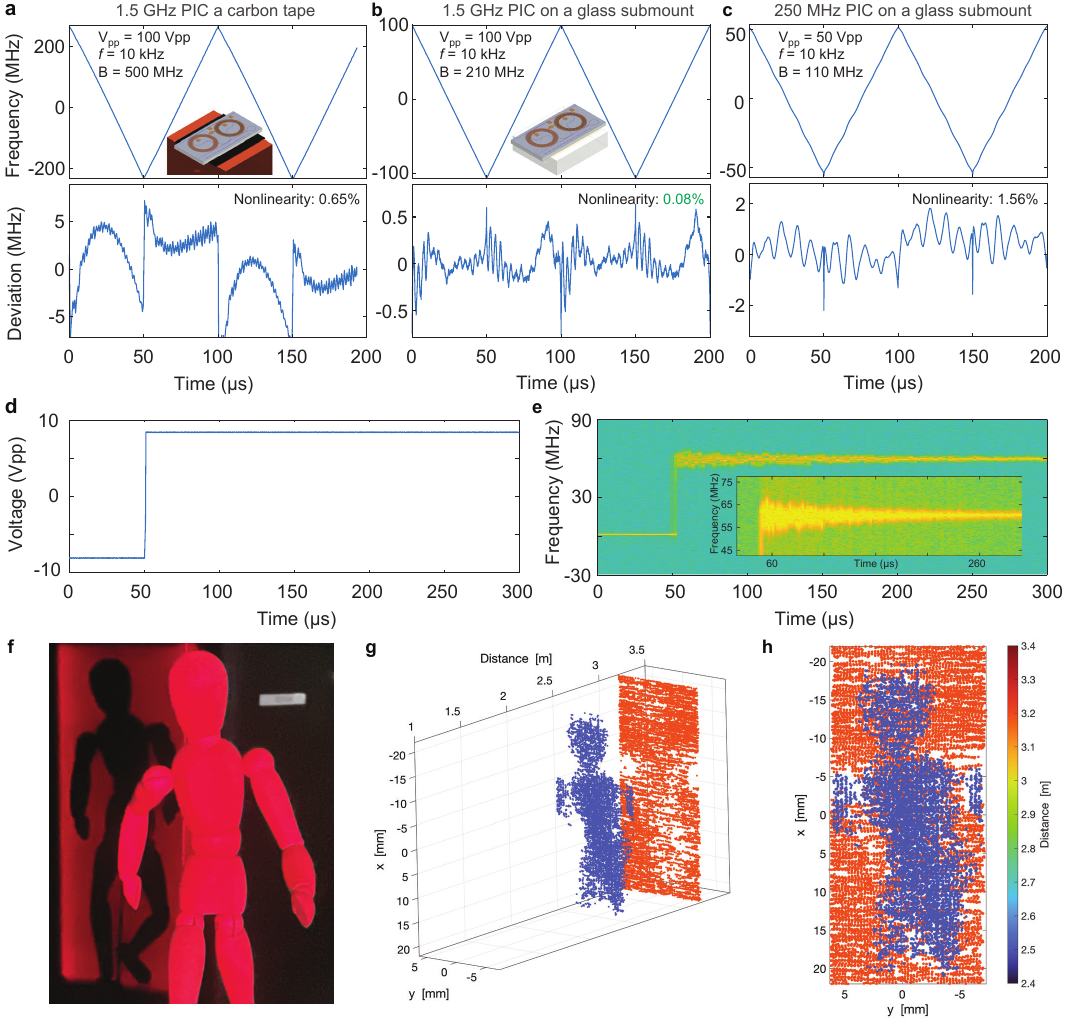}
		\caption{
			\footnotesize
			\textbf{Frequency-Modulated Continuous Wave (FMCW) operation of the hybrid integrated laser.}
			(a, b, c) Demonstration of frequency modulated operation with sawtooth signals (sweep rate 10 kHz).
			(a) shows a chirped laser frequency in a case of an unpackaged laser which is characterized by high non-linearity of 0.65\% and high 					tuning efficiency of 5 MHz/V (due to the non-local stress caused by flapping and bulk modes).
			(b) the same chip after packaging supports frequency modulated operation with reduced actuation efficiency but the non-linearity of 					0.08\% without any linearization or pre-distortion and low tuning effiency of 2.1 MHz/V.
			(c) Frequency-modulated continuous wave operation of the 250 MHz SIL laser with non-linearity of 1.65\%.
			(d, e) shows the step response of the hybrid integrated laser. When the step signal of 16 V is applied across the piezoactuator, the laser frequency changes by 60 MHz. The settling time is 200 $\mu$s and is limited by the mechanical modes of the chip. This performance can be enhanced through improved acoustic and mechanical packaging.
			(f) Photograph of the target scanned using a red laser for visualization purposes. 
			(g) Three-dimensional reconstruction of the ranged 3D target and scene performed without any linearization of the laser chirp.
			(h) Orthographic projection of the reconstructed 3D target and scene with color-coded ranging information.
			\label{fig:fmcw}
		}
	\end{figure*}
	
	\section{Frequency-Modulated Continuous Wave operation of the hybrid integrated laser}
	
	Photonic packaging can also play a crucial role in influencing tuning efficiency. In previous work \cite{lihachev2022low}, we demonstrated that a compact ring resonator with a free spectral range (FSR) of 200 GHz, made from a thick (800 nm) \SiN layer, could achieve a tuning efficiency of 20 MHz/V with a 240 $\mu$m diameter actuator—an efficiency that remained stable post-packaging.
	These piezoelectric transducers can be approximated as point sources of stress, where local rigidity is predominantly governed by the bulk stiffness of the chip. For a thin (200 nm) \SiN ring with a 50 GHz FSR, a tuning efficiency of 10 MHz/V was achieved with a piezo actuator. In contrast, for a large spiral cavity, simulations indicate a lower tuning efficiency, below 3 MHz/V, due to a suboptimal actuator choice (Fig. \ref{fig:concept}(e)). However, experimental observations show broadband tuning efficiencies of 5 MHz/V for an unanchored or taped chip, which reduce to 2.4 MHz/V following photonic packaging.
	
	In this work, the piezoelectric transducer occupies 20\% of the chip area, functioning as a distributed transducer that couples to low-frequency, low spring-constant flexural modes of the chip \cite{Demirci03IEEE}. Although boundary conditions provided by carbon tape or adhesive assembly are relatively rigid, they are inadequate to fully suppress flapping modes, particularly the third harmonic ($f_3$), which interacts with the chip overhang required for fiber coupling (cf. Fig. \ref{fig:actuation}(g)). Consequently, the stress induced by the piezoelectric transducer is distributed across the chip surface, limiting the overall tuning efficiency. In this configuration, the refractive index modulation is governed not by localized stress applied by the piezo-actuators, as demonstrated in prior work \cite{lihachev2022low, tian2020bulk, liu2020monolithic}, but by stress resulting from geometric distortions caused by flapping and other mechanical modes. This distributed stress pattern enhances tuning efficiency to 5 MHz/V, exceeding the simulated tuning efficiency of less than 3 MHz/V; however, it introduces significant nonlinearity. After photonic packaging, the tuning efficiency of the laser is markedly reduced, as the spiral cavity is now primarily actuated by the localized stress from the piezo-actuator. This leads to an observed tuning efficiency of 2.4 MHz/V.
	
	We also evaluated the frequency-modulated continuous-wave (FMCW) performance of the laser using the setup depicted in Fig. \ref{fig:actuation}(e) with 10 kHz sawtooth modulation signals. In Fig. \ref{fig:fmcw}(a), the frequency chirping of the SIL 1.5 GHz laser is shown prior to photonic packaging. In this unpackaged state, the laser achieves a high modulation efficiency of 5 MHz/V due to the influence of the chip’s flexural and bulk mechanical modes. However, the presence of these multiple mechanical modes results in a relatively high non-linearity of 0.65\%.
	After photonic packaging, the tuning efficiency decreases to 2.4 MHz/V, as illustrated in Fig. \ref{fig:fmcw}(b). Nonetheless, the non-linearity of the FMCW signal improves significantly, dropping to 0.08\%, even without applying any linearization or pre-distortion techniques.
	%This becomes possible since the refractive index of a spiral cavity is primarily controlled by local stress applied by a piezo-actuator.
%	The FMCW nonlinearity of the SIL 250 MHz laser is 1.56\% at 10 kHz Fig. \ref{fig:fmcw}(c). However, while the laser has the same tuning efficiency of 2.5 MHz/V, the laser chirp distorts a lot with voltages higher than 50 $\rm{V}_{\rm{pp}}$. This is a demonstration that small signal modulation bandwidth can be different from large signal modulation.
	We also measure the laser frequency's settling time in response to an applied step voltage function (cf. Fig. \ref{fig:fmcw}(d, e)). The step voltage, shifting from -8 V to 8 V, induces a frequency change of approximately 60~MHz. This abrupt input excites multiple mechanical modes within the chip, resulting in mechanical oscillations and a settling time of around 200 $\mu$s.
	
	To showcase the tuning bandwidth and chirp linearity (without any linearization) of the photonic integrated circuit-based lasers, we conducted an optical coherent ranging demo in a lab environment using the FMCW LiDAR scheme with the fully hybrid integrated SIL 1.5 GHz laser.
	A wooden mannequin placed at 2.5 m from the collimator output, with a scattering cardboard wall placed behind it at around 3.2 m from the collimator is used as a target, as shown in \ref{fig:fmcw}(f). The fundamental depth resolution of the measurement is determined by the 200-MHz tuning range of the laser at this applied voltage, as given in eq. \ref{fig:fmcw} (g), resulting in around 75 cm. Figure \ref{fig:fmcw} (h) shows point cloud representations of the reconstructed target scene with color-coded distance information: the mannequin profile is mostly visible in blue, and the wall behind in orange. The detailed methodology of the experiment is presented in the Supplementary Information.	

	\section*{Conclusion}
	
	%In this work, we investigated the limitations of laser systems based on the ultra-low loss \SiN photonic platform  monolithically integrated with AlN micro-mechanical piezo-electric actuators.
	We have developed a frequency-agile laser that surpasses conventional laser systems by addressing the longstanding trade-off between ultra-low noise and high tuning capabilities \cite{bowers2021reaching, lihachev2022low, siddharth2023hertz}. Operating at output powers of up to 30 mW, this laser exhibits frequency noise lower than that of standard fiber lasers, with an impressively low integrated linewidth of a few kHz and a Lorentzian linewidth of just 3 Hz. The laser design enables rapid frequency actuation through the stress-optic effect, with an actuation bandwidth extending from DC to the MHz range and maintaining a flat response up to 400~kHz. This high performance allows for precise linear frequency modulation sweeps, crucial for coherent sensing applications, achieving a residual nonlinearity of only 0.08\%—a capability realized only after photonic packaging.
	Our work highlights one of the largest piezo-actuators capable of MHz-level operation, demonstrating the robustness and uniformity of monolithic integration in piezo-MEMS-photonic chip technology. These MEMS actuators, with a footprint of 4.1 $\times$ 1.8 mm, are integrated with low-loss \SiN spirals, occupying over 20\% of the chip surface. Moving forward, we aim to develop an optimized packaging strategy to suppress parasitic bending and length extension modes, enabling the film bulk acoustic mode \cite{tian2020bulk} to dominate. This mode would provide uniform mechanical stress across the optical cavity, both near DC and at mechanical resonance.
	%The laser is realized as a fully packaged photonic assembly, significantly enhancing its performance. The laser's noise is less susceptible to external acoustics and vibrations, allowing us to explore the limitations of the SIL architecture.
	%We find that a simple theory describing an oscillator SIL to a high-Q cavity precisely predicts the frequency noise of the locked laser, offering insights into the behavior of ultra-low-noise optically locked systems.
	
	%We observed that tuning efficiency can be enhanced from 2 MHz/V to 5 MHz/V over a broad frequency range. However, this improvement introduces increased nonlinearity and not flat frequency response, suggesting a trade-off between tuning efficiency and actuation linearity.
	%Despite rigorous attempts at anchoring, the chip could still be utilized for packaged stress monitoring in a multi-chip module \cite{Temples23}.
	
	This laser system concept offers key major advancements. It overcomes the limitations of ultra-low noise laser systems, paving the way toward fully integrated Hz-level linewidth lasers and provides insights into laser behavior under active MHz-speed actuation within the laser cavity. 
	Applications in advanced optical and quantum metrology, coherent sensing (including coherent LiDAR, gyroscopes, and accelerometers), photonic and quantum computing, and other fields are heavily constrained by the performance and size of laser systems. Finally, silicon photonic FMCW LiDAR systems which currently rely on lasers based on discrete components can utilize photonic chip-based lasers. This work lays the groundwork for high-performance, chip-based optical coherent lasers suitable for both research and industrial applications, including fiber optic distributed sensing. Such advancements hold significant potential for industrial applications where laser noise performance and frequency agility are critical.
	
	\section*{Methods}
	
	%\subsection{Choice of a DFB laser}
	%The choice of a DFB laser diode depends on the targeted performance and trade-offs. High-power laser DFB diodes \cite{35lab_dfb, freedom_dfb} can deliver power up to 200 mW but they exhibit small mode field diameters (1-2 $\times$ 3-5 $\mu$m) of the emitter, which challenges the photonic packaging. On the other hand, low-power DFB diodes \cite{sivers_dfb, seminex_dfb} have powers up to 70 mW and a bigger mode field diameter (3-4 $\times$ 3-4 $\mu$m) which relaxes requirements for the active alignment and even allows passive alignment.    
	
	\subsection{Photonic packaging}
	
	All lasers, including the SIL 250 MHz and SIL 1.5 GHz models, are packaged through an active alignment process, beginning with the precise mounting of chips onto submounts. This step is essential for providing mechanical stability and acoustic isolation for the MEMS actuators, contributing directly to achieving a flat actuation response for the frequency-agile laser.
The photonic packaging itself is conducted using a custom-designed setup tailored specifically for these laser systems. This custom process enables optimization of the output power while preserving the unique characteristics of the SIL laser. High-precision epoxies, selected based on specific requirements, are used for various parts of the assembly to ensure durability and precision. Additionally, a single-channel fiber array unit is employed, incorporating a waveguide-array-to-fiber-transposer (WAFT) component. This component integrates a polarization-maintaining fiber, supports high power handling, and exhibits low internal insertion losses (0.4-0.6~dB). Some of the laser packages are hermetically sealed after packaging to enhance reliability and longevity.
	
	\subsection{Frequency response $\rm{S}_{21}$}
	In our setup, we measured the optomechanical $\rm{S}_{21}$ response using the slope of an spiral cavity optical mode. A continuous-wave (CW) laser is tuned to the linear slope of the cavity mode. The vector network analyzer (VNA) was configured with its output port connected to the piezoactuator, driving it across a range of frequencies, while its input port was connected to a photodetector that monitored the transmitted power along the cavity mode’s linear slope. As the optical frequency of the cavity shifted, this linear slope translated the frequency shift into a measurable variation in output power, enabling precise response characterization.
	
	\subsection{Frequency noise measurement}
	All frequency noise (FN) spectra are measured using a commercial laser phase noise analyzing system, OEWaves OE4000. The available system had an ultra-low noise option with the noise levels: 2500 \HzsHz at 10 Hz, 100~\HzsHz at 100~Hz, 4 \HzsHz at 1 kHz, 0.04 \HzsHz at 1 MHz. Alternatively, we measured the noise with two other systems. We performed IQ detection of the radio frequency beatnote signal of two lasers (heterodyne technique). First, we used a Menlo 1 Hz laser for heterodyne detection, but the measurement is limited by factors: (i) the Menlo laser had a fixed wavelength 1554.0 nm; (ii) the Menlo 1 Hz laser provides excellent FN below 10-100 kHz of frequencies offsets, i.e., is limited by a servo-bump of the electronic locking. Second, we developed an in-house external cavity diode laser locked using the Pound-Drever-Hall technique to an ultra-low expansion (ULE) cavity. This laser provides wide tunability and the locking scheme is optimized to reduce the servo bump. Thus, the FN at low frequency offsets (below 1 kHz) is higher than measured in this work. FN results obtained with different techniques are consistent and prove the limitations of different approaches. However, none of them allows measuring ultra-low FN level, i.e. 0.1-1 \HzsHz at all frequency offsets from 1 Hz to 1-10 GHz and they should be combined in the future.
\\	
	
\begin{footnotesize}
		
		%\noindent \textbf{Supplementary Material}: 
		
		\noindent \textbf{Author Contributions}:
		A.S., V.S., A.V., A.B., J.R. simulated and designed the devices.
		R.N.W. fabricated PICs. A.A. fabricated piezo-actuators.
		A.B., A.V., S.L., A.S., S.B. characterized the devices. S.L., A.B. performed the photonic packaging. 
		A.B., A.V., S.L., A.S., S.B. did the experiments and analyzed the data with the help of V.S., J.R., A.A.\,. A.V. wrote the manuscript with input from all authors. 
		T.J.K., S.B., and A.V. supervised the project.
		
		\noindent \textbf{Funding Information and Disclaimer}: This work was supported by was supported by the Swiss National Science Foundation under grant agreement No. 211728 (BRIDGE) and by the Horizon Europe EIC transition programme under grant No. 101137084 (FORTE). EPFL’s and Purdue’s contribution towards this work was also supported by contract W911NF2120248 (NINJA LASER) from the Defense Advanced Research Projects Agency (DARPA), Microsystems Technology Office (MTO). The views, opinions and/or findings expressed are those of the authors and should not be interpreted as representing the official views or policies of the Department of Defense or the U.S. Government. Dr. Sunil Bhave performed this research at Purdue University prior to becoming a DARPA program manager.
		
		\noindent \textbf{Acknowledgments}:
		The chip samples are fabricated in the EPFL center of MicroNanoTechnology (CMi), and in the Birck Nanotechnology Center at Purdue University. Aluminum Nitride was deposited at Advanced Modular Systems, Inc.
		
		\noindent \textbf{Data Availability Statement}: The code and data used to produce the plots within this work will be released on the repository \texttt{Zenodo} upon publication of this preprint.
		
		\noindent\textbf{Correspondence and requests for materials} should be addressed to T.J.K.
\end{footnotesize}
\bibliography{sample}

\begin{thebibliography}{56}%
\makeatletter
\providecommand \@ifxundefined [1]{%
 \@ifx{#1\undefined}
}%
\providecommand \@ifnum [1]{%
 \ifnum #1\expandafter \@firstoftwo
 \else \expandafter \@secondoftwo
 \fi
}%
\providecommand \@ifx [1]{%
 \ifx #1\expandafter \@firstoftwo
 \else \expandafter \@secondoftwo
 \fi
}%
\providecommand \natexlab [1]{#1}%
\providecommand \enquote  [1]{``#1''}%
\providecommand \bibnamefont  [1]{#1}%
\providecommand \bibfnamefont [1]{#1}%
\providecommand \citenamefont [1]{#1}%
\providecommand \href@noop [0]{\@secondoftwo}%
\providecommand \href [0]{\begingroup \@sanitize@url \@href}%
\providecommand \@href[1]{\@@startlink{#1}\@@href}%
\providecommand \@@href[1]{\endgroup#1\@@endlink}%
\providecommand \@sanitize@url [0]{\catcode `\\12\catcode `\$12\catcode
  `\&12\catcode `\#12\catcode `\^12\catcode `\_12\catcode `\%12\relax}%
\providecommand \@@startlink[1]{}%
\providecommand \@@endlink[0]{}%
\providecommand \url  [0]{\begingroup\@sanitize@url \@url }%
\providecommand \@url [1]{\endgroup\@href {#1}{\urlprefix }}%
\providecommand \urlprefix  [0]{URL }%
\providecommand \Eprint [0]{\href }%
\providecommand \doibase [0]{https://doi.org/}%
\providecommand \selectlanguage [0]{\@gobble}%
\providecommand \bibinfo  [0]{\@secondoftwo}%
\providecommand \bibfield  [0]{\@secondoftwo}%
\providecommand \translation [1]{[#1]}%
\providecommand \BibitemOpen [0]{}%
\providecommand \bibitemStop [0]{}%
\providecommand \bibitemNoStop [0]{.\EOS\space}%
\providecommand \EOS [0]{\spacefactor3000\relax}%
\providecommand \BibitemShut  [1]{\csname bibitem#1\endcsname}%
\let\auto@bib@innerbib\@empty
%</preamble>
\bibitem [{\citenamefont {Kikuchi}(2016)}]{Kikuchi_fundametals_communications}%
  \BibitemOpen
  \bibfield  {author} {\bibinfo {author} {\bibfnamefont {K.}~\bibnamefont
  {Kikuchi}},\ }\bibfield  {title} {\bibinfo {title} {Fundamentals of coherent
  optical fiber communications},\ }\href
  {https://opg.optica.org/jlt/abstract.cfm?URI=jlt-34-1-157} {\bibfield
  {journal} {\bibinfo  {journal} {J. Lightwave Technol.}\ }\textbf {\bibinfo
  {volume} {34}},\ \bibinfo {pages} {157} (\bibinfo {year} {2016})}\BibitemShut
  {NoStop}%
\bibitem [{\citenamefont {Armstrong}(2009)}]{OFDM}%
  \BibitemOpen
  \bibfield  {author} {\bibinfo {author} {\bibfnamefont {J.}~\bibnamefont
  {Armstrong}},\ }\bibfield  {title} {\bibinfo {title} {Ofdm for optical
  communications},\ }\href {https://doi.org/10.1109/JLT.2008.2010061}
  {\bibfield  {journal} {\bibinfo  {journal} {Journal of Lightwave Technology}\
  }\textbf {\bibinfo {volume} {27}},\ \bibinfo {pages} {189} (\bibinfo {year}
  {2009})}\BibitemShut {NoStop}%
\bibitem [{\citenamefont {MacDonald}(1981)}]{FDOR_1981}%
  \BibitemOpen
  \bibfield  {author} {\bibinfo {author} {\bibfnamefont {R.~I.}\ \bibnamefont
  {MacDonald}},\ }\bibfield  {title} {\bibinfo {title} {Frequency domain
  optical reflectometer},\ }\href {https://doi.org/10.1364/AO.20.001840}
  {\bibfield  {journal} {\bibinfo  {journal} {Appl. Opt.}\ }\textbf {\bibinfo
  {volume} {20}},\ \bibinfo {pages} {1840} (\bibinfo {year}
  {1981})}\BibitemShut {NoStop}%
\bibitem [{\citenamefont {Ip}\ and\ \citenamefont
  {Kahn}(2007)}]{Carrier_Recovery_2007}%
  \BibitemOpen
  \bibfield  {author} {\bibinfo {author} {\bibfnamefont {E.}~\bibnamefont
  {Ip}}\ and\ \bibinfo {author} {\bibfnamefont {J.~M.}\ \bibnamefont {Kahn}},\
  }\bibfield  {title} {\bibinfo {title} {Feedforward carrier recovery for
  coherent optical communications},\ }\href
  {https://doi.org/10.1109/JLT.2007.902118} {\bibfield  {journal} {\bibinfo
  {journal} {Journal of Lightwave Technology}\ }\textbf {\bibinfo {volume}
  {25}},\ \bibinfo {pages} {2675} (\bibinfo {year} {2007})}\BibitemShut
  {NoStop}%
\bibitem [{\citenamefont {Degen}\ \emph {et~al.}(2017)\citenamefont {Degen},
  \citenamefont {Reinhard},\ and\ \citenamefont
  {Cappellaro}}]{Cappellaro_quantum_sensing}%
  \BibitemOpen
  \bibfield  {author} {\bibinfo {author} {\bibfnamefont {C.~L.}\ \bibnamefont
  {Degen}}, \bibinfo {author} {\bibfnamefont {F.}~\bibnamefont {Reinhard}},\
  and\ \bibinfo {author} {\bibfnamefont {P.}~\bibnamefont {Cappellaro}},\
  }\bibfield  {title} {\bibinfo {title} {Quantum sensing},\ }\href
  {https://doi.org/10.1103/RevModPhys.89.035002} {\bibfield  {journal}
  {\bibinfo  {journal} {Rev. Mod. Phys.}\ }\textbf {\bibinfo {volume} {89}},\
  \bibinfo {pages} {035002} (\bibinfo {year} {2017})}\BibitemShut {NoStop}%
\bibitem [{\citenamefont {Ludlow}\ \emph {et~al.}(2015)\citenamefont {Ludlow},
  \citenamefont {Boyd}, \citenamefont {Ye}, \citenamefont {Peik},\ and\
  \citenamefont {Schmidt}}]{Ludlow_atomic_clocks}%
  \BibitemOpen
  \bibfield  {author} {\bibinfo {author} {\bibfnamefont {A.~D.}\ \bibnamefont
  {Ludlow}}, \bibinfo {author} {\bibfnamefont {M.~M.}\ \bibnamefont {Boyd}},
  \bibinfo {author} {\bibfnamefont {J.}~\bibnamefont {Ye}}, \bibinfo {author}
  {\bibfnamefont {E.}~\bibnamefont {Peik}},\ and\ \bibinfo {author}
  {\bibfnamefont {P.~O.}\ \bibnamefont {Schmidt}},\ }\bibfield  {title}
  {\bibinfo {title} {Optical atomic clocks},\ }\href
  {https://doi.org/10.1103/RevModPhys.87.637} {\bibfield  {journal} {\bibinfo
  {journal} {Rev. Mod. Phys.}\ }\textbf {\bibinfo {volume} {87}},\ \bibinfo
  {pages} {637} (\bibinfo {year} {2015})}\BibitemShut {NoStop}%
\bibitem [{\citenamefont {Bostick}(1967)}]{laser_radar}%
  \BibitemOpen
  \bibfield  {author} {\bibinfo {author} {\bibfnamefont {H.}~\bibnamefont
  {Bostick}},\ }\bibfield  {title} {\bibinfo {title} {A carbon dioxide laser
  radar system},\ }\href {https://doi.org/10.1109/JQE.1967.1074540} {\bibfield
  {journal} {\bibinfo  {journal} {IEEE Journal of Quantum Electronics}\
  }\textbf {\bibinfo {volume} {3}},\ \bibinfo {pages} {232} (\bibinfo {year}
  {1967})}\BibitemShut {NoStop}%
\bibitem [{\citenamefont {Rogers}(1999)}]{rogers1999distributed}%
  \BibitemOpen
  \bibfield  {author} {\bibinfo {author} {\bibfnamefont {A.}~\bibnamefont
  {Rogers}},\ }\bibfield  {title} {\bibinfo {title} {Distributed optical-fibre
  sensing},\ }\href@noop {} {\bibfield  {journal} {\bibinfo  {journal}
  {Measurement Science and Technology}\ }\textbf {\bibinfo {volume} {10}},\
  \bibinfo {pages} {R75} (\bibinfo {year} {1999})}\BibitemShut {NoStop}%
\bibitem [{\citenamefont {Dell’Olio}\ \emph {et~al.}(2023)\citenamefont
  {Dell’Olio}, \citenamefont {Natale}, \citenamefont {Wang},\ and\
  \citenamefont {Hung}}]{optical_gyro_review}%
  \BibitemOpen
  \bibfield  {author} {\bibinfo {author} {\bibfnamefont {F.}~\bibnamefont
  {Dell’Olio}}, \bibinfo {author} {\bibfnamefont {T.}~\bibnamefont {Natale}},
  \bibinfo {author} {\bibfnamefont {Y.-C.}\ \bibnamefont {Wang}},\ and\
  \bibinfo {author} {\bibfnamefont {Y.-J.}\ \bibnamefont {Hung}},\ }\bibfield
  {title} {\bibinfo {title} {Miniaturization of interferometric optical
  gyroscopes: A review},\ }\href {https://doi.org/10.1109/JSEN.2023.3327217}
  {\bibfield  {journal} {\bibinfo  {journal} {IEEE Sensors Journal}\ }\textbf
  {\bibinfo {volume} {PP}},\ \bibinfo {pages} {1} (\bibinfo {year}
  {2023})}\BibitemShut {NoStop}%
\bibitem [{\citenamefont {Zhang}\ \emph
  {et~al.}(2023{\natexlab{a}})\citenamefont {Zhang}, \citenamefont {Nagata},
  \citenamefont {Sharifuddin}, \citenamefont {Ito}, \citenamefont {Nakamura},\
  and\ \citenamefont {Koshikiya}}]{zhang2023long}%
  \BibitemOpen
  \bibfield  {author} {\bibinfo {author} {\bibfnamefont {C.}~\bibnamefont
  {Zhang}}, \bibinfo {author} {\bibfnamefont {T.}~\bibnamefont {Nagata}},
  \bibinfo {author} {\bibfnamefont {M.~S. B.~A.}\ \bibnamefont {Sharifuddin}},
  \bibinfo {author} {\bibfnamefont {F.}~\bibnamefont {Ito}}, \bibinfo {author}
  {\bibfnamefont {A.}~\bibnamefont {Nakamura}},\ and\ \bibinfo {author}
  {\bibfnamefont {Y.}~\bibnamefont {Koshikiya}},\ }\bibfield  {title} {\bibinfo
  {title} {Long-range frequency-modulated continuous-wave lidar employing
  wavelength-swept optical frequency comb},\ }\href@noop {} {\bibfield
  {journal} {\bibinfo  {journal} {Optics Communications}\ }\textbf {\bibinfo
  {volume} {545}},\ \bibinfo {pages} {129702} (\bibinfo {year}
  {2023}{\natexlab{a}})}\BibitemShut {NoStop}%
\bibitem [{\citenamefont {Martin}\ \emph {et~al.}(2018)\citenamefont {Martin},
  \citenamefont {Dodane}, \citenamefont {Leviandier}, \citenamefont {Dolfi},
  \citenamefont {Naughton}, \citenamefont {O'Brien}, \citenamefont {Spuessens},
  \citenamefont {Baets}, \citenamefont {Lepage}, \citenamefont {Verheyen},
  \citenamefont {Heyn}, \citenamefont {Absil}, \citenamefont {Feneyrou},\ and\
  \citenamefont {Bourderionnet}}]{Martin18}%
  \BibitemOpen
  \bibfield  {author} {\bibinfo {author} {\bibfnamefont {A.}~\bibnamefont
  {Martin}}, \bibinfo {author} {\bibfnamefont {D.}~\bibnamefont {Dodane}},
  \bibinfo {author} {\bibfnamefont {L.}~\bibnamefont {Leviandier}}, \bibinfo
  {author} {\bibfnamefont {D.}~\bibnamefont {Dolfi}}, \bibinfo {author}
  {\bibfnamefont {A.}~\bibnamefont {Naughton}}, \bibinfo {author}
  {\bibfnamefont {P.}~\bibnamefont {O'Brien}}, \bibinfo {author} {\bibfnamefont
  {T.}~\bibnamefont {Spuessens}}, \bibinfo {author} {\bibfnamefont
  {R.}~\bibnamefont {Baets}}, \bibinfo {author} {\bibfnamefont
  {G.}~\bibnamefont {Lepage}}, \bibinfo {author} {\bibfnamefont
  {P.}~\bibnamefont {Verheyen}}, \bibinfo {author} {\bibfnamefont {P.~D.}\
  \bibnamefont {Heyn}}, \bibinfo {author} {\bibfnamefont {P.}~\bibnamefont
  {Absil}}, \bibinfo {author} {\bibfnamefont {P.}~\bibnamefont {Feneyrou}},\
  and\ \bibinfo {author} {\bibfnamefont {J.}~\bibnamefont {Bourderionnet}},\
  }\bibfield  {title} {\bibinfo {title} {Photonic integrated circuit-based fmcw
  coherent lidar},\ }\href
  {https://opg.optica.org/jlt/abstract.cfm?URI=jlt-36-19-4640} {\bibfield
  {journal} {\bibinfo  {journal} {J. Lightwave Technol.}\ }\textbf {\bibinfo
  {volume} {36}},\ \bibinfo {pages} {4640} (\bibinfo {year}
  {2018})}\BibitemShut {NoStop}%
\bibitem [{\citenamefont {Rogers}\ \emph {et~al.}(2021)\citenamefont {Rogers},
  \citenamefont {Piggott}, \citenamefont {Thomson}, \citenamefont {Wiser},
  \citenamefont {Opris}, \citenamefont {Fortune}, \citenamefont {Compston},
  \citenamefont {Gondarenko}, \citenamefont {Meng}, \citenamefont {Chen} \emph
  {et~al.}}]{rogers2021universal}%
  \BibitemOpen
  \bibfield  {author} {\bibinfo {author} {\bibfnamefont {C.}~\bibnamefont
  {Rogers}}, \bibinfo {author} {\bibfnamefont {A.~Y.}\ \bibnamefont {Piggott}},
  \bibinfo {author} {\bibfnamefont {D.~J.}\ \bibnamefont {Thomson}}, \bibinfo
  {author} {\bibfnamefont {R.~F.}\ \bibnamefont {Wiser}}, \bibinfo {author}
  {\bibfnamefont {I.~E.}\ \bibnamefont {Opris}}, \bibinfo {author}
  {\bibfnamefont {S.~A.}\ \bibnamefont {Fortune}}, \bibinfo {author}
  {\bibfnamefont {A.~J.}\ \bibnamefont {Compston}}, \bibinfo {author}
  {\bibfnamefont {A.}~\bibnamefont {Gondarenko}}, \bibinfo {author}
  {\bibfnamefont {F.}~\bibnamefont {Meng}}, \bibinfo {author} {\bibfnamefont
  {X.}~\bibnamefont {Chen}}, \emph {et~al.},\ }\bibfield  {title} {\bibinfo
  {title} {A universal 3d imaging sensor on a silicon photonics platform},\
  }\href@noop {} {\bibfield  {journal} {\bibinfo  {journal} {Nature}\ }\textbf
  {\bibinfo {volume} {590}},\ \bibinfo {pages} {256} (\bibinfo {year}
  {2021})}\BibitemShut {NoStop}%
\bibitem [{\citenamefont {Pfeiffer}\ \emph {et~al.}(2016)\citenamefont
  {Pfeiffer}, \citenamefont {Kordts}, \citenamefont {Brasch}, \citenamefont
  {Zervas}, \citenamefont {Geiselmann}, \citenamefont {Jost},\ and\
  \citenamefont {Kippenberg}}]{pfeiffer2016damascene}%
  \BibitemOpen
  \bibfield  {author} {\bibinfo {author} {\bibfnamefont {M.~H.~P.}\
  \bibnamefont {Pfeiffer}}, \bibinfo {author} {\bibfnamefont {A.}~\bibnamefont
  {Kordts}}, \bibinfo {author} {\bibfnamefont {V.}~\bibnamefont {Brasch}},
  \bibinfo {author} {\bibfnamefont {M.}~\bibnamefont {Zervas}}, \bibinfo
  {author} {\bibfnamefont {M.}~\bibnamefont {Geiselmann}}, \bibinfo {author}
  {\bibfnamefont {J.~D.}\ \bibnamefont {Jost}},\ and\ \bibinfo {author}
  {\bibfnamefont {T.~J.}\ \bibnamefont {Kippenberg}},\ }\bibfield  {title}
  {\bibinfo {title} {Photonic damascene process for integrated high-q
  microresonator based nonlinear photonics},\ }\href
  {https://doi.org/10.1364/OPTICA.3.000020} {\bibfield  {journal} {\bibinfo
  {journal} {Optica}\ }\textbf {\bibinfo {volume} {3}},\ \bibinfo {pages} {20}
  (\bibinfo {year} {2016})}\BibitemShut {NoStop}%
\bibitem [{\citenamefont {Liu}\ \emph {et~al.}(2021)\citenamefont {Liu},
  \citenamefont {Huang}, \citenamefont {Wang}, \citenamefont {He},
  \citenamefont {Raja}, \citenamefont {Liu}, \citenamefont {Engelsen},\ and\
  \citenamefont {Kippenberg}}]{liu2021highyield}%
  \BibitemOpen
  \bibfield  {author} {\bibinfo {author} {\bibfnamefont {J.}~\bibnamefont
  {Liu}}, \bibinfo {author} {\bibfnamefont {G.}~\bibnamefont {Huang}}, \bibinfo
  {author} {\bibfnamefont {R.~N.}\ \bibnamefont {Wang}}, \bibinfo {author}
  {\bibfnamefont {J.}~\bibnamefont {He}}, \bibinfo {author} {\bibfnamefont
  {A.~S.}\ \bibnamefont {Raja}}, \bibinfo {author} {\bibfnamefont
  {T.}~\bibnamefont {Liu}}, \bibinfo {author} {\bibfnamefont {N.~J.}\
  \bibnamefont {Engelsen}},\ and\ \bibinfo {author} {\bibfnamefont {T.~J.}\
  \bibnamefont {Kippenberg}},\ }\bibfield  {title} {\bibinfo {title}
  {High-yield, wafer-scale fabrication of ultralow-loss, dispersion-engineered
  silicon nitride photonic circuits},\ }\href@noop {} {\bibfield  {journal}
  {\bibinfo  {journal} {Nature communications}\ }\textbf {\bibinfo {volume}
  {12}},\ \bibinfo {pages} {2236} (\bibinfo {year} {2021})}\BibitemShut
  {NoStop}%
\bibitem [{\citenamefont {Bauters}\ \emph {et~al.}(2011)\citenamefont
  {Bauters}, \citenamefont {Heck}, \citenamefont {John}, \citenamefont {Dai},
  \citenamefont {Tien}, \citenamefont {Barton}, \citenamefont {Leinse},
  \citenamefont {Heideman}, \citenamefont {Blumenthal},\ and\ \citenamefont
  {Bowers}}]{blumenthal2011lowloss}%
  \BibitemOpen
  \bibfield  {author} {\bibinfo {author} {\bibfnamefont {J.~F.}\ \bibnamefont
  {Bauters}}, \bibinfo {author} {\bibfnamefont {M.~J.~R.}\ \bibnamefont
  {Heck}}, \bibinfo {author} {\bibfnamefont {D.}~\bibnamefont {John}}, \bibinfo
  {author} {\bibfnamefont {D.}~\bibnamefont {Dai}}, \bibinfo {author}
  {\bibfnamefont {M.-C.}\ \bibnamefont {Tien}}, \bibinfo {author}
  {\bibfnamefont {J.~S.}\ \bibnamefont {Barton}}, \bibinfo {author}
  {\bibfnamefont {A.}~\bibnamefont {Leinse}}, \bibinfo {author} {\bibfnamefont
  {R.~G.}\ \bibnamefont {Heideman}}, \bibinfo {author} {\bibfnamefont {D.~J.}\
  \bibnamefont {Blumenthal}},\ and\ \bibinfo {author} {\bibfnamefont {J.~E.}\
  \bibnamefont {Bowers}},\ }\bibfield  {title} {\bibinfo {title}
  {Ultra-low-loss high-aspect-ratio si3n4 waveguides},\ }\href
  {https://doi.org/10.1364/OE.19.003163} {\bibfield  {journal} {\bibinfo
  {journal} {Opt. Express}\ }\textbf {\bibinfo {volume} {19}},\ \bibinfo
  {pages} {3163} (\bibinfo {year} {2011})}\BibitemShut {NoStop}%
\bibitem [{\citenamefont {Siddharth}\ \emph {et~al.}(2022)\citenamefont
  {Siddharth}, \citenamefont {Wunderer}, \citenamefont {Lihachev},
  \citenamefont {Voloshin}, \citenamefont {Haller}, \citenamefont {Wang},
  \citenamefont {Teepe}, \citenamefont {Yang}, \citenamefont {Liu},
  \citenamefont {Riemensberger}, \citenamefont {Grandjean}, \citenamefont
  {Johnson},\ and\ \citenamefont {Kippenberg}}]{anat2022bluelaser}%
  \BibitemOpen
  \bibfield  {author} {\bibinfo {author} {\bibfnamefont {A.}~\bibnamefont
  {Siddharth}}, \bibinfo {author} {\bibfnamefont {T.}~\bibnamefont {Wunderer}},
  \bibinfo {author} {\bibfnamefont {G.}~\bibnamefont {Lihachev}}, \bibinfo
  {author} {\bibfnamefont {A.~S.}\ \bibnamefont {Voloshin}}, \bibinfo {author}
  {\bibfnamefont {C.}~\bibnamefont {Haller}}, \bibinfo {author} {\bibfnamefont
  {R.~N.}\ \bibnamefont {Wang}}, \bibinfo {author} {\bibfnamefont
  {M.}~\bibnamefont {Teepe}}, \bibinfo {author} {\bibfnamefont
  {Z.}~\bibnamefont {Yang}}, \bibinfo {author} {\bibfnamefont {J.}~\bibnamefont
  {Liu}}, \bibinfo {author} {\bibfnamefont {J.}~\bibnamefont {Riemensberger}},
  \bibinfo {author} {\bibfnamefont {N.}~\bibnamefont {Grandjean}}, \bibinfo
  {author} {\bibfnamefont {N.}~\bibnamefont {Johnson}},\ and\ \bibinfo {author}
  {\bibfnamefont {T.~J.}\ \bibnamefont {Kippenberg}},\ }\bibfield  {title}
  {\bibinfo {title} {{Near ultraviolet photonic integrated lasers based on
  silicon nitride}},\ }\href {https://doi.org/10.1063/5.0081660} {\bibfield
  {journal} {\bibinfo  {journal} {APL Photonics}\ }\textbf {\bibinfo {volume}
  {7}},\ \bibinfo {pages} {046108} (\bibinfo {year} {2022})}\BibitemShut
  {NoStop}%
\bibitem [{\citenamefont {Winkler}\ \emph {et~al.}(2024)\citenamefont
  {Winkler}, \citenamefont {Gerritsma}, \citenamefont {van Rees}, \citenamefont
  {Schrinner}, \citenamefont {Hoekman}, \citenamefont {Dekker}, \citenamefont
  {Nascimento~Jr}, \citenamefont {van~der Slot}, \citenamefont {N{\"o}lleke},\
  and\ \citenamefont {Boller}}]{winkler2024redlaser}%
  \BibitemOpen
  \bibfield  {author} {\bibinfo {author} {\bibfnamefont {L.~V.}\ \bibnamefont
  {Winkler}}, \bibinfo {author} {\bibfnamefont {K.}~\bibnamefont {Gerritsma}},
  \bibinfo {author} {\bibfnamefont {A.}~\bibnamefont {van Rees}}, \bibinfo
  {author} {\bibfnamefont {P.~P.}\ \bibnamefont {Schrinner}}, \bibinfo {author}
  {\bibfnamefont {M.}~\bibnamefont {Hoekman}}, \bibinfo {author} {\bibfnamefont
  {R.}~\bibnamefont {Dekker}}, \bibinfo {author} {\bibfnamefont {A.~R.~d.}\
  \bibnamefont {Nascimento~Jr}}, \bibinfo {author} {\bibfnamefont {P.~J.}\
  \bibnamefont {van~der Slot}}, \bibinfo {author} {\bibfnamefont
  {C.}~\bibnamefont {N{\"o}lleke}},\ and\ \bibinfo {author} {\bibfnamefont
  {K.-J.}\ \bibnamefont {Boller}},\ }\bibfield  {title} {\bibinfo {title}
  {Widely-tunable and narrow-linewidth hybrid-integrated diode laser at 637
  nm},\ }\href@noop {} {\bibfield  {journal} {\bibinfo  {journal} {arXiv
  preprint arXiv:2404.05325}\ } (\bibinfo {year} {2024})}\BibitemShut {NoStop}%
\bibitem [{\citenamefont {Isichenko}\ \emph {et~al.}(2023)\citenamefont
  {Isichenko}, \citenamefont {Chauhan}, \citenamefont {Liu}, \citenamefont
  {Harrington},\ and\ \citenamefont {Blumenthal}}]{isichenko2023chip}%
  \BibitemOpen
  \bibfield  {author} {\bibinfo {author} {\bibfnamefont {A.}~\bibnamefont
  {Isichenko}}, \bibinfo {author} {\bibfnamefont {N.}~\bibnamefont {Chauhan}},
  \bibinfo {author} {\bibfnamefont {K.}~\bibnamefont {Liu}}, \bibinfo {author}
  {\bibfnamefont {M.~W.}\ \bibnamefont {Harrington}},\ and\ \bibinfo {author}
  {\bibfnamefont {D.~J.}\ \bibnamefont {Blumenthal}},\ }\bibfield  {title}
  {\bibinfo {title} {Chip-scale, sub-hz fundamental sub-khz integral linewidth
  780 nm laser through self-injection-locking a fabry-p$\backslash$'erot laser
  to an ultra-high q integrated resonator},\ }\href@noop {} {\bibfield
  {journal} {\bibinfo  {journal} {arXiv preprint arXiv:2307.04947}\ } (\bibinfo
  {year} {2023})}\BibitemShut {NoStop}%
\bibitem [{\citenamefont {Li}\ \emph {et~al.}(2023{\natexlab{a}})\citenamefont
  {Li}, \citenamefont {Yuan}, \citenamefont {Jin}, \citenamefont {Wu},
  \citenamefont {Guo}, \citenamefont {Ji}, \citenamefont {Feshali},
  \citenamefont {Paniccia}, \citenamefont {Bowers},\ and\ \citenamefont
  {Vahala}}]{780Li23}%
  \BibitemOpen
  \bibfield  {author} {\bibinfo {author} {\bibfnamefont {B.}~\bibnamefont
  {Li}}, \bibinfo {author} {\bibfnamefont {Z.}~\bibnamefont {Yuan}}, \bibinfo
  {author} {\bibfnamefont {W.}~\bibnamefont {Jin}}, \bibinfo {author}
  {\bibfnamefont {L.}~\bibnamefont {Wu}}, \bibinfo {author} {\bibfnamefont
  {J.}~\bibnamefont {Guo}}, \bibinfo {author} {\bibfnamefont {Q.-X.}\
  \bibnamefont {Ji}}, \bibinfo {author} {\bibfnamefont {A.}~\bibnamefont
  {Feshali}}, \bibinfo {author} {\bibfnamefont {M.}~\bibnamefont {Paniccia}},
  \bibinfo {author} {\bibfnamefont {J.~E.}\ \bibnamefont {Bowers}},\ and\
  \bibinfo {author} {\bibfnamefont {K.~J.}\ \bibnamefont {Vahala}},\ }\bibfield
   {title} {\bibinfo {title} {High-coherence hybrid-integrated
  780\&\#x2009;\&\#x2009;nm source by self-injection-locked second-harmonic
  generation in a high-q silicon-nitride resonator},\ }\href
  {https://doi.org/10.1364/OPTICA.498391} {\bibfield  {journal} {\bibinfo
  {journal} {Optica}\ }\textbf {\bibinfo {volume} {10}},\ \bibinfo {pages}
  {1241} (\bibinfo {year} {2023}{\natexlab{a}})}\BibitemShut {NoStop}%
\bibitem [{\citenamefont {Prokoshin}\ \emph {et~al.}(2024)\citenamefont
  {Prokoshin}, \citenamefont {Gehl}, \citenamefont {Madaras}, \citenamefont
  {Chow},\ and\ \citenamefont {Wan}}]{780Prokoshin24}%
  \BibitemOpen
  \bibfield  {author} {\bibinfo {author} {\bibfnamefont {A.}~\bibnamefont
  {Prokoshin}}, \bibinfo {author} {\bibfnamefont {M.}~\bibnamefont {Gehl}},
  \bibinfo {author} {\bibfnamefont {S.}~\bibnamefont {Madaras}}, \bibinfo
  {author} {\bibfnamefont {W.~W.}\ \bibnamefont {Chow}},\ and\ \bibinfo
  {author} {\bibfnamefont {Y.}~\bibnamefont {Wan}},\ }\bibfield  {title}
  {\bibinfo {title} {Ultra-narrow-linewidth hybrid-integrated self-injection
  locked laser at 780 nm},\ }\href {https://doi.org/10.1364/OPTICA.531152}
  {\bibfield  {journal} {\bibinfo  {journal} {Optica}\ }\textbf {\bibinfo
  {volume} {11}},\ \bibinfo {pages} {1024} (\bibinfo {year}
  {2024})}\BibitemShut {NoStop}%
\bibitem [{\citenamefont {Franken}\ \emph {et~al.}(2021)\citenamefont
  {Franken}, \citenamefont {van Rees}, \citenamefont {Winkler}, \citenamefont
  {Fan}, \citenamefont {Geskus}, \citenamefont {Dekker}, \citenamefont
  {Geuzebroek}, \citenamefont {Fallnich}, \citenamefont {van~der Slot},\ and\
  \citenamefont {Boller}}]{Franken:21}%
  \BibitemOpen
  \bibfield  {author} {\bibinfo {author} {\bibfnamefont {C.~A.~A.}\
  \bibnamefont {Franken}}, \bibinfo {author} {\bibfnamefont {A.}~\bibnamefont
  {van Rees}}, \bibinfo {author} {\bibfnamefont {L.~V.}\ \bibnamefont
  {Winkler}}, \bibinfo {author} {\bibfnamefont {Y.}~\bibnamefont {Fan}},
  \bibinfo {author} {\bibfnamefont {D.}~\bibnamefont {Geskus}}, \bibinfo
  {author} {\bibfnamefont {R.}~\bibnamefont {Dekker}}, \bibinfo {author}
  {\bibfnamefont {D.~H.}\ \bibnamefont {Geuzebroek}}, \bibinfo {author}
  {\bibfnamefont {C.}~\bibnamefont {Fallnich}}, \bibinfo {author}
  {\bibfnamefont {P.~J.~M.}\ \bibnamefont {van~der Slot}},\ and\ \bibinfo
  {author} {\bibfnamefont {K.-J.}\ \bibnamefont {Boller}},\ }\bibfield  {title}
  {\bibinfo {title} {Hybrid-integrated diode laser in the visible spectral
  range},\ }\href {https://doi.org/10.1364/OL.433636} {\bibfield  {journal}
  {\bibinfo  {journal} {Opt. Lett.}\ }\textbf {\bibinfo {volume} {46}},\
  \bibinfo {pages} {4904} (\bibinfo {year} {2021})}\BibitemShut {NoStop}%
\bibitem [{\citenamefont {Frentrop}\ \emph {et~al.}(2023)\citenamefont
  {Frentrop}, \citenamefont {Schilder}, \citenamefont {Hegeman}, \citenamefont
  {Everhardt}, \citenamefont {Klein}, \citenamefont {Geuzebroek}, \citenamefont
  {Winkler}, \citenamefont {Ensher}, \citenamefont {Heideman},\ and\
  \citenamefont {Kelly}}]{frentrop2023800}%
  \BibitemOpen
  \bibfield  {author} {\bibinfo {author} {\bibfnamefont {R.}~\bibnamefont
  {Frentrop}}, \bibinfo {author} {\bibfnamefont {N.~A.}\ \bibnamefont
  {Schilder}}, \bibinfo {author} {\bibfnamefont {I.}~\bibnamefont {Hegeman}},
  \bibinfo {author} {\bibfnamefont {A.~S.}\ \bibnamefont {Everhardt}}, \bibinfo
  {author} {\bibfnamefont {E.~J.}\ \bibnamefont {Klein}}, \bibinfo {author}
  {\bibfnamefont {D.~H.}\ \bibnamefont {Geuzebroek}}, \bibinfo {author}
  {\bibfnamefont {L.~V.}\ \bibnamefont {Winkler}}, \bibinfo {author}
  {\bibfnamefont {J.}~\bibnamefont {Ensher}}, \bibinfo {author} {\bibfnamefont
  {R.~G.}\ \bibnamefont {Heideman}},\ and\ \bibinfo {author} {\bibfnamefont
  {C.}~\bibnamefont {Kelly}},\ }\bibfield  {title} {\bibinfo {title} {800 nm
  narrow linewidth tunable hybrid laser based on a dual micro-ring external
  cavity},\ }in\ \href@noop {} {\emph {\bibinfo {booktitle} {Novel In-Plane
  Semiconductor Lasers XXII}}},\ Vol.\ \bibinfo {volume} {12440}\ (\bibinfo
  {organization} {SPIE},\ \bibinfo {year} {2023})\ pp.\ \bibinfo {pages}
  {5--8}\BibitemShut {NoStop}%
\bibitem [{\citenamefont {Guo}\ \emph {et~al.}(2022)\citenamefont {Guo},
  \citenamefont {McLemore}, \citenamefont {Xiang}, \citenamefont {Lee},
  \citenamefont {Wu}, \citenamefont {Jin}, \citenamefont {Kelleher},
  \citenamefont {Jin}, \citenamefont {Mason}, \citenamefont {Chang} \emph
  {et~al.}}]{bower_hertz_integrated}%
  \BibitemOpen
  \bibfield  {author} {\bibinfo {author} {\bibfnamefont {J.}~\bibnamefont
  {Guo}}, \bibinfo {author} {\bibfnamefont {C.~A.}\ \bibnamefont {McLemore}},
  \bibinfo {author} {\bibfnamefont {C.}~\bibnamefont {Xiang}}, \bibinfo
  {author} {\bibfnamefont {D.}~\bibnamefont {Lee}}, \bibinfo {author}
  {\bibfnamefont {L.}~\bibnamefont {Wu}}, \bibinfo {author} {\bibfnamefont
  {W.}~\bibnamefont {Jin}}, \bibinfo {author} {\bibfnamefont {M.}~\bibnamefont
  {Kelleher}}, \bibinfo {author} {\bibfnamefont {N.}~\bibnamefont {Jin}},
  \bibinfo {author} {\bibfnamefont {D.}~\bibnamefont {Mason}}, \bibinfo
  {author} {\bibfnamefont {L.}~\bibnamefont {Chang}}, \emph {et~al.},\
  }\bibfield  {title} {\bibinfo {title} {Chip-based laser with 1-hertz
  integrated linewidth},\ }\href@noop {} {\bibfield  {journal} {\bibinfo
  {journal} {Science advances}\ }\textbf {\bibinfo {volume} {8}},\ \bibinfo
  {pages} {eabp9006} (\bibinfo {year} {2022})}\BibitemShut {NoStop}%
\bibitem [{\citenamefont {Li}\ \emph {et~al.}(2021)\citenamefont {Li},
  \citenamefont {Jin}, \citenamefont {Wu}, \citenamefont {Chang}, \citenamefont
  {Wang}, \citenamefont {Shen}, \citenamefont {Yuan}, \citenamefont {Feshali},
  \citenamefont {Paniccia}, \citenamefont {Vahala},\ and\ \citenamefont
  {Bowers}}]{bowers2021reaching}%
  \BibitemOpen
  \bibfield  {author} {\bibinfo {author} {\bibfnamefont {B.}~\bibnamefont
  {Li}}, \bibinfo {author} {\bibfnamefont {W.}~\bibnamefont {Jin}}, \bibinfo
  {author} {\bibfnamefont {L.}~\bibnamefont {Wu}}, \bibinfo {author}
  {\bibfnamefont {L.}~\bibnamefont {Chang}}, \bibinfo {author} {\bibfnamefont
  {H.}~\bibnamefont {Wang}}, \bibinfo {author} {\bibfnamefont {B.}~\bibnamefont
  {Shen}}, \bibinfo {author} {\bibfnamefont {Z.}~\bibnamefont {Yuan}}, \bibinfo
  {author} {\bibfnamefont {A.}~\bibnamefont {Feshali}}, \bibinfo {author}
  {\bibfnamefont {M.}~\bibnamefont {Paniccia}}, \bibinfo {author}
  {\bibfnamefont {K.~J.}\ \bibnamefont {Vahala}},\ and\ \bibinfo {author}
  {\bibfnamefont {J.~E.}\ \bibnamefont {Bowers}},\ }\bibfield  {title}
  {\bibinfo {title} {Reaching fiber-laser coherence in integrated photonics},\
  }\href {https://doi.org/10.1364/OL.439720} {\bibfield  {journal} {\bibinfo
  {journal} {Opt. Lett.}\ }\textbf {\bibinfo {volume} {46}},\ \bibinfo {pages}
  {5201} (\bibinfo {year} {2021})}\BibitemShut {NoStop}%
\bibitem [{\citenamefont {Siddharth}\ \emph
  {et~al.}(2024{\natexlab{a}})\citenamefont {Siddharth}, \citenamefont
  {Attanasio}, \citenamefont {Bianconi}, \citenamefont {Lihachev},
  \citenamefont {Zhang}, \citenamefont {Qiu}, \citenamefont {Bancora},
  \citenamefont {Kenning}, \citenamefont {Wang}, \citenamefont {Voloshin},
  \citenamefont {Bhave}, \citenamefont {Riemensberger},\ and\ \citenamefont
  {Kippenberg}}]{siddharth2023hertz}%
  \BibitemOpen
  \bibfield  {author} {\bibinfo {author} {\bibfnamefont {A.}~\bibnamefont
  {Siddharth}}, \bibinfo {author} {\bibfnamefont {A.}~\bibnamefont
  {Attanasio}}, \bibinfo {author} {\bibfnamefont {S.}~\bibnamefont {Bianconi}},
  \bibinfo {author} {\bibfnamefont {G.}~\bibnamefont {Lihachev}}, \bibinfo
  {author} {\bibfnamefont {J.}~\bibnamefont {Zhang}}, \bibinfo {author}
  {\bibfnamefont {Z.}~\bibnamefont {Qiu}}, \bibinfo {author} {\bibfnamefont
  {A.}~\bibnamefont {Bancora}}, \bibinfo {author} {\bibfnamefont
  {S.}~\bibnamefont {Kenning}}, \bibinfo {author} {\bibfnamefont {R.~N.}\
  \bibnamefont {Wang}}, \bibinfo {author} {\bibfnamefont {A.~S.}\ \bibnamefont
  {Voloshin}}, \bibinfo {author} {\bibfnamefont {S.~A.}\ \bibnamefont {Bhave}},
  \bibinfo {author} {\bibfnamefont {J.}~\bibnamefont {Riemensberger}},\ and\
  \bibinfo {author} {\bibfnamefont {T.~J.}\ \bibnamefont {Kippenberg}},\
  }\bibfield  {title} {\bibinfo {title} {Piezoelectrically tunable, narrow
  linewidth photonic integrated extended-dbr lasers},\ }\href
  {https://doi.org/10.1364/OPTICA.524703} {\bibfield  {journal} {\bibinfo
  {journal} {Optica}\ }\textbf {\bibinfo {volume} {11}},\ \bibinfo {pages}
  {1062} (\bibinfo {year} {2024}{\natexlab{a}})}\BibitemShut {NoStop}%
\bibitem [{\citenamefont {Snigirev}\ \emph {et~al.}(2023)\citenamefont
  {Snigirev}, \citenamefont {Riedhauser}, \citenamefont {Lihachev},
  \citenamefont {Churaev}, \citenamefont {Riemensberger}, \citenamefont {Wang},
  \citenamefont {Siddharth}, \citenamefont {Huang}, \citenamefont {M{\"o}hl},
  \citenamefont {Popoff} \emph {et~al.}}]{snigirev2023lnodlaser}%
  \BibitemOpen
  \bibfield  {author} {\bibinfo {author} {\bibfnamefont {V.}~\bibnamefont
  {Snigirev}}, \bibinfo {author} {\bibfnamefont {A.}~\bibnamefont
  {Riedhauser}}, \bibinfo {author} {\bibfnamefont {G.}~\bibnamefont
  {Lihachev}}, \bibinfo {author} {\bibfnamefont {M.}~\bibnamefont {Churaev}},
  \bibinfo {author} {\bibfnamefont {J.}~\bibnamefont {Riemensberger}}, \bibinfo
  {author} {\bibfnamefont {R.~N.}\ \bibnamefont {Wang}}, \bibinfo {author}
  {\bibfnamefont {A.}~\bibnamefont {Siddharth}}, \bibinfo {author}
  {\bibfnamefont {G.}~\bibnamefont {Huang}}, \bibinfo {author} {\bibfnamefont
  {C.}~\bibnamefont {M{\"o}hl}}, \bibinfo {author} {\bibfnamefont
  {Y.}~\bibnamefont {Popoff}}, \emph {et~al.},\ }\bibfield  {title} {\bibinfo
  {title} {Ultrafast tunable lasers using lithium niobate integrated
  photonics},\ }\href@noop {} {\bibfield  {journal} {\bibinfo  {journal}
  {Nature}\ }\textbf {\bibinfo {volume} {615}},\ \bibinfo {pages} {411}
  (\bibinfo {year} {2023})}\BibitemShut {NoStop}%
\bibitem [{\citenamefont {Li}\ \emph {et~al.}(2022)\citenamefont {Li},
  \citenamefont {Chang}, \citenamefont {Wu}, \citenamefont {Staffa},
  \citenamefont {Ling}, \citenamefont {Javid}, \citenamefont {Xue},
  \citenamefont {He}, \citenamefont {Lopez-Rios}, \citenamefont {Morin} \emph
  {et~al.}}]{li2022pockels}%
  \BibitemOpen
  \bibfield  {author} {\bibinfo {author} {\bibfnamefont {M.}~\bibnamefont
  {Li}}, \bibinfo {author} {\bibfnamefont {L.}~\bibnamefont {Chang}}, \bibinfo
  {author} {\bibfnamefont {L.}~\bibnamefont {Wu}}, \bibinfo {author}
  {\bibfnamefont {J.}~\bibnamefont {Staffa}}, \bibinfo {author} {\bibfnamefont
  {J.}~\bibnamefont {Ling}}, \bibinfo {author} {\bibfnamefont {U.~A.}\
  \bibnamefont {Javid}}, \bibinfo {author} {\bibfnamefont {S.}~\bibnamefont
  {Xue}}, \bibinfo {author} {\bibfnamefont {Y.}~\bibnamefont {He}}, \bibinfo
  {author} {\bibfnamefont {R.}~\bibnamefont {Lopez-Rios}}, \bibinfo {author}
  {\bibfnamefont {T.~J.}\ \bibnamefont {Morin}}, \emph {et~al.},\ }\bibfield
  {title} {\bibinfo {title} {Integrated pockels laser},\ }\href@noop {}
  {\bibfield  {journal} {\bibinfo  {journal} {Nature communications}\ }\textbf
  {\bibinfo {volume} {13}},\ \bibinfo {pages} {5344} (\bibinfo {year}
  {2022})}\BibitemShut {NoStop}%
\bibitem [{\citenamefont {Lihachev}\ \emph {et~al.}(2022)\citenamefont
  {Lihachev}, \citenamefont {Riemensberger}, \citenamefont {Weng},
  \citenamefont {Liu}, \citenamefont {Tian}, \citenamefont {Siddharth},
  \citenamefont {Snigirev}, \citenamefont {Shadymov}, \citenamefont {Voloshin},
  \citenamefont {Wang} \emph {et~al.}}]{lihachev2022low}%
  \BibitemOpen
  \bibfield  {author} {\bibinfo {author} {\bibfnamefont {G.}~\bibnamefont
  {Lihachev}}, \bibinfo {author} {\bibfnamefont {J.}~\bibnamefont
  {Riemensberger}}, \bibinfo {author} {\bibfnamefont {W.}~\bibnamefont {Weng}},
  \bibinfo {author} {\bibfnamefont {J.}~\bibnamefont {Liu}}, \bibinfo {author}
  {\bibfnamefont {H.}~\bibnamefont {Tian}}, \bibinfo {author} {\bibfnamefont
  {A.}~\bibnamefont {Siddharth}}, \bibinfo {author} {\bibfnamefont
  {V.}~\bibnamefont {Snigirev}}, \bibinfo {author} {\bibfnamefont
  {V.}~\bibnamefont {Shadymov}}, \bibinfo {author} {\bibfnamefont
  {A.}~\bibnamefont {Voloshin}}, \bibinfo {author} {\bibfnamefont {R.~N.}\
  \bibnamefont {Wang}}, \emph {et~al.},\ }\bibfield  {title} {\bibinfo {title}
  {Low-noise frequency-agile photonic integrated lasers for coherent ranging},\
  }\href@noop {} {\bibfield  {journal} {\bibinfo  {journal} {Nature
  communications}\ }\textbf {\bibinfo {volume} {13}},\ \bibinfo {pages} {3522}
  (\bibinfo {year} {2022})}\BibitemShut {NoStop}%
\bibitem [{\citenamefont {Wang}\ \emph {et~al.}(2024)\citenamefont {Wang},
  \citenamefont {Li}, \citenamefont {Riemensberger}, \citenamefont {Lihachev},
  \citenamefont {Churaev}, \citenamefont {Kao}, \citenamefont {Ji},
  \citenamefont {Zhang}, \citenamefont {Blesin}, \citenamefont {Davydova} \emph
  {et~al.}}]{wang2024tantalate}%
  \BibitemOpen
  \bibfield  {author} {\bibinfo {author} {\bibfnamefont {C.}~\bibnamefont
  {Wang}}, \bibinfo {author} {\bibfnamefont {Z.}~\bibnamefont {Li}}, \bibinfo
  {author} {\bibfnamefont {J.}~\bibnamefont {Riemensberger}}, \bibinfo {author}
  {\bibfnamefont {G.}~\bibnamefont {Lihachev}}, \bibinfo {author}
  {\bibfnamefont {M.}~\bibnamefont {Churaev}}, \bibinfo {author} {\bibfnamefont
  {W.}~\bibnamefont {Kao}}, \bibinfo {author} {\bibfnamefont {X.}~\bibnamefont
  {Ji}}, \bibinfo {author} {\bibfnamefont {J.}~\bibnamefont {Zhang}}, \bibinfo
  {author} {\bibfnamefont {T.}~\bibnamefont {Blesin}}, \bibinfo {author}
  {\bibfnamefont {A.}~\bibnamefont {Davydova}}, \emph {et~al.},\ }\bibfield
  {title} {\bibinfo {title} {Lithium tantalate photonic integrated circuits for
  volume manufacturing},\ }\href@noop {} {\bibfield  {journal} {\bibinfo
  {journal} {Nature}\ ,\ \bibinfo {pages} {1}} (\bibinfo {year}
  {2024})}\BibitemShut {NoStop}%
\bibitem [{\citenamefont {Li}\ \emph {et~al.}(2023{\natexlab{b}})\citenamefont
  {Li}, \citenamefont {Wang}, \citenamefont {Lihachev}, \citenamefont {Zhang},
  \citenamefont {Tan}, \citenamefont {Churaev}, \citenamefont {Kuznetsov},
  \citenamefont {Siddharth}, \citenamefont {Bereyhi}, \citenamefont
  {Riemensberger} \emph {et~al.}}]{li2023lnoi}%
  \BibitemOpen
  \bibfield  {author} {\bibinfo {author} {\bibfnamefont {Z.}~\bibnamefont
  {Li}}, \bibinfo {author} {\bibfnamefont {R.~N.}\ \bibnamefont {Wang}},
  \bibinfo {author} {\bibfnamefont {G.}~\bibnamefont {Lihachev}}, \bibinfo
  {author} {\bibfnamefont {J.}~\bibnamefont {Zhang}}, \bibinfo {author}
  {\bibfnamefont {Z.}~\bibnamefont {Tan}}, \bibinfo {author} {\bibfnamefont
  {M.}~\bibnamefont {Churaev}}, \bibinfo {author} {\bibfnamefont
  {N.}~\bibnamefont {Kuznetsov}}, \bibinfo {author} {\bibfnamefont
  {A.}~\bibnamefont {Siddharth}}, \bibinfo {author} {\bibfnamefont {M.~J.}\
  \bibnamefont {Bereyhi}}, \bibinfo {author} {\bibfnamefont {J.}~\bibnamefont
  {Riemensberger}}, \emph {et~al.},\ }\bibfield  {title} {\bibinfo {title}
  {High density lithium niobate photonic integrated circuits},\ }\href@noop {}
  {\bibfield  {journal} {\bibinfo  {journal} {Nature Communications}\ }\textbf
  {\bibinfo {volume} {14}},\ \bibinfo {pages} {4856} (\bibinfo {year}
  {2023}{\natexlab{b}})}\BibitemShut {NoStop}%
\bibitem [{\citenamefont {Siddharth}\ \emph
  {et~al.}(2024{\natexlab{b}})\citenamefont {Siddharth}, \citenamefont
  {Bianconi}, \citenamefont {Wang}, \citenamefont {Qiu}, \citenamefont
  {Voloshin}, \citenamefont {Bereyhi}, \citenamefont {Riemensberger},\ and\
  \citenamefont {Kippenberg}}]{siddharth2024ultrafast}%
  \BibitemOpen
  \bibfield  {author} {\bibinfo {author} {\bibfnamefont {A.}~\bibnamefont
  {Siddharth}}, \bibinfo {author} {\bibfnamefont {S.}~\bibnamefont {Bianconi}},
  \bibinfo {author} {\bibfnamefont {R.~N.}\ \bibnamefont {Wang}}, \bibinfo
  {author} {\bibfnamefont {Z.}~\bibnamefont {Qiu}}, \bibinfo {author}
  {\bibfnamefont {A.~S.}\ \bibnamefont {Voloshin}}, \bibinfo {author}
  {\bibfnamefont {M.~J.}\ \bibnamefont {Bereyhi}}, \bibinfo {author}
  {\bibfnamefont {J.}~\bibnamefont {Riemensberger}},\ and\ \bibinfo {author}
  {\bibfnamefont {T.~J.}\ \bibnamefont {Kippenberg}},\ }\bibfield  {title}
  {\bibinfo {title} {Ultrafast tunable photonic integrated pockels extended-dbr
  laser},\ }\href@noop {} {\bibfield  {journal} {\bibinfo  {journal} {arXiv
  preprint arXiv:2408.01743}\ } (\bibinfo {year}
  {2024}{\natexlab{b}})}\BibitemShut {NoStop}%
\bibitem [{\citenamefont {Xue}\ \emph {et~al.}(2024)\citenamefont {Xue},
  \citenamefont {Li}, \citenamefont {Lopez-rios}, \citenamefont {Ling},
  \citenamefont {Gao}, \citenamefont {Hu}, \citenamefont {Qiu}, \citenamefont
  {Staffa}, \citenamefont {Chang}, \citenamefont {Wang} \emph
  {et~al.}}]{xue2024pockels}%
  \BibitemOpen
  \bibfield  {author} {\bibinfo {author} {\bibfnamefont {S.}~\bibnamefont
  {Xue}}, \bibinfo {author} {\bibfnamefont {M.}~\bibnamefont {Li}}, \bibinfo
  {author} {\bibfnamefont {R.}~\bibnamefont {Lopez-rios}}, \bibinfo {author}
  {\bibfnamefont {J.}~\bibnamefont {Ling}}, \bibinfo {author} {\bibfnamefont
  {Z.}~\bibnamefont {Gao}}, \bibinfo {author} {\bibfnamefont {Q.}~\bibnamefont
  {Hu}}, \bibinfo {author} {\bibfnamefont {T.}~\bibnamefont {Qiu}}, \bibinfo
  {author} {\bibfnamefont {J.}~\bibnamefont {Staffa}}, \bibinfo {author}
  {\bibfnamefont {L.}~\bibnamefont {Chang}}, \bibinfo {author} {\bibfnamefont
  {H.}~\bibnamefont {Wang}}, \emph {et~al.},\ }\bibfield  {title} {\bibinfo
  {title} {Pockels laser directly driving ultrafast optical metrology},\
  }\href@noop {} {\bibfield  {journal} {\bibinfo  {journal} {arXiv preprint
  arXiv:2410.07482}\ } (\bibinfo {year} {2024})}\BibitemShut {NoStop}%
\bibitem [{\citenamefont {Shams-Ansari}\ \emph {et~al.}(2022)\citenamefont
  {Shams-Ansari}, \citenamefont {Renaud}, \citenamefont {Cheng}, \citenamefont
  {Shao}, \citenamefont {He}, \citenamefont {Zhu}, \citenamefont {Yu},
  \citenamefont {Grant}, \citenamefont {Johansson}, \citenamefont {Zhang},\
  and\ \citenamefont {Lon\v{c}ar}}]{Shams-Ansari:22}%
  \BibitemOpen
  \bibfield  {author} {\bibinfo {author} {\bibfnamefont {A.}~\bibnamefont
  {Shams-Ansari}}, \bibinfo {author} {\bibfnamefont {D.}~\bibnamefont
  {Renaud}}, \bibinfo {author} {\bibfnamefont {R.}~\bibnamefont {Cheng}},
  \bibinfo {author} {\bibfnamefont {L.}~\bibnamefont {Shao}}, \bibinfo {author}
  {\bibfnamefont {L.}~\bibnamefont {He}}, \bibinfo {author} {\bibfnamefont
  {D.}~\bibnamefont {Zhu}}, \bibinfo {author} {\bibfnamefont {M.}~\bibnamefont
  {Yu}}, \bibinfo {author} {\bibfnamefont {H.~R.}\ \bibnamefont {Grant}},
  \bibinfo {author} {\bibfnamefont {L.}~\bibnamefont {Johansson}}, \bibinfo
  {author} {\bibfnamefont {M.}~\bibnamefont {Zhang}},\ and\ \bibinfo {author}
  {\bibfnamefont {M.}~\bibnamefont {Lon\v{c}ar}},\ }\bibfield  {title}
  {\bibinfo {title} {Electrically pumped laser transmitter integrated on
  thin-film lithium niobate},\ }\href {https://doi.org/10.1364/OPTICA.448617}
  {\bibfield  {journal} {\bibinfo  {journal} {Optica}\ }\textbf {\bibinfo
  {volume} {9}},\ \bibinfo {pages} {408} (\bibinfo {year} {2022})}\BibitemShut
  {NoStop}%
\bibitem [{\citenamefont {Franken}\ \emph {et~al.}(2024)\citenamefont
  {Franken}, \citenamefont {Cheng}, \citenamefont {Powell}, \citenamefont
  {Kyriazidis}, \citenamefont {Rosborough}, \citenamefont {Musolf},
  \citenamefont {Shah}, \citenamefont {Barton~III}, \citenamefont {Hills},
  \citenamefont {Johansson} \emph {et~al.}}]{franken2024high}%
  \BibitemOpen
  \bibfield  {author} {\bibinfo {author} {\bibfnamefont {C.~A.}\ \bibnamefont
  {Franken}}, \bibinfo {author} {\bibfnamefont {R.}~\bibnamefont {Cheng}},
  \bibinfo {author} {\bibfnamefont {K.}~\bibnamefont {Powell}}, \bibinfo
  {author} {\bibfnamefont {G.}~\bibnamefont {Kyriazidis}}, \bibinfo {author}
  {\bibfnamefont {V.}~\bibnamefont {Rosborough}}, \bibinfo {author}
  {\bibfnamefont {J.}~\bibnamefont {Musolf}}, \bibinfo {author} {\bibfnamefont
  {M.}~\bibnamefont {Shah}}, \bibinfo {author} {\bibfnamefont {D.~R.}\
  \bibnamefont {Barton~III}}, \bibinfo {author} {\bibfnamefont
  {G.}~\bibnamefont {Hills}}, \bibinfo {author} {\bibfnamefont
  {L.}~\bibnamefont {Johansson}}, \emph {et~al.},\ }\bibfield  {title}
  {\bibinfo {title} {High-power and narrow-linewidth laser on thin-film lithium
  niobate enabled by photonic wire bonding},\ }\href@noop {} {\bibfield
  {journal} {\bibinfo  {journal} {arXiv preprint arXiv:2407.00269}\ } (\bibinfo
  {year} {2024})}\BibitemShut {NoStop}%
\bibitem [{\citenamefont {Zhang}\ \emph
  {et~al.}(2023{\natexlab{b}})\citenamefont {Zhang}, \citenamefont {Li},
  \citenamefont {Riemensberger}, \citenamefont {Lihachev}, \citenamefont
  {Huang},\ and\ \citenamefont {Kippenberg}}]{zhang2023fundamental}%
  \BibitemOpen
  \bibfield  {author} {\bibinfo {author} {\bibfnamefont {J.}~\bibnamefont
  {Zhang}}, \bibinfo {author} {\bibfnamefont {Z.}~\bibnamefont {Li}}, \bibinfo
  {author} {\bibfnamefont {J.}~\bibnamefont {Riemensberger}}, \bibinfo {author}
  {\bibfnamefont {G.}~\bibnamefont {Lihachev}}, \bibinfo {author}
  {\bibfnamefont {G.}~\bibnamefont {Huang}},\ and\ \bibinfo {author}
  {\bibfnamefont {T.~J.}\ \bibnamefont {Kippenberg}},\ }\href@noop {} {\bibinfo
  {title} {Fundamental charge noise in electro-optic photonic integrated
  circuits}} (\bibinfo {year} {2023}{\natexlab{b}}),\ \Eprint
  {https://arxiv.org/abs/2308.15404} {arXiv:2308.15404} \BibitemShut {NoStop}%
\bibitem [{\citenamefont {Liang}\ \emph {et~al.}(2015)\citenamefont {Liang},
  \citenamefont {Ilchenko}, \citenamefont {Eliyahu}, \citenamefont
  {Savchenkov}, \citenamefont {Matsko}, \citenamefont {Seidel},\ and\
  \citenamefont {Maleki}}]{liang2015ultralow}%
  \BibitemOpen
  \bibfield  {author} {\bibinfo {author} {\bibfnamefont {W.}~\bibnamefont
  {Liang}}, \bibinfo {author} {\bibfnamefont {V.~S.}\ \bibnamefont {Ilchenko}},
  \bibinfo {author} {\bibfnamefont {D.}~\bibnamefont {Eliyahu}}, \bibinfo
  {author} {\bibfnamefont {A.~A.}\ \bibnamefont {Savchenkov}}, \bibinfo
  {author} {\bibfnamefont {A.~B.}\ \bibnamefont {Matsko}}, \bibinfo {author}
  {\bibfnamefont {D.}~\bibnamefont {Seidel}},\ and\ \bibinfo {author}
  {\bibfnamefont {L.}~\bibnamefont {Maleki}},\ }\bibfield  {title} {\bibinfo
  {title} {Ultralow noise miniature external cavity semiconductor laser},\
  }\href {https://doi.org/10.1038/ncomms8371} {\bibfield  {journal} {\bibinfo
  {journal} {Nature Communications}\ }\textbf {\bibinfo {volume} {6}},\
  \bibinfo {pages} {7371} (\bibinfo {year} {2015})}\BibitemShut {NoStop}%
\bibitem [{\citenamefont {Choudhary}\ and\ \citenamefont
  {Iniewski}(2017)}]{choudhary2017mems}%
  \BibitemOpen
  \bibfield  {author} {\bibinfo {author} {\bibfnamefont {V.}~\bibnamefont
  {Choudhary}}\ and\ \bibinfo {author} {\bibfnamefont {K.}~\bibnamefont
  {Iniewski}},\ }\href@noop {} {\emph {\bibinfo {title} {Mems: fundamental
  technology and applications}}}\ (\bibinfo  {publisher} {CRC Press},\ \bibinfo
  {year} {2017})\BibitemShut {NoStop}%
\bibitem [{\citenamefont {Gao}\ \emph {et~al.}(2020)\citenamefont {Gao},
  \citenamefont {Liu}, \citenamefont {Liang},\ and\ \citenamefont
  {Wu}}]{gao2020aln}%
  \BibitemOpen
  \bibfield  {author} {\bibinfo {author} {\bibfnamefont {A.}~\bibnamefont
  {Gao}}, \bibinfo {author} {\bibfnamefont {K.}~\bibnamefont {Liu}}, \bibinfo
  {author} {\bibfnamefont {J.}~\bibnamefont {Liang}},\ and\ \bibinfo {author}
  {\bibfnamefont {T.}~\bibnamefont {Wu}},\ }\bibfield  {title} {\bibinfo
  {title} {Aln mems filters with extremely high bandwidth widening
  capability},\ }\href@noop {} {\bibfield  {journal} {\bibinfo  {journal}
  {Microsystems \& Nanoengineering}\ }\textbf {\bibinfo {volume} {6}},\
  \bibinfo {pages} {74} (\bibinfo {year} {2020})}\BibitemShut {NoStop}%
\bibitem [{\citenamefont {Zou}\ \emph {et~al.}(2022)\citenamefont {Zou},
  \citenamefont {Gao}, \citenamefont {Zhou}, \citenamefont {Liu}, \citenamefont
  {Xu}, \citenamefont {Qu}, \citenamefont {Liu}, \citenamefont {Soon},
  \citenamefont {Cai},\ and\ \citenamefont {Sun}}]{zou2022aluminum}%
  \BibitemOpen
  \bibfield  {author} {\bibinfo {author} {\bibfnamefont {Y.}~\bibnamefont
  {Zou}}, \bibinfo {author} {\bibfnamefont {C.}~\bibnamefont {Gao}}, \bibinfo
  {author} {\bibfnamefont {J.}~\bibnamefont {Zhou}}, \bibinfo {author}
  {\bibfnamefont {Y.}~\bibnamefont {Liu}}, \bibinfo {author} {\bibfnamefont
  {Q.}~\bibnamefont {Xu}}, \bibinfo {author} {\bibfnamefont {Y.}~\bibnamefont
  {Qu}}, \bibinfo {author} {\bibfnamefont {W.}~\bibnamefont {Liu}}, \bibinfo
  {author} {\bibfnamefont {J.~B.~W.}\ \bibnamefont {Soon}}, \bibinfo {author}
  {\bibfnamefont {Y.}~\bibnamefont {Cai}},\ and\ \bibinfo {author}
  {\bibfnamefont {C.}~\bibnamefont {Sun}},\ }\bibfield  {title} {\bibinfo
  {title} {Aluminum scandium nitride thin-film bulk acoustic resonators for 5g
  wideband applications},\ }\href@noop {} {\bibfield  {journal} {\bibinfo
  {journal} {Microsystems \& Nanoengineering}\ }\textbf {\bibinfo {volume}
  {8}},\ \bibinfo {pages} {124} (\bibinfo {year} {2022})}\BibitemShut {NoStop}%
\bibitem [{\citenamefont {Polcawich}\ \emph {et~al.}(2010)\citenamefont
  {Polcawich}, \citenamefont {Pulskamp}, \citenamefont {Bedair}, \citenamefont
  {Smith}, \citenamefont {Kaul}, \citenamefont {Kroninger}, \citenamefont
  {Wetzel}, \citenamefont {Chandrahalim},\ and\ \citenamefont
  {Bhave}}]{PZT_wings1}%
  \BibitemOpen
  \bibfield  {author} {\bibinfo {author} {\bibfnamefont {R.~G.}\ \bibnamefont
  {Polcawich}}, \bibinfo {author} {\bibfnamefont {J.~S.}\ \bibnamefont
  {Pulskamp}}, \bibinfo {author} {\bibfnamefont {S.}~\bibnamefont {Bedair}},
  \bibinfo {author} {\bibfnamefont {G.}~\bibnamefont {Smith}}, \bibinfo
  {author} {\bibfnamefont {R.}~\bibnamefont {Kaul}}, \bibinfo {author}
  {\bibfnamefont {C.}~\bibnamefont {Kroninger}}, \bibinfo {author}
  {\bibfnamefont {E.}~\bibnamefont {Wetzel}}, \bibinfo {author} {\bibfnamefont
  {H.}~\bibnamefont {Chandrahalim}},\ and\ \bibinfo {author} {\bibfnamefont
  {S.~A.}\ \bibnamefont {Bhave}},\ }\bibfield  {title} {\bibinfo {title}
  {Integrated piezomems actuators and sensors},\ }in\ \href
  {https://doi.org/10.1109/ICSENS.2010.5690603} {\emph {\bibinfo {booktitle}
  {SENSORS, 2010 IEEE}}}\ (\bibinfo {year} {2010})\ pp.\ \bibinfo {pages}
  {2193--2196}\BibitemShut {NoStop}%
\bibitem [{\citenamefont {Kroninger}\ \emph {et~al.}(2009)\citenamefont
  {Kroninger}, \citenamefont {Pulskamp}, \citenamefont {Bronson}, \citenamefont
  {Polcawich},\ and\ \citenamefont {Wetzel}}]{PZT_wings2}%
  \BibitemOpen
  \bibfield  {author} {\bibinfo {author} {\bibfnamefont {C.}~\bibnamefont
  {Kroninger}}, \bibinfo {author} {\bibfnamefont {J.}~\bibnamefont {Pulskamp}},
  \bibinfo {author} {\bibfnamefont {J.}~\bibnamefont {Bronson}}, \bibinfo
  {author} {\bibfnamefont {R.}~\bibnamefont {Polcawich}},\ and\ \bibinfo
  {author} {\bibfnamefont {E.}~\bibnamefont {Wetzel}},\ }\href
  {https://citeseerx.ist.psu.edu/document?repid=rep1&type=pdf&doi=dd388a64163de15fc36ccc110a3f984974ffc1b0}
  {\emph {\bibinfo {title} {Bio-mimetic millimeter-scale flapping wings for
  micro air vehicles}}},\ \bibinfo {number} {ARL-TR-4729}\ (\bibinfo
  {publisher} {Army Research Laboratory},\ \bibinfo {year} {2009})\BibitemShut
  {NoStop}%
\bibitem [{\citenamefont {Jimbo}\ \emph {et~al.}(2020)\citenamefont {Jimbo},
  \citenamefont {Ozaki}, \citenamefont {Amano},\ and\ \citenamefont
  {Fujimoto}}]{jimbo2020flight}%
  \BibitemOpen
  \bibfield  {author} {\bibinfo {author} {\bibfnamefont {T.}~\bibnamefont
  {Jimbo}}, \bibinfo {author} {\bibfnamefont {T.}~\bibnamefont {Ozaki}},
  \bibinfo {author} {\bibfnamefont {Y.}~\bibnamefont {Amano}},\ and\ \bibinfo
  {author} {\bibfnamefont {K.}~\bibnamefont {Fujimoto}},\ }\bibfield  {title}
  {\bibinfo {title} {Flight control of flapping-wing robot with three paired
  direct-driven piezoelectric actuators},\ }\href@noop {} {\bibfield  {journal}
  {\bibinfo  {journal} {IFAC-PapersOnLine}\ }\textbf {\bibinfo {volume} {53}},\
  \bibinfo {pages} {9391} (\bibinfo {year} {2020})}\BibitemShut {NoStop}%
\bibitem [{\citenamefont {Han}\ \emph {et~al.}(2022)\citenamefont {Han},
  \citenamefont {Colburn}, \citenamefont {Majumdar},\ and\ \citenamefont
  {B{\"o}hringer}}]{han2022millimeter}%
  \BibitemOpen
  \bibfield  {author} {\bibinfo {author} {\bibfnamefont {Z.}~\bibnamefont
  {Han}}, \bibinfo {author} {\bibfnamefont {S.}~\bibnamefont {Colburn}},
  \bibinfo {author} {\bibfnamefont {A.}~\bibnamefont {Majumdar}},\ and\
  \bibinfo {author} {\bibfnamefont {K.~F.}\ \bibnamefont {B{\"o}hringer}},\
  }\bibfield  {title} {\bibinfo {title} {Millimeter-scale focal length tuning
  with mems-integrated meta-optics employing high-throughput fabrication},\
  }\href@noop {} {\bibfield  {journal} {\bibinfo  {journal} {Scientific
  Reports}\ }\textbf {\bibinfo {volume} {12}},\ \bibinfo {pages} {5385}
  (\bibinfo {year} {2022})}\BibitemShut {NoStop}%
\bibitem [{\citenamefont {Seo}\ \emph {et~al.}(2010)\citenamefont {Seo},
  \citenamefont {Park}, \citenamefont {Kim}, \citenamefont {Kim}, \citenamefont
  {Ur},\ and\ \citenamefont {Yi}}]{seo10speakers}%
  \BibitemOpen
  \bibfield  {author} {\bibinfo {author} {\bibfnamefont {K.}~\bibnamefont
  {Seo}}, \bibinfo {author} {\bibfnamefont {J.}~\bibnamefont {Park}}, \bibinfo
  {author} {\bibfnamefont {H.}~\bibnamefont {Kim}}, \bibinfo {author}
  {\bibfnamefont {D.}~\bibnamefont {Kim}}, \bibinfo {author} {\bibfnamefont
  {S.}~\bibnamefont {Ur}},\ and\ \bibinfo {author} {\bibfnamefont
  {S.}~\bibnamefont {Yi}},\ }\bibfield  {title} {\bibinfo {title}
  {Micromachined piezoelectric microspeakers fabricated with high quality aln
  thin film},\ }\href {https://doi.org/10.1080/10584580701756524} {\bibfield
  {journal} {\bibinfo  {journal} {Integrated Ferroelectrics}\ }\textbf
  {\bibinfo {volume} {95}},\ \bibinfo {pages} {74} (\bibinfo {year}
  {2010})}\BibitemShut {NoStop}%
\bibitem [{\citenamefont {Wang}\ \emph {et~al.}(2000)\citenamefont {Wang},
  \citenamefont {Wong},\ and\ \citenamefont {Nguyen}}]{870061}%
  \BibitemOpen
  \bibfield  {author} {\bibinfo {author} {\bibfnamefont {K.}~\bibnamefont
  {Wang}}, \bibinfo {author} {\bibfnamefont {A.-C.}\ \bibnamefont {Wong}},\
  and\ \bibinfo {author} {\bibfnamefont {C.-C.}\ \bibnamefont {Nguyen}},\
  }\bibfield  {title} {\bibinfo {title} {Vhf free-free beam high-q
  micromechanical resonators},\ }\href {https://doi.org/10.1109/84.870061}
  {\bibfield  {journal} {\bibinfo  {journal} {Journal of Microelectromechanical
  Systems}\ }\textbf {\bibinfo {volume} {9}},\ \bibinfo {pages} {347} (\bibinfo
  {year} {2000})}\BibitemShut {NoStop}%
\bibitem [{\citenamefont {Demirci}\ and\ \citenamefont
  {Nguyen}(2003)}]{Demirci03IEEE}%
  \BibitemOpen
  \bibfield  {author} {\bibinfo {author} {\bibfnamefont {M.}~\bibnamefont
  {Demirci}}\ and\ \bibinfo {author} {\bibfnamefont {C.-C.}\ \bibnamefont
  {Nguyen}},\ }\bibfield  {title} {\bibinfo {title} {Higher-mode free-free beam
  micromechanical resonators},\ }in\ \href
  {https://doi.org/10.1109/FREQ.2003.1275195} {\emph {\bibinfo {booktitle}
  {IEEE International Frequency Control Symposium and PDA Exhibition Jointly
  with the 17th European Frequency and Time Forum, 2003. Proceedings of the
  2003}}}\ (\bibinfo {year} {2003})\ pp.\ \bibinfo {pages}
  {810--818}\BibitemShut {NoStop}%
\bibitem [{\citenamefont {Clark}\ \emph {et~al.}(2013)\citenamefont {Clark},
  \citenamefont {Brown}, \citenamefont {He},\ and\ \citenamefont
  {Hsu}}]{clark2013temperature}%
  \BibitemOpen
  \bibfield  {author} {\bibinfo {author} {\bibfnamefont {J.~R.}\ \bibnamefont
  {Clark}}, \bibinfo {author} {\bibfnamefont {A.~R.}\ \bibnamefont {Brown}},
  \bibinfo {author} {\bibfnamefont {G.}~\bibnamefont {He}},\ and\ \bibinfo
  {author} {\bibfnamefont {W.-T.}\ \bibnamefont {Hsu}},\ }\bibfield  {title}
  {\bibinfo {title} {Temperature compensated overtone resonators},\ }in\ \href
  {https://ieeexplore.ieee.org/abstract/document/6626886} {\emph {\bibinfo
  {booktitle} {2013 Transducers \& Eurosensors XXVII: The 17th International
  Conference on Solid-State Sensors, Actuators and Microsystems (TRANSDUCERS \&
  EUROSENSORS XXVII)}}}\ (\bibinfo {organization} {IEEE},\ \bibinfo {year}
  {2013})\ pp.\ \bibinfo {pages} {794--797}\BibitemShut {NoStop}%
\bibitem [{\citenamefont {Clark}\ and\ \citenamefont
  {Mostyn}(2017)}]{clark2017microchip}%
  \BibitemOpen
  \bibfield  {author} {\bibinfo {author} {\bibfnamefont {J.}~\bibnamefont
  {Clark}}\ and\ \bibinfo {author} {\bibfnamefont {G.}~\bibnamefont {Mostyn}},\
  }\href
  {https://ww1.microchip.com/downloads/aemtest/OTH/ProductDocuments/SupportingCollateral/00002344A.pdf}
  {\emph {\bibinfo {title} {Microchip Oscillators and Clocks Using
  Microelectromechanical Systems (MEMS) Technology}}},\ \bibinfo {type} {Tech.
  Rep.}\ \bibinfo {number} {DS00002344A-page 1}\ (\bibinfo  {institution}
  {Microchip Technology Inc.},\ \bibinfo {year} {2017})\BibitemShut {NoStop}%
\bibitem [{\citenamefont {Jin}\ \emph {et~al.}(2021)\citenamefont {Jin},
  \citenamefont {Yang}, \citenamefont {Chang}, \citenamefont {Shen},
  \citenamefont {Wang}, \citenamefont {Leal}, \citenamefont {Wu}, \citenamefont
  {Gao}, \citenamefont {Feshali}, \citenamefont {Paniccia} \emph
  {et~al.}}]{jin2021hertz}%
  \BibitemOpen
  \bibfield  {author} {\bibinfo {author} {\bibfnamefont {W.}~\bibnamefont
  {Jin}}, \bibinfo {author} {\bibfnamefont {Q.-F.}\ \bibnamefont {Yang}},
  \bibinfo {author} {\bibfnamefont {L.}~\bibnamefont {Chang}}, \bibinfo
  {author} {\bibfnamefont {B.}~\bibnamefont {Shen}}, \bibinfo {author}
  {\bibfnamefont {H.}~\bibnamefont {Wang}}, \bibinfo {author} {\bibfnamefont
  {M.~A.}\ \bibnamefont {Leal}}, \bibinfo {author} {\bibfnamefont
  {L.}~\bibnamefont {Wu}}, \bibinfo {author} {\bibfnamefont {M.}~\bibnamefont
  {Gao}}, \bibinfo {author} {\bibfnamefont {A.}~\bibnamefont {Feshali}},
  \bibinfo {author} {\bibfnamefont {M.}~\bibnamefont {Paniccia}}, \emph
  {et~al.},\ }\bibfield  {title} {\bibinfo {title} {Hertz-linewidth
  semiconductor lasers using cmos-ready ultra-high-q microresonators},\
  }\href@noop {} {\bibfield  {journal} {\bibinfo  {journal} {Nature Photonics}\
  }\textbf {\bibinfo {volume} {15}},\ \bibinfo {pages} {346} (\bibinfo {year}
  {2021})}\BibitemShut {NoStop}%
\bibitem [{\citenamefont {Kondratiev}\ \emph {et~al.}(2017)\citenamefont
  {Kondratiev}, \citenamefont {Lobanov}, \citenamefont {Cherenkov},
  \citenamefont {Voloshin}, \citenamefont {Pavlov}, \citenamefont {Koptyaev},\
  and\ \citenamefont {Gorodetsky}}]{Kondratiev17}%
  \BibitemOpen
  \bibfield  {author} {\bibinfo {author} {\bibfnamefont {N.~M.}\ \bibnamefont
  {Kondratiev}}, \bibinfo {author} {\bibfnamefont {V.~E.}\ \bibnamefont
  {Lobanov}}, \bibinfo {author} {\bibfnamefont {A.~V.}\ \bibnamefont
  {Cherenkov}}, \bibinfo {author} {\bibfnamefont {A.~S.}\ \bibnamefont
  {Voloshin}}, \bibinfo {author} {\bibfnamefont {N.~G.}\ \bibnamefont
  {Pavlov}}, \bibinfo {author} {\bibfnamefont {S.}~\bibnamefont {Koptyaev}},\
  and\ \bibinfo {author} {\bibfnamefont {M.~L.}\ \bibnamefont {Gorodetsky}},\
  }\bibfield  {title} {\bibinfo {title} {Self-injection locking of a laser
  diode to a high-{Q} {W}{G}{M} microresonator},\ }\href
  {https://doi.org/10.1364/OE.25.028167} {\bibfield  {journal} {\bibinfo
  {journal} {Opt. Express}\ }\textbf {\bibinfo {volume} {25}},\ \bibinfo
  {pages} {28167} (\bibinfo {year} {2017})}\BibitemShut {NoStop}%
\bibitem [{\citenamefont {Ulanov}\ \emph
  {et~al.}(2024{\natexlab{a}})\citenamefont {Ulanov}, \citenamefont {Wildi},
  \citenamefont {Pavlov}, \citenamefont {Jost}, \citenamefont {Karpov},\ and\
  \citenamefont {Herr}}]{ulanov2024synthetic}%
  \BibitemOpen
  \bibfield  {author} {\bibinfo {author} {\bibfnamefont {A.~E.}\ \bibnamefont
  {Ulanov}}, \bibinfo {author} {\bibfnamefont {T.}~\bibnamefont {Wildi}},
  \bibinfo {author} {\bibfnamefont {N.~G.}\ \bibnamefont {Pavlov}}, \bibinfo
  {author} {\bibfnamefont {J.~D.}\ \bibnamefont {Jost}}, \bibinfo {author}
  {\bibfnamefont {M.}~\bibnamefont {Karpov}},\ and\ \bibinfo {author}
  {\bibfnamefont {T.}~\bibnamefont {Herr}},\ }\bibfield  {title} {\bibinfo
  {title} {Synthetic reflection self-injection-locked microcombs},\ }\href@noop
  {} {\bibfield  {journal} {\bibinfo  {journal} {Nature Photonics}\ }\textbf
  {\bibinfo {volume} {18}},\ \bibinfo {pages} {294} (\bibinfo {year}
  {2024}{\natexlab{a}})}\BibitemShut {NoStop}%
\bibitem [{\citenamefont {Ulanov}\ \emph
  {et~al.}(2024{\natexlab{b}})\citenamefont {Ulanov}, \citenamefont {Wildi},
  \citenamefont {Bhatnagar},\ and\ \citenamefont {Herr}}]{ulanov2024laser}%
  \BibitemOpen
  \bibfield  {author} {\bibinfo {author} {\bibfnamefont {A.~E.}\ \bibnamefont
  {Ulanov}}, \bibinfo {author} {\bibfnamefont {T.}~\bibnamefont {Wildi}},
  \bibinfo {author} {\bibfnamefont {U.}~\bibnamefont {Bhatnagar}},\ and\
  \bibinfo {author} {\bibfnamefont {T.}~\bibnamefont {Herr}},\ }\bibfield
  {title} {\bibinfo {title} {Laser diode self-injection locking to an
  integrated high-q fabry-p$\backslash$'erot microresonator},\ }\href@noop {}
  {\bibfield  {journal} {\bibinfo  {journal} {arXiv preprint arXiv:2408.08679}\
  } (\bibinfo {year} {2024}{\natexlab{b}})}\BibitemShut {NoStop}%
\bibitem [{\citenamefont {Liu}\ \emph {et~al.}(2020)\citenamefont {Liu},
  \citenamefont {Tian}, \citenamefont {Lucas}, \citenamefont {Raja},
  \citenamefont {Lihachev}, \citenamefont {Wang}, \citenamefont {He},
  \citenamefont {Liu}, \citenamefont {Anderson}, \citenamefont {Weng} \emph
  {et~al.}}]{liu2020monolithic}%
  \BibitemOpen
  \bibfield  {author} {\bibinfo {author} {\bibfnamefont {J.}~\bibnamefont
  {Liu}}, \bibinfo {author} {\bibfnamefont {H.}~\bibnamefont {Tian}}, \bibinfo
  {author} {\bibfnamefont {E.}~\bibnamefont {Lucas}}, \bibinfo {author}
  {\bibfnamefont {A.~S.}\ \bibnamefont {Raja}}, \bibinfo {author}
  {\bibfnamefont {G.}~\bibnamefont {Lihachev}}, \bibinfo {author}
  {\bibfnamefont {R.~N.}\ \bibnamefont {Wang}}, \bibinfo {author}
  {\bibfnamefont {J.}~\bibnamefont {He}}, \bibinfo {author} {\bibfnamefont
  {T.}~\bibnamefont {Liu}}, \bibinfo {author} {\bibfnamefont {M.~H.}\
  \bibnamefont {Anderson}}, \bibinfo {author} {\bibfnamefont {W.}~\bibnamefont
  {Weng}}, \emph {et~al.},\ }\bibfield  {title} {\bibinfo {title} {Monolithic
  piezoelectric control of soliton microcombs},\ }\href@noop {} {\bibfield
  {journal} {\bibinfo  {journal} {Nature}\ }\textbf {\bibinfo {volume} {583}},\
  \bibinfo {pages} {385} (\bibinfo {year} {2020})}\BibitemShut {NoStop}%
\bibitem [{\citenamefont {Inc.}(2023)}]{oewaves2023}%
  \BibitemOpen
  \bibfield  {author} {\bibinfo {author} {\bibfnamefont {O.}~\bibnamefont
  {Inc.}},\ }\bibfield  {title} {\bibinfo {title} {Specifications of
  oe4040-xln},\ }\href@noop {} {\bibfield  {journal} {\bibinfo  {journal}
  {Specifications}\ } (\bibinfo {year} {2023})},\ \bibinfo {note}
  {\url{https://morephotonics.com/wp-content/uploads/2023/02/OE4040.pdf}}\BibitemShut
  {NoStop}%
\bibitem [{\citenamefont {Ousaid}\ \emph {et~al.}(2024)\citenamefont {Ousaid},
  \citenamefont {Bourcier}, \citenamefont {Fernandez}, \citenamefont {Llopis},
  \citenamefont {Lumeau}, \citenamefont {Moreau}, \citenamefont {Bunel},
  \citenamefont {Conforti}, \citenamefont {Mussot}, \citenamefont {Crozatier},\
  and\ \citenamefont {Balac}}]{ousaid2024SIL}%
  \BibitemOpen
  \bibfield  {author} {\bibinfo {author} {\bibfnamefont {S.~M.}\ \bibnamefont
  {Ousaid}}, \bibinfo {author} {\bibfnamefont {G.}~\bibnamefont {Bourcier}},
  \bibinfo {author} {\bibfnamefont {A.}~\bibnamefont {Fernandez}}, \bibinfo
  {author} {\bibfnamefont {O.}~\bibnamefont {Llopis}}, \bibinfo {author}
  {\bibfnamefont {J.}~\bibnamefont {Lumeau}}, \bibinfo {author} {\bibfnamefont
  {A.}~\bibnamefont {Moreau}}, \bibinfo {author} {\bibfnamefont
  {T.}~\bibnamefont {Bunel}}, \bibinfo {author} {\bibfnamefont
  {M.}~\bibnamefont {Conforti}}, \bibinfo {author} {\bibfnamefont
  {A.}~\bibnamefont {Mussot}}, \bibinfo {author} {\bibfnamefont
  {V.}~\bibnamefont {Crozatier}},\ and\ \bibinfo {author} {\bibfnamefont
  {S.}~\bibnamefont {Balac}},\ }\bibfield  {title} {\bibinfo {title} {Low phase
  noise self-injection-locked diode laser with a high-q fiber resonator: model
  and experiment},\ }\href {https://doi.org/10.1364/OL.514778} {\bibfield
  {journal} {\bibinfo  {journal} {Opt. Lett.}\ }\textbf {\bibinfo {volume}
  {49}},\ \bibinfo {pages} {1933} (\bibinfo {year} {2024})}\BibitemShut
  {NoStop}%
\bibitem [{\citenamefont {Tian}\ \emph {et~al.}(2020)\citenamefont {Tian},
  \citenamefont {Liu}, \citenamefont {Dong}, \citenamefont {Skehan},
  \citenamefont {Zervas}, \citenamefont {Kippenberg},\ and\ \citenamefont
  {Bhave}}]{tian2020bulk}%
  \BibitemOpen
  \bibfield  {author} {\bibinfo {author} {\bibfnamefont {H.}~\bibnamefont
  {Tian}}, \bibinfo {author} {\bibfnamefont {J.}~\bibnamefont {Liu}}, \bibinfo
  {author} {\bibfnamefont {B.}~\bibnamefont {Dong}}, \bibinfo {author}
  {\bibfnamefont {J.~C.}\ \bibnamefont {Skehan}}, \bibinfo {author}
  {\bibfnamefont {M.}~\bibnamefont {Zervas}}, \bibinfo {author} {\bibfnamefont
  {T.~J.}\ \bibnamefont {Kippenberg}},\ and\ \bibinfo {author} {\bibfnamefont
  {S.~A.}\ \bibnamefont {Bhave}},\ }\bibfield  {title} {\bibinfo {title}
  {Hybrid integrated photonics using bulk acoustic resonators},\ }\href@noop {}
  {\bibfield  {journal} {\bibinfo  {journal} {Nature communications}\ }\textbf
  {\bibinfo {volume} {11}},\ \bibinfo {pages} {3073} (\bibinfo {year}
  {2020})}\BibitemShut {NoStop}%
\end{thebibliography}%


\begin{thebibliography}{29}%
\makeatletter
\providecommand \@ifxundefined [1]{%
 \@ifx{#1\undefined}
}%
\providecommand \@ifnum [1]{%
 \ifnum #1\expandafter \@firstoftwo
 \else \expandafter \@secondoftwo
 \fi
}%
\providecommand \@ifx [1]{%
 \ifx #1\expandafter \@firstoftwo
 \else \expandafter \@secondoftwo
 \fi
}%
\providecommand \natexlab [1]{#1}%
\providecommand \enquote  [1]{``#1''}%
\providecommand \bibnamefont  [1]{#1}%
\providecommand \bibfnamefont [1]{#1}%
\providecommand \citenamefont [1]{#1}%
\providecommand \href@noop [0]{\@secondoftwo}%
\providecommand \href [0]{\begingroup \@sanitize@url \@href}%
\providecommand \@href[1]{\@@startlink{#1}\@@href}%
\providecommand \@@href[1]{\endgroup#1\@@endlink}%
\providecommand \@sanitize@url [0]{\catcode `\\12\catcode `\$12\catcode
  `\&12\catcode `\#12\catcode `\^12\catcode `\_12\catcode `\%12\relax}%
\providecommand \@@startlink[1]{}%
\providecommand \@@endlink[0]{}%
\providecommand \url  [0]{\begingroup\@sanitize@url \@url }%
\providecommand \@url [1]{\endgroup\@href {#1}{\urlprefix }}%
\providecommand \urlprefix  [0]{URL }%
\providecommand \Eprint [0]{\href }%
\providecommand \doibase [0]{https://doi.org/}%
\providecommand \selectlanguage [0]{\@gobble}%
\providecommand \bibinfo  [0]{\@secondoftwo}%
\providecommand \bibfield  [0]{\@secondoftwo}%
\providecommand \translation [1]{[#1]}%
\providecommand \BibitemOpen [0]{}%
\providecommand \bibitemStop [0]{}%
\providecommand \bibitemNoStop [0]{.\EOS\space}%
\providecommand \EOS [0]{\spacefactor3000\relax}%
\providecommand \BibitemShut  [1]{\csname bibitem#1\endcsname}%
\let\auto@bib@innerbib\@empty
%</preamble>
\bibitem [{\citenamefont {Tran}\ \emph {et~al.}(2019)\citenamefont {Tran},
  \citenamefont {Huang},\ and\ \citenamefont {Bowers}}]{bowers2019tutorial}%
  \BibitemOpen
  \bibfield  {author} {\bibinfo {author} {\bibfnamefont {M.~A.}\ \bibnamefont
  {Tran}}, \bibinfo {author} {\bibfnamefont {D.}~\bibnamefont {Huang}},\ and\
  \bibinfo {author} {\bibfnamefont {J.~E.}\ \bibnamefont {Bowers}},\ }\href
  {https://doi.org/10.1063/1.5124254} {\bibfield  {journal} {\bibinfo
  {journal} {APL Photonics}\ }\textbf {\bibinfo {volume} {4}},\ \bibinfo
  {pages} {111101} (\bibinfo {year} {2019})}\BibitemShut {NoStop}%
\bibitem [{\citenamefont {Fan}\ \emph {et~al.}(2020)\citenamefont {Fan},
  \citenamefont {van Rees}, \citenamefont {Van~der Slot}, \citenamefont {Mak},
  \citenamefont {Oldenbeuving}, \citenamefont {Hoekman}, \citenamefont
  {Geskus}, \citenamefont {Roeloffzen},\ and\ \citenamefont
  {Boller}}]{fan2020hybrid}%
  \BibitemOpen
  \bibfield  {author} {\bibinfo {author} {\bibfnamefont {Y.}~\bibnamefont
  {Fan}}, \bibinfo {author} {\bibfnamefont {A.}~\bibnamefont {van Rees}},
  \bibinfo {author} {\bibfnamefont {P.~J.}\ \bibnamefont {Van~der Slot}},
  \bibinfo {author} {\bibfnamefont {J.}~\bibnamefont {Mak}}, \bibinfo {author}
  {\bibfnamefont {R.~M.}\ \bibnamefont {Oldenbeuving}}, \bibinfo {author}
  {\bibfnamefont {M.}~\bibnamefont {Hoekman}}, \bibinfo {author} {\bibfnamefont
  {D.}~\bibnamefont {Geskus}}, \bibinfo {author} {\bibfnamefont {C.~G.}\
  \bibnamefont {Roeloffzen}},\ and\ \bibinfo {author} {\bibfnamefont {K.-J.}\
  \bibnamefont {Boller}},\ }\href
  {https://opg.optica.org/oe/fulltext.cfm?uri=oe-28-15-21713&id=433285}
  {\bibfield  {journal} {\bibinfo  {journal} {Optics express}\ }\textbf
  {\bibinfo {volume} {28}},\ \bibinfo {pages} {21713} (\bibinfo {year}
  {2020})}\BibitemShut {NoStop}%
\bibitem [{\citenamefont {van Rees}\ \emph {et~al.}(2023)\citenamefont {van
  Rees}, \citenamefont {Winkler}, \citenamefont {Brochard}, \citenamefont
  {Geskus}, \citenamefont {van~der Slot}, \citenamefont {Nölleke},\ and\
  \citenamefont {Boller}}]{rees2023chilas}%
  \BibitemOpen
  \bibfield  {author} {\bibinfo {author} {\bibfnamefont {A.}~\bibnamefont {van
  Rees}}, \bibinfo {author} {\bibfnamefont {L.~V.}\ \bibnamefont {Winkler}},
  \bibinfo {author} {\bibfnamefont {P.}~\bibnamefont {Brochard}}, \bibinfo
  {author} {\bibfnamefont {D.}~\bibnamefont {Geskus}}, \bibinfo {author}
  {\bibfnamefont {P.~J.~M.}\ \bibnamefont {van~der Slot}}, \bibinfo {author}
  {\bibfnamefont {C.}~\bibnamefont {Nölleke}},\ and\ \bibinfo {author}
  {\bibfnamefont {K.-J.}\ \bibnamefont {Boller}},\ }\href
  {https://doi.org/10.1109/JPHOT.2023.3320393} {\bibfield  {journal} {\bibinfo
  {journal} {IEEE Photonics Journal}\ }\textbf {\bibinfo {volume} {15}},\
  \bibinfo {pages} {1} (\bibinfo {year} {2023})}\BibitemShut {NoStop}%
\bibitem [{\citenamefont {Lihachev}\ \emph {et~al.}(2022)\citenamefont
  {Lihachev}, \citenamefont {Riemensberger}, \citenamefont {Weng},
  \citenamefont {Liu}, \citenamefont {Tian}, \citenamefont {Siddharth},
  \citenamefont {Snigirev}, \citenamefont {Shadymov}, \citenamefont {Voloshin},
  \citenamefont {Wang}, \citenamefont {He}, \citenamefont {Bhave},\ and\
  \citenamefont {Kippenberg}}]{lihachev_low-noise_2022}%
  \BibitemOpen
  \bibfield  {author} {\bibinfo {author} {\bibfnamefont {G.}~\bibnamefont
  {Lihachev}}, \bibinfo {author} {\bibfnamefont {J.}~\bibnamefont
  {Riemensberger}}, \bibinfo {author} {\bibfnamefont {W.}~\bibnamefont {Weng}},
  \bibinfo {author} {\bibfnamefont {J.}~\bibnamefont {Liu}}, \bibinfo {author}
  {\bibfnamefont {H.}~\bibnamefont {Tian}}, \bibinfo {author} {\bibfnamefont
  {A.}~\bibnamefont {Siddharth}}, \bibinfo {author} {\bibfnamefont
  {V.}~\bibnamefont {Snigirev}}, \bibinfo {author} {\bibfnamefont
  {V.}~\bibnamefont {Shadymov}}, \bibinfo {author} {\bibfnamefont
  {A.}~\bibnamefont {Voloshin}}, \bibinfo {author} {\bibfnamefont {R.~N.}\
  \bibnamefont {Wang}}, \bibinfo {author} {\bibfnamefont {J.}~\bibnamefont
  {He}}, \bibinfo {author} {\bibfnamefont {S.~A.}\ \bibnamefont {Bhave}},\ and\
  \bibinfo {author} {\bibfnamefont {T.~J.}\ \bibnamefont {Kippenberg}},\ }\href
  {https://doi.org/10.1038/s41467-022-30911-6} {\bibfield  {journal} {\bibinfo
  {journal} {Nature Communications}\ }\textbf {\bibinfo {volume} {13}},\
  \bibinfo {pages} {3522} (\bibinfo {year} {2022})}\BibitemShut {NoStop}%
\bibitem [{\citenamefont {Li}\ \emph {et~al.}(2021)\citenamefont {Li},
  \citenamefont {Jin}, \citenamefont {Wu}, \citenamefont {Chang}, \citenamefont
  {Wang}, \citenamefont {Shen}, \citenamefont {Yuan}, \citenamefont {Feshali},
  \citenamefont {Paniccia}, \citenamefont {Vahala},\ and\ \citenamefont
  {Bowers}}]{bowers2021reaching}%
  \BibitemOpen
  \bibfield  {author} {\bibinfo {author} {\bibfnamefont {B.}~\bibnamefont
  {Li}}, \bibinfo {author} {\bibfnamefont {W.}~\bibnamefont {Jin}}, \bibinfo
  {author} {\bibfnamefont {L.}~\bibnamefont {Wu}}, \bibinfo {author}
  {\bibfnamefont {L.}~\bibnamefont {Chang}}, \bibinfo {author} {\bibfnamefont
  {H.}~\bibnamefont {Wang}}, \bibinfo {author} {\bibfnamefont {B.}~\bibnamefont
  {Shen}}, \bibinfo {author} {\bibfnamefont {Z.}~\bibnamefont {Yuan}}, \bibinfo
  {author} {\bibfnamefont {A.}~\bibnamefont {Feshali}}, \bibinfo {author}
  {\bibfnamefont {M.}~\bibnamefont {Paniccia}}, \bibinfo {author}
  {\bibfnamefont {K.~J.}\ \bibnamefont {Vahala}},\ and\ \bibinfo {author}
  {\bibfnamefont {J.~E.}\ \bibnamefont {Bowers}},\ }\href
  {https://doi.org/10.1364/OL.439720} {\bibfield  {journal} {\bibinfo
  {journal} {Opt. Lett.}\ }\textbf {\bibinfo {volume} {46}},\ \bibinfo {pages}
  {5201} (\bibinfo {year} {2021})}\BibitemShut {NoStop}%
\bibitem [{\citenamefont {Lihachev}\ \emph {et~al.}(2023)\citenamefont
  {Lihachev}, \citenamefont {Bancora}, \citenamefont {Snigirev}, \citenamefont
  {Tian}, \citenamefont {Riemensberger}, \citenamefont {Shadymov},
  \citenamefont {Siddharth}, \citenamefont {Attanasio}, \citenamefont {Wang},
  \citenamefont {Visani} \emph {et~al.}}]{lihachev2023frequency}%
  \BibitemOpen
  \bibfield  {author} {\bibinfo {author} {\bibfnamefont {G.}~\bibnamefont
  {Lihachev}}, \bibinfo {author} {\bibfnamefont {A.}~\bibnamefont {Bancora}},
  \bibinfo {author} {\bibfnamefont {V.}~\bibnamefont {Snigirev}}, \bibinfo
  {author} {\bibfnamefont {H.}~\bibnamefont {Tian}}, \bibinfo {author}
  {\bibfnamefont {J.}~\bibnamefont {Riemensberger}}, \bibinfo {author}
  {\bibfnamefont {V.}~\bibnamefont {Shadymov}}, \bibinfo {author}
  {\bibfnamefont {A.}~\bibnamefont {Siddharth}}, \bibinfo {author}
  {\bibfnamefont {A.}~\bibnamefont {Attanasio}}, \bibinfo {author}
  {\bibfnamefont {R.~N.}\ \bibnamefont {Wang}}, \bibinfo {author}
  {\bibfnamefont {D.}~\bibnamefont {Visani}}, \emph {et~al.},\ }\href
  {https://arxiv.org/abs/2303.00425} {\bibfield  {journal} {\bibinfo  {journal}
  {arXiv preprint arXiv:2303.00425}\ } (\bibinfo {year} {2023})}\BibitemShut
  {NoStop}%
\bibitem [{\citenamefont {Siddharth}\ \emph
  {et~al.}(2024{\natexlab{a}})\citenamefont {Siddharth}, \citenamefont
  {Attanasio}, \citenamefont {Bianconi}, \citenamefont {Lihachev},
  \citenamefont {Zhang}, \citenamefont {Qiu}, \citenamefont {Bancora},
  \citenamefont {Kenning}, \citenamefont {Wang}, \citenamefont {Voloshin},
  \citenamefont {Bhave}, \citenamefont {Riemensberger},\ and\ \citenamefont
  {Kippenberg}}]{siddharth2023hertz}%
  \BibitemOpen
  \bibfield  {author} {\bibinfo {author} {\bibfnamefont {A.}~\bibnamefont
  {Siddharth}}, \bibinfo {author} {\bibfnamefont {A.}~\bibnamefont
  {Attanasio}}, \bibinfo {author} {\bibfnamefont {S.}~\bibnamefont {Bianconi}},
  \bibinfo {author} {\bibfnamefont {G.}~\bibnamefont {Lihachev}}, \bibinfo
  {author} {\bibfnamefont {J.}~\bibnamefont {Zhang}}, \bibinfo {author}
  {\bibfnamefont {Z.}~\bibnamefont {Qiu}}, \bibinfo {author} {\bibfnamefont
  {A.}~\bibnamefont {Bancora}}, \bibinfo {author} {\bibfnamefont
  {S.}~\bibnamefont {Kenning}}, \bibinfo {author} {\bibfnamefont {R.~N.}\
  \bibnamefont {Wang}}, \bibinfo {author} {\bibfnamefont {A.~S.}\ \bibnamefont
  {Voloshin}}, \bibinfo {author} {\bibfnamefont {S.~A.}\ \bibnamefont {Bhave}},
  \bibinfo {author} {\bibfnamefont {J.}~\bibnamefont {Riemensberger}},\ and\
  \bibinfo {author} {\bibfnamefont {T.~J.}\ \bibnamefont {Kippenberg}},\ }\href
  {https://doi.org/10.1364/OPTICA.524703} {\bibfield  {journal} {\bibinfo
  {journal} {Optica}\ }\textbf {\bibinfo {volume} {11}},\ \bibinfo {pages}
  {1062} (\bibinfo {year} {2024}{\natexlab{a}})}\BibitemShut {NoStop}%
\bibitem [{\citenamefont {Li}\ \emph {et~al.}(2022)\citenamefont {Li},
  \citenamefont {Chang}, \citenamefont {Wu}, \citenamefont {Staffa},
  \citenamefont {Ling}, \citenamefont {Javid}, \citenamefont {Xue},
  \citenamefont {He}, \citenamefont {Lopez-Rios}, \citenamefont {Morin} \emph
  {et~al.}}]{li2022pockels}%
  \BibitemOpen
  \bibfield  {author} {\bibinfo {author} {\bibfnamefont {M.}~\bibnamefont
  {Li}}, \bibinfo {author} {\bibfnamefont {L.}~\bibnamefont {Chang}}, \bibinfo
  {author} {\bibfnamefont {L.}~\bibnamefont {Wu}}, \bibinfo {author}
  {\bibfnamefont {J.}~\bibnamefont {Staffa}}, \bibinfo {author} {\bibfnamefont
  {J.}~\bibnamefont {Ling}}, \bibinfo {author} {\bibfnamefont {U.~A.}\
  \bibnamefont {Javid}}, \bibinfo {author} {\bibfnamefont {S.}~\bibnamefont
  {Xue}}, \bibinfo {author} {\bibfnamefont {Y.}~\bibnamefont {He}}, \bibinfo
  {author} {\bibfnamefont {R.}~\bibnamefont {Lopez-Rios}}, \bibinfo {author}
  {\bibfnamefont {T.~J.}\ \bibnamefont {Morin}}, \emph {et~al.},\ }\href@noop
  {} {\bibfield  {journal} {\bibinfo  {journal} {Nature communications}\
  }\textbf {\bibinfo {volume} {13}},\ \bibinfo {pages} {5344} (\bibinfo {year}
  {2022})}\BibitemShut {NoStop}%
\bibitem [{\citenamefont {Li}\ \emph {et~al.}(2023)\citenamefont {Li},
  \citenamefont {Wang}, \citenamefont {Lihachev}, \citenamefont {Zhang},
  \citenamefont {Tan}, \citenamefont {Churaev}, \citenamefont {Kuznetsov},
  \citenamefont {Siddharth}, \citenamefont {Bereyhi}, \citenamefont
  {Riemensberger} \emph {et~al.}}]{li2023lnoi}%
  \BibitemOpen
  \bibfield  {author} {\bibinfo {author} {\bibfnamefont {Z.}~\bibnamefont
  {Li}}, \bibinfo {author} {\bibfnamefont {R.~N.}\ \bibnamefont {Wang}},
  \bibinfo {author} {\bibfnamefont {G.}~\bibnamefont {Lihachev}}, \bibinfo
  {author} {\bibfnamefont {J.}~\bibnamefont {Zhang}}, \bibinfo {author}
  {\bibfnamefont {Z.}~\bibnamefont {Tan}}, \bibinfo {author} {\bibfnamefont
  {M.}~\bibnamefont {Churaev}}, \bibinfo {author} {\bibfnamefont
  {N.}~\bibnamefont {Kuznetsov}}, \bibinfo {author} {\bibfnamefont
  {A.}~\bibnamefont {Siddharth}}, \bibinfo {author} {\bibfnamefont {M.~J.}\
  \bibnamefont {Bereyhi}}, \bibinfo {author} {\bibfnamefont {J.}~\bibnamefont
  {Riemensberger}}, \emph {et~al.},\ }\href@noop {} {\bibfield  {journal}
  {\bibinfo  {journal} {Nature Communications}\ }\textbf {\bibinfo {volume}
  {14}},\ \bibinfo {pages} {4856} (\bibinfo {year} {2023})}\BibitemShut
  {NoStop}%
\bibitem [{\citenamefont {Xue}\ \emph {et~al.}(2024)\citenamefont {Xue},
  \citenamefont {Li}, \citenamefont {Lopez-rios}, \citenamefont {Ling},
  \citenamefont {Gao}, \citenamefont {Hu}, \citenamefont {Qiu}, \citenamefont
  {Staffa}, \citenamefont {Chang}, \citenamefont {Wang} \emph
  {et~al.}}]{xue2024pockels}%
  \BibitemOpen
  \bibfield  {author} {\bibinfo {author} {\bibfnamefont {S.}~\bibnamefont
  {Xue}}, \bibinfo {author} {\bibfnamefont {M.}~\bibnamefont {Li}}, \bibinfo
  {author} {\bibfnamefont {R.}~\bibnamefont {Lopez-rios}}, \bibinfo {author}
  {\bibfnamefont {J.}~\bibnamefont {Ling}}, \bibinfo {author} {\bibfnamefont
  {Z.}~\bibnamefont {Gao}}, \bibinfo {author} {\bibfnamefont {Q.}~\bibnamefont
  {Hu}}, \bibinfo {author} {\bibfnamefont {T.}~\bibnamefont {Qiu}}, \bibinfo
  {author} {\bibfnamefont {J.}~\bibnamefont {Staffa}}, \bibinfo {author}
  {\bibfnamefont {L.}~\bibnamefont {Chang}}, \bibinfo {author} {\bibfnamefont
  {H.}~\bibnamefont {Wang}}, \emph {et~al.},\ }\href@noop {} {\bibfield
  {journal} {\bibinfo  {journal} {arXiv preprint arXiv:2410.07482}\ } (\bibinfo
  {year} {2024})}\BibitemShut {NoStop}%
\bibitem [{\citenamefont {Siddharth}\ \emph
  {et~al.}(2024{\natexlab{b}})\citenamefont {Siddharth}, \citenamefont
  {Bianconi}, \citenamefont {Wang}, \citenamefont {Qiu}, \citenamefont
  {Voloshin}, \citenamefont {Bereyhi}, \citenamefont {Riemensberger},\ and\
  \citenamefont {Kippenberg}}]{siddharth2024ultrafast}%
  \BibitemOpen
  \bibfield  {author} {\bibinfo {author} {\bibfnamefont {A.}~\bibnamefont
  {Siddharth}}, \bibinfo {author} {\bibfnamefont {S.}~\bibnamefont {Bianconi}},
  \bibinfo {author} {\bibfnamefont {R.~N.}\ \bibnamefont {Wang}}, \bibinfo
  {author} {\bibfnamefont {Z.}~\bibnamefont {Qiu}}, \bibinfo {author}
  {\bibfnamefont {A.~S.}\ \bibnamefont {Voloshin}}, \bibinfo {author}
  {\bibfnamefont {M.~J.}\ \bibnamefont {Bereyhi}}, \bibinfo {author}
  {\bibfnamefont {J.}~\bibnamefont {Riemensberger}},\ and\ \bibinfo {author}
  {\bibfnamefont {T.~J.}\ \bibnamefont {Kippenberg}},\ }\href@noop {}
  {\bibfield  {journal} {\bibinfo  {journal} {arXiv preprint arXiv:2408.01743}\
  } (\bibinfo {year} {2024}{\natexlab{b}})}\BibitemShut {NoStop}%
\bibitem [{\citenamefont {Franken}\ \emph {et~al.}(2024)\citenamefont
  {Franken}, \citenamefont {Cheng}, \citenamefont {Powell}, \citenamefont
  {Kyriazidis}, \citenamefont {Rosborough}, \citenamefont {Musolf},
  \citenamefont {Shah}, \citenamefont {Barton~III}, \citenamefont {Hills},
  \citenamefont {Johansson} \emph {et~al.}}]{franken2024high}%
  \BibitemOpen
  \bibfield  {author} {\bibinfo {author} {\bibfnamefont {C.~A.}\ \bibnamefont
  {Franken}}, \bibinfo {author} {\bibfnamefont {R.}~\bibnamefont {Cheng}},
  \bibinfo {author} {\bibfnamefont {K.}~\bibnamefont {Powell}}, \bibinfo
  {author} {\bibfnamefont {G.}~\bibnamefont {Kyriazidis}}, \bibinfo {author}
  {\bibfnamefont {V.}~\bibnamefont {Rosborough}}, \bibinfo {author}
  {\bibfnamefont {J.}~\bibnamefont {Musolf}}, \bibinfo {author} {\bibfnamefont
  {M.}~\bibnamefont {Shah}}, \bibinfo {author} {\bibfnamefont {D.~R.}\
  \bibnamefont {Barton~III}}, \bibinfo {author} {\bibfnamefont
  {G.}~\bibnamefont {Hills}}, \bibinfo {author} {\bibfnamefont
  {L.}~\bibnamefont {Johansson}}, \emph {et~al.},\ }\href@noop {} {\bibfield
  {journal} {\bibinfo  {journal} {arXiv preprint arXiv:2407.00269}\ } (\bibinfo
  {year} {2024})}\BibitemShut {NoStop}%
\bibitem [{\citenamefont {{RIO - Redfern Integrated
  Optics}}(2014)}]{rio_planex_2014}%
  \BibitemOpen
  \bibfield  {author} {\bibinfo {author} {\bibnamefont {{RIO - Redfern
  Integrated Optics}}},\ }\href
  {https://morephotonics.com/wp-content/uploads/2020/02/Rio_Planex_Product-Brief_1.24.14.pdf}
  {\bibinfo {title} {Rio planex product brief}} (\bibinfo {year} {2014}),\
  \bibinfo {note} {accessed: 2024-09-09}\BibitemShut {NoStop}%
\bibitem [{thi(2024)}]{this_work}%
  \BibitemOpen
  \href@noop {} {\bibinfo {title} {This work}} (\bibinfo {year} {2024}),\
  \bibinfo {note} {this work presents original research conducted by the
  authors.}\BibitemShut {Stop}%
\bibitem [{\citenamefont {Inc.}(2023)}]{oewaves2023}%
  \BibitemOpen
  \bibfield  {author} {\bibinfo {author} {\bibfnamefont {O.}~\bibnamefont
  {Inc.}},\ }\href@noop {} {\bibfield  {journal} {\bibinfo  {journal}
  {Specifications}\ } (\bibinfo {year} {2023})},\ \bibinfo {note}
  {\url{https://morephotonics.com/wp-content/uploads/2023/02/OE4040.pdf}}\BibitemShut
  {NoStop}%
\bibitem [{\citenamefont {Liang}\ \emph {et~al.}(2015)\citenamefont {Liang},
  \citenamefont {Ilchenko}, \citenamefont {Eliyahu}, \citenamefont
  {Savchenkov}, \citenamefont {Matsko}, \citenamefont {Seidel},\ and\
  \citenamefont {Maleki}}]{liang2015ultralow}%
  \BibitemOpen
  \bibfield  {author} {\bibinfo {author} {\bibfnamefont {W.}~\bibnamefont
  {Liang}}, \bibinfo {author} {\bibfnamefont {V.~S.}\ \bibnamefont {Ilchenko}},
  \bibinfo {author} {\bibfnamefont {D.}~\bibnamefont {Eliyahu}}, \bibinfo
  {author} {\bibfnamefont {A.~A.}\ \bibnamefont {Savchenkov}}, \bibinfo
  {author} {\bibfnamefont {A.~B.}\ \bibnamefont {Matsko}}, \bibinfo {author}
  {\bibfnamefont {D.}~\bibnamefont {Seidel}},\ and\ \bibinfo {author}
  {\bibfnamefont {L.}~\bibnamefont {Maleki}},\ }\href
  {https://doi.org/10.1038/ncomms8371} {\bibfield  {journal} {\bibinfo
  {journal} {Nature Communications}\ }\textbf {\bibinfo {volume} {6}},\
  \bibinfo {pages} {7371} (\bibinfo {year} {2015})}\BibitemShut {NoStop}%
\bibitem [{\citenamefont {Voloshin}\ \emph {et~al.}(2021)\citenamefont
  {Voloshin}, \citenamefont {Kondratiev}, \citenamefont {Lihachev},
  \citenamefont {Liu}, \citenamefont {Lobanov}, \citenamefont {Dmitriev},
  \citenamefont {Weng}, \citenamefont {Kippenberg},\ and\ \citenamefont
  {Bilenko}}]{voloshin2021dynamics}%
  \BibitemOpen
  \bibfield  {author} {\bibinfo {author} {\bibfnamefont {A.~S.}\ \bibnamefont
  {Voloshin}}, \bibinfo {author} {\bibfnamefont {N.~M.}\ \bibnamefont
  {Kondratiev}}, \bibinfo {author} {\bibfnamefont {G.~V.}\ \bibnamefont
  {Lihachev}}, \bibinfo {author} {\bibfnamefont {J.}~\bibnamefont {Liu}},
  \bibinfo {author} {\bibfnamefont {V.~E.}\ \bibnamefont {Lobanov}}, \bibinfo
  {author} {\bibfnamefont {N.~Y.}\ \bibnamefont {Dmitriev}}, \bibinfo {author}
  {\bibfnamefont {W.}~\bibnamefont {Weng}}, \bibinfo {author} {\bibfnamefont
  {T.~J.}\ \bibnamefont {Kippenberg}},\ and\ \bibinfo {author} {\bibfnamefont
  {I.~A.}\ \bibnamefont {Bilenko}},\ }\href@noop {} {\bibfield  {journal}
  {\bibinfo  {journal} {Nature communications}\ }\textbf {\bibinfo {volume}
  {12}},\ \bibinfo {pages} {235} (\bibinfo {year} {2021})}\BibitemShut
  {NoStop}%
\bibitem [{\citenamefont {Ulanov}\ \emph {et~al.}(2024)\citenamefont {Ulanov},
  \citenamefont {Wildi}, \citenamefont {Bhatnagar},\ and\ \citenamefont
  {Herr}}]{ulanov2024laser}%
  \BibitemOpen
  \bibfield  {author} {\bibinfo {author} {\bibfnamefont {A.~E.}\ \bibnamefont
  {Ulanov}}, \bibinfo {author} {\bibfnamefont {T.}~\bibnamefont {Wildi}},
  \bibinfo {author} {\bibfnamefont {U.}~\bibnamefont {Bhatnagar}},\ and\
  \bibinfo {author} {\bibfnamefont {T.}~\bibnamefont {Herr}},\ }\href@noop {}
  {\bibfield  {journal} {\bibinfo  {journal} {arXiv preprint arXiv:2408.08679}\
  } (\bibinfo {year} {2024})}\BibitemShut {NoStop}%
\bibitem [{\citenamefont {Jin}\ \emph {et~al.}(2021)\citenamefont {Jin},
  \citenamefont {Yang}, \citenamefont {Chang}, \citenamefont {Shen},
  \citenamefont {Wang}, \citenamefont {Leal}, \citenamefont {Wu}, \citenamefont
  {Gao}, \citenamefont {Feshali}, \citenamefont {Paniccia} \emph
  {et~al.}}]{jin2021hertz}%
  \BibitemOpen
  \bibfield  {author} {\bibinfo {author} {\bibfnamefont {W.}~\bibnamefont
  {Jin}}, \bibinfo {author} {\bibfnamefont {Q.-F.}\ \bibnamefont {Yang}},
  \bibinfo {author} {\bibfnamefont {L.}~\bibnamefont {Chang}}, \bibinfo
  {author} {\bibfnamefont {B.}~\bibnamefont {Shen}}, \bibinfo {author}
  {\bibfnamefont {H.}~\bibnamefont {Wang}}, \bibinfo {author} {\bibfnamefont
  {M.~A.}\ \bibnamefont {Leal}}, \bibinfo {author} {\bibfnamefont
  {L.}~\bibnamefont {Wu}}, \bibinfo {author} {\bibfnamefont {M.}~\bibnamefont
  {Gao}}, \bibinfo {author} {\bibfnamefont {A.}~\bibnamefont {Feshali}},
  \bibinfo {author} {\bibfnamefont {M.}~\bibnamefont {Paniccia}}, \emph
  {et~al.},\ }\href@noop {} {\bibfield  {journal} {\bibinfo  {journal} {Nature
  Photonics}\ }\textbf {\bibinfo {volume} {15}},\ \bibinfo {pages} {346}
  (\bibinfo {year} {2021})}\BibitemShut {NoStop}%
\bibitem [{\citenamefont {Ousaid}\ \emph {et~al.}(2024)\citenamefont {Ousaid},
  \citenamefont {Bourcier}, \citenamefont {Fernandez}, \citenamefont {Llopis},
  \citenamefont {Lumeau}, \citenamefont {Moreau}, \citenamefont {Bunel},
  \citenamefont {Conforti}, \citenamefont {Mussot}, \citenamefont {Crozatier},\
  and\ \citenamefont {Balac}}]{ousaid2024SIL}%
  \BibitemOpen
  \bibfield  {author} {\bibinfo {author} {\bibfnamefont {S.~M.}\ \bibnamefont
  {Ousaid}}, \bibinfo {author} {\bibfnamefont {G.}~\bibnamefont {Bourcier}},
  \bibinfo {author} {\bibfnamefont {A.}~\bibnamefont {Fernandez}}, \bibinfo
  {author} {\bibfnamefont {O.}~\bibnamefont {Llopis}}, \bibinfo {author}
  {\bibfnamefont {J.}~\bibnamefont {Lumeau}}, \bibinfo {author} {\bibfnamefont
  {A.}~\bibnamefont {Moreau}}, \bibinfo {author} {\bibfnamefont
  {T.}~\bibnamefont {Bunel}}, \bibinfo {author} {\bibfnamefont
  {M.}~\bibnamefont {Conforti}}, \bibinfo {author} {\bibfnamefont
  {A.}~\bibnamefont {Mussot}}, \bibinfo {author} {\bibfnamefont
  {V.}~\bibnamefont {Crozatier}},\ and\ \bibinfo {author} {\bibfnamefont
  {S.}~\bibnamefont {Balac}},\ }\href {https://doi.org/10.1364/OL.514778}
  {\bibfield  {journal} {\bibinfo  {journal} {Opt. Lett.}\ }\textbf {\bibinfo
  {volume} {49}},\ \bibinfo {pages} {1933} (\bibinfo {year}
  {2024})}\BibitemShut {NoStop}%
\bibitem [{\citenamefont {Adler}(1946)}]{adler1946study}%
  \BibitemOpen
  \bibfield  {author} {\bibinfo {author} {\bibfnamefont {R.}~\bibnamefont
  {Adler}},\ }\href@noop {} {\bibfield  {journal} {\bibinfo  {journal}
  {Proceedings of the IRE}\ }\textbf {\bibinfo {volume} {34}},\ \bibinfo
  {pages} {351} (\bibinfo {year} {1946})}\BibitemShut {NoStop}%
\bibitem [{\citenamefont {Chang}(2003)}]{chang2003phase}%
  \BibitemOpen
  \bibfield  {author} {\bibinfo {author} {\bibfnamefont {H.-C.}\ \bibnamefont
  {Chang}},\ }\href@noop {} {\bibfield  {journal} {\bibinfo  {journal} {IEEE
  Transactions on microwave theory and techniques}\ }\textbf {\bibinfo {volume}
  {51}},\ \bibinfo {pages} {1994} (\bibinfo {year} {2003})}\BibitemShut
  {NoStop}%
\bibitem [{\citenamefont {Torres-Company}\ \emph {et~al.}(2022)\citenamefont
  {Torres-Company}, \citenamefont {Ye}, \citenamefont {Zhao}, \citenamefont
  {Karlsson},\ and\ \citenamefont {Andrekson}}]{torres2022ultralow}%
  \BibitemOpen
  \bibfield  {author} {\bibinfo {author} {\bibfnamefont {V.}~\bibnamefont
  {Torres-Company}}, \bibinfo {author} {\bibfnamefont {Z.}~\bibnamefont {Ye}},
  \bibinfo {author} {\bibfnamefont {P.}~\bibnamefont {Zhao}}, \bibinfo {author}
  {\bibfnamefont {M.}~\bibnamefont {Karlsson}},\ and\ \bibinfo {author}
  {\bibfnamefont {P.~A.}\ \bibnamefont {Andrekson}},\ }in\ \href
  {https://opg.optica.org/abstract.cfm?uri=OFC-2022-W4J.3} {\emph {\bibinfo
  {booktitle} {2022 Optical Fiber Communications Conference and Exhibition
  (OFC)}}}\ (\bibinfo {organization} {IEEE},\ \bibinfo {year} {2022})\ pp.\
  \bibinfo {pages} {1--3}\BibitemShut {NoStop}%
\bibitem [{\citenamefont {Puckett}\ \emph {et~al.}(2021)\citenamefont
  {Puckett}, \citenamefont {Liu}, \citenamefont {Chauhan}, \citenamefont
  {Zhao}, \citenamefont {Jin}, \citenamefont {Cheng}, \citenamefont {Wu},
  \citenamefont {Behunin}, \citenamefont {Rakich}, \citenamefont {Nelson} \emph
  {et~al.}}]{puckett2021422}%
  \BibitemOpen
  \bibfield  {author} {\bibinfo {author} {\bibfnamefont {M.~W.}\ \bibnamefont
  {Puckett}}, \bibinfo {author} {\bibfnamefont {K.}~\bibnamefont {Liu}},
  \bibinfo {author} {\bibfnamefont {N.}~\bibnamefont {Chauhan}}, \bibinfo
  {author} {\bibfnamefont {Q.}~\bibnamefont {Zhao}}, \bibinfo {author}
  {\bibfnamefont {N.}~\bibnamefont {Jin}}, \bibinfo {author} {\bibfnamefont
  {H.}~\bibnamefont {Cheng}}, \bibinfo {author} {\bibfnamefont
  {J.}~\bibnamefont {Wu}}, \bibinfo {author} {\bibfnamefont {R.~O.}\
  \bibnamefont {Behunin}}, \bibinfo {author} {\bibfnamefont {P.~T.}\
  \bibnamefont {Rakich}}, \bibinfo {author} {\bibfnamefont {K.~D.}\
  \bibnamefont {Nelson}}, \emph {et~al.},\ }\href
  {https://opg.optica.org/oe/fulltext.cfm?uri=oe-20-20-22819&id=242392}
  {\bibfield  {journal} {\bibinfo  {journal} {Nature communications}\ }\textbf
  {\bibinfo {volume} {12}},\ \bibinfo {pages} {934} (\bibinfo {year}
  {2021})}\BibitemShut {NoStop}%
\bibitem [{\citenamefont {Tian}\ \emph {et~al.}(2020)\citenamefont {Tian},
  \citenamefont {Liu}, \citenamefont {Dong}, \citenamefont {Skehan},
  \citenamefont {Zervas}, \citenamefont {Kippenberg},\ and\ \citenamefont
  {Bhave}}]{tian2020hybrid}%
  \BibitemOpen
  \bibfield  {author} {\bibinfo {author} {\bibfnamefont {H.}~\bibnamefont
  {Tian}}, \bibinfo {author} {\bibfnamefont {J.}~\bibnamefont {Liu}}, \bibinfo
  {author} {\bibfnamefont {B.}~\bibnamefont {Dong}}, \bibinfo {author}
  {\bibfnamefont {J.~C.}\ \bibnamefont {Skehan}}, \bibinfo {author}
  {\bibfnamefont {M.}~\bibnamefont {Zervas}}, \bibinfo {author} {\bibfnamefont
  {T.~J.}\ \bibnamefont {Kippenberg}},\ and\ \bibinfo {author} {\bibfnamefont
  {S.~A.}\ \bibnamefont {Bhave}},\ }\href
  {https://doi.org/10.1038/s41467-020-16812-6} {\bibfield  {journal} {\bibinfo
  {journal} {Nature Communications}\ }\textbf {\bibinfo {volume} {11}},\
  \bibinfo {pages} {3073} (\bibinfo {year} {2020})}\BibitemShut {NoStop}%
\bibitem [{\citenamefont {Liu}\ \emph {et~al.}(2016)\citenamefont {Liu},
  \citenamefont {Brasch}, \citenamefont {Pfeiffer}, \citenamefont {Kordts},
  \citenamefont {Kamel}, \citenamefont {Guo}, \citenamefont {Geiselmann},\ and\
  \citenamefont {Kippenberg}}]{Liu_char_setup}%
  \BibitemOpen
  \bibfield  {author} {\bibinfo {author} {\bibfnamefont {J.}~\bibnamefont
  {Liu}}, \bibinfo {author} {\bibfnamefont {V.}~\bibnamefont {Brasch}},
  \bibinfo {author} {\bibfnamefont {M.~H.~P.}\ \bibnamefont {Pfeiffer}},
  \bibinfo {author} {\bibfnamefont {A.}~\bibnamefont {Kordts}}, \bibinfo
  {author} {\bibfnamefont {A.~N.}\ \bibnamefont {Kamel}}, \bibinfo {author}
  {\bibfnamefont {H.}~\bibnamefont {Guo}}, \bibinfo {author} {\bibfnamefont
  {M.}~\bibnamefont {Geiselmann}},\ and\ \bibinfo {author} {\bibfnamefont
  {T.~J.}\ \bibnamefont {Kippenberg}},\ }\href
  {https://doi.org/10.1364/OL.41.003134} {\bibfield  {journal} {\bibinfo
  {journal} {Opt. Lett.}\ }\textbf {\bibinfo {volume} {41}},\ \bibinfo {pages}
  {3134} (\bibinfo {year} {2016})}\BibitemShut {NoStop}%
\bibitem [{\citenamefont {Riemensberger}\ \emph {et~al.}(2022)\citenamefont
  {Riemensberger}, \citenamefont {Kuznetsov}, \citenamefont {Liu},
  \citenamefont {He}, \citenamefont {Wang},\ and\ \citenamefont
  {Kippenberg}}]{riemensberger2022TWPA}%
  \BibitemOpen
  \bibfield  {author} {\bibinfo {author} {\bibfnamefont {J.}~\bibnamefont
  {Riemensberger}}, \bibinfo {author} {\bibfnamefont {N.}~\bibnamefont
  {Kuznetsov}}, \bibinfo {author} {\bibfnamefont {J.}~\bibnamefont {Liu}},
  \bibinfo {author} {\bibfnamefont {J.}~\bibnamefont {He}}, \bibinfo {author}
  {\bibfnamefont {R.~N.}\ \bibnamefont {Wang}},\ and\ \bibinfo {author}
  {\bibfnamefont {T.~J.}\ \bibnamefont {Kippenberg}},\ }\href@noop {}
  {\bibfield  {journal} {\bibinfo  {journal} {Nature}\ }\textbf {\bibinfo
  {volume} {612}},\ \bibinfo {pages} {56} (\bibinfo {year} {2022})}\BibitemShut
  {NoStop}%
\bibitem [{\citenamefont {Demirci}\ and\ \citenamefont
  {Nguyen}(2003)}]{Demirci03IEEE}%
  \BibitemOpen
  \bibfield  {author} {\bibinfo {author} {\bibfnamefont {M.~U.}\ \bibnamefont
  {Demirci}}\ and\ \bibinfo {author} {\bibfnamefont {C.~T.-C.}\ \bibnamefont
  {Nguyen}},\ }in\ \href@noop {} {\emph {\bibinfo {booktitle} {IEEE
  International Frequency Symposium and PDA Exhibition Jointly with the 17th
  European Freqeuncy and Time Forum}}}\ (\bibinfo {year} {2003})\BibitemShut
  {NoStop}%
\bibitem [{\citenamefont {Behroozpour}\ \emph {et~al.}(2017)\citenamefont
  {Behroozpour}, \citenamefont {Sandborn}, \citenamefont {Wu},\ and\
  \citenamefont {Boser}}]{behroozpour2017lidar}%
  \BibitemOpen
  \bibfield  {author} {\bibinfo {author} {\bibfnamefont {B.}~\bibnamefont
  {Behroozpour}}, \bibinfo {author} {\bibfnamefont {P.~A.}\ \bibnamefont
  {Sandborn}}, \bibinfo {author} {\bibfnamefont {M.~C.}\ \bibnamefont {Wu}},\
  and\ \bibinfo {author} {\bibfnamefont {B.~E.}\ \bibnamefont {Boser}},\
  }\href@noop {} {\bibfield  {journal} {\bibinfo  {journal} {IEEE
  Communications Magazine}\ }\textbf {\bibinfo {volume} {55}},\ \bibinfo
  {pages} {135} (\bibinfo {year} {2017})}\BibitemShut {NoStop}%
\end{thebibliography}%
	
\end{document}

	%\includepdf[landscape=false]{Science_SM_Cover.pdf}
	\title{Supplementary Information for: Monolithic piezoelectrically tunable hybrid integrated laser with sub-fiber laser coherence}
	
\author{Andrey Voloshin$^{1,2,3,4}$}\thanks{authors contributed equally}
\email[]{andrey.voloshin@epfl.ch}

\author{Anat Siddharth$^{1,2,3}$}\thanks{authors contributed equally}

\author{Simone Bianconi$^{1,2,3}$}\thanks{authors contributed equally}

\author{Alaina Attanasio$^{5}$}\thanks{authors contributed equally}

\author{Andrea Bancora$^{1,2,3,4}$}\thanks{authors contributed equally}

\author{Vladimir Shadymov$^{1,2,3,4}$}\thanks{authors contributed equally}

\author{Sebastien Leni$^{4}$}

\author{Rui Ning Wang$^{1,2,3}$}

\author{Johann Riemensberger$^{1,2,3}$}

\author{Sunil A. Bhave$^{5}$}
\email[]{sunil.bhave@epfl.ch}

\author{Tobias J. Kippenberg$^{1,2,3,4}$}
\email[]{tobias.kippenberg@epfl.ch}

\affiliation{
	\mbox{$^1$Institute of Physics, Swiss Federal Institute of Technology Lausanne (EPFL), CH-1015 Lausanne, Switzerland}
	\mbox{$^2$Center for Quantum Science and Engineering, EPFL, CH-1015 Lausanne, Switzerland}
	\mbox{$^3$Institute of Electrical and Micro-Engineering, EPFL, CH-1015 Lausanne, Switzerland}
	\mbox{$^4$Deeplight SA, St Sulpice CH-1025, Switzerland}\\
	\mbox{$^5$OxideMEMS Lab, Purdue University, 47907 West Lafayette, IN, USA}\\
}
	
	%%%%%% RESET EQUATION NUMBERS ETC. %%%%%%%%
	\setcounter{equation}{0}
	\setcounter{figure}{0}
	\setcounter{table}{0}
	
	\setcounter{subsection}{0}
	\setcounter{section}{0}
	\setcounter{secnumdepth}{3}
	
	\maketitle
	{\hypersetup{linkcolor=blue}\tableofcontents}
	\newpage

	\section{Benchmark of fully integrated photonic chip-based ultra-low noise lasers}
	
	\begin{figure*}[htb]
		\centering
		\includegraphics[width=\textwidth]{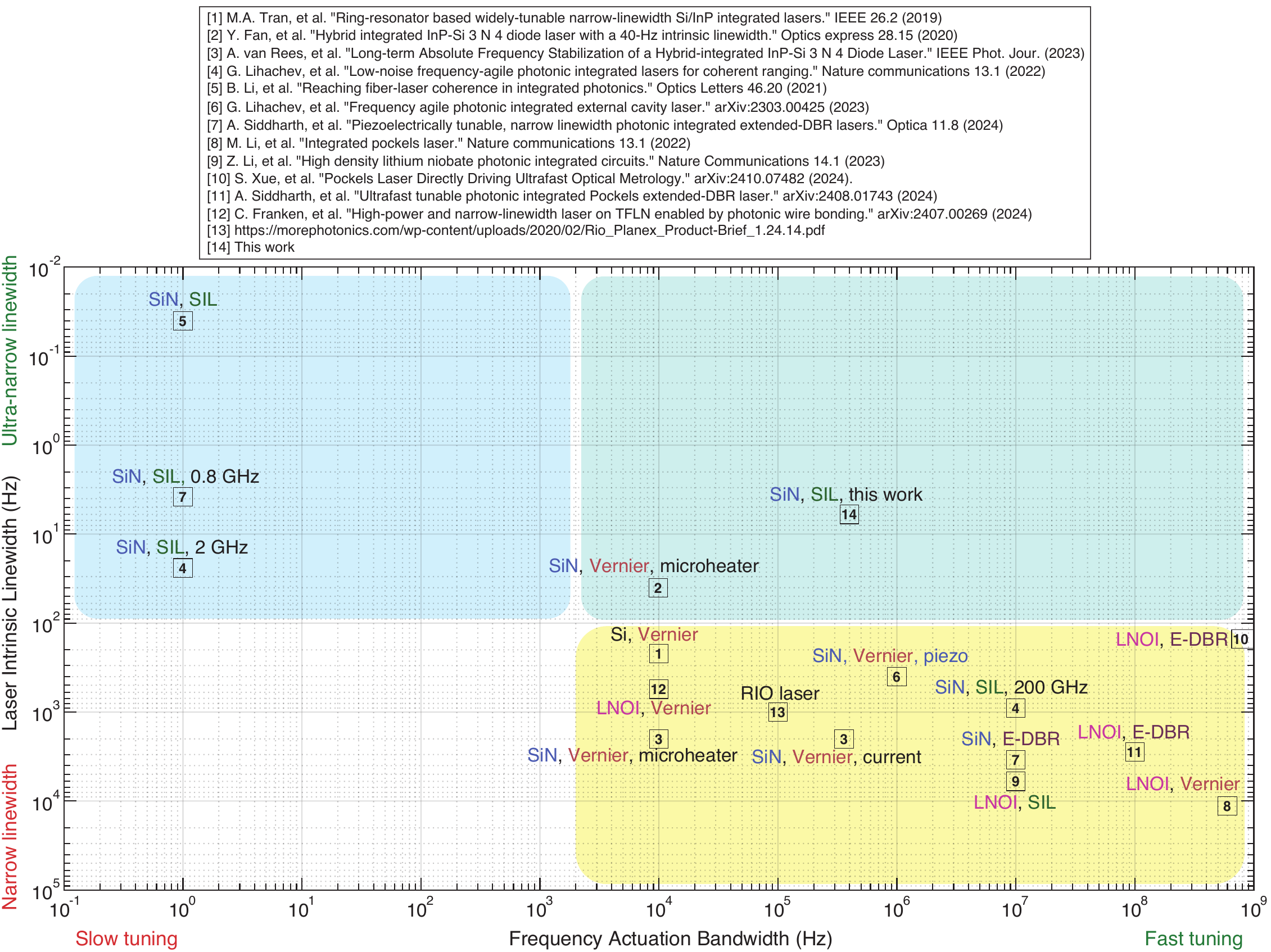}
		\caption{ 
			\footnotesize
			\textbf{Benchmark of fully integrated photonic chip-based ultra-low noise laser sources.} Each point corresponds to different publications with reported Lorentzian linewidth and actuation bandwidth.}
		\label{Fig:benchmark}
	\end{figure*}
	
	We benchmarked all recent advances in the development of photonic chip-based ultra-low noise lasers. It is important to mention that we do not consider systems  stabilized to external cavities using electronic feedback. The first work \cite{bowers2019tutorial} features an a dual-ring Vernier laser with an InP gain section based on a silicon waveguide platform. The intrinsic linewidth is 220 Hz, while the tuneability is enabled by microheaters resulting in a relatively slow tuning of 10 kHz. The next work \cite{fan2020hybrid} demonstrates a remarkable result of a \SiN photonic chip-based laser with a InP gain chip and a dual/triple ring Vernier filter with 40 kHz intrinsic linewidth. The researchers do not report on the actuation bandwidth but it is possible to estimate to be 10 kH, which is reasonable for microheater performance.
	
	The work \cite{rees2023chilas} demonstrates the state-of-the-art performance of commercially available \SiN photonic-chip based Vernier lasers with 2 kHz linewidth. The authors report on the actuation bandwidth of the laser using different mechanisms. They find that the bandwidth for thermal tuning by the phase section is limited to approximately 50 kHz, while tuning the diode laser frequency by the amplifier current provides a higher bandwidth of at least 0.33 MHz, which can be limited by the equipment used for that measurement.
	
	The early demonstration of a \SiN self-injection locked (SIL) laser \cite{lihachev_low-noise_2022} with piezo-actuators is represented by two points. The first one corresponds to an ultra-low noise SIL laser based on 2.45 GHz ring without any actuator and the intrinsic linewidth of 24 Hz with, presumably, 1 Hz actuation bandwidth by a temperature control of the entire laser package. The second point corresponds to a 200 GHz ring-based SIL laser with 900 Hz of the intrinsic linewidth and 10 MHz actuation bandwidth. However, the same architecture enables an exceptionally low intrinsic linewidth of 0.04 Hz using a 125 MHz spiral cavity \cite{bowers2021reaching} without any fast tunability (the actuation bandwidth of 1 Hz at the plot).
	
	The recent result \cite{lihachev2023frequency} shows 1 MHz actuation bandwidth of a \SiN photonic-chip based laser with dual-ring Vernier filter and PZT actuators and 400 Hz linewidth.
	
	Another platform which can demonstrate exceptional performance is a laser with a gain chip and an extended distributed Bragg reflector (E-DBR). A free-running E-DBR laser \cite{siddharth2023hertz} has a linewidth of 3.45 kHz and 10 MHz actuation bandwidth. The combination of two platform, SIL and E-DBR, significantly improves the linewidth down to 3.8 Hz. However, the ultra-low noise SIL E-DBR laser has no tunability at this demonstration, the actuation bandwidth is 1 Hz .
	
	An emerging platform based on thin-film lithium niobate allows for the unprecedentedly fast tuning: 600 MHz \cite{li2022pockels}, 10 MHz \cite{li2023lnoi}, 1-10 GHz \cite{xue2024pockels}, 100 MHz \cite{siddharth2024ultrafast}, 10 kHz with microheaters \cite{franken2024high}. However, the laser intrinsic linewidth does not go below 167 Hz.
	
	In the end, we include the data from a specifications sheet of a commercial Si chip-based E-DBR laser \cite{rio_planex_2014} with 1 kHz linewidth and 100 kHz actuation bandwidth. In this work \cite{this_work}, we demonstrate a 1.5 SIL laser with 6 Hz intrinsic linewidth and at least 400 kHz actuation bandwidth. These results are shown in Fig. \ref{Fig:benchmark} demonstrating how it differentiates from the state-of-the-art. All lasers are measured using the OEWaves OE4000 laser frequency noise analyzer except two curves: the whispering gallery mode-based SIL laser OEWaves OE4040-XLN \cite{oewaves2023, liang2015ultralow} and the 125 MHz SIL laser \cite{bowers2021reaching}. Please note that the SIL 250 MHz laser laser still demonstrates the kHz-level of intrinsic linewidth even outside the locked state (Fig. \ref{Fig:benchmark}, black solid curve).
	
	The fiber laser Koheras Adjustik E15 has a FN with a relaxation oscillations feature at approximately 400 kHz and ultra-low white noise floor reaching 0.2 $ \rm{Hz}^\textsuperscript{2}/\rm{Hz}$ after a 3 MHz frequency offset (red curve). The fiber laser $\beta$-linewidth with an integration time of $\tau_0 =$ 0.1 s is 2.17 kHz. The SIL lasers allow us to have similar levels of FN and integrated linewidth.
	
	The state-of-the-art OEWaves laser OE4040-XLN FN reaches the level of 10$^{-2}$ Hz$^{2}/$ Hz after 1 MHz. But the noise at lower frequency offsets is higher than SIL 250 MHz resulting in an integrated linewidth of 2.47 kHz ($\tau_0 = $ 0.1 s). The SIL 135 MHz laser \cite{bowers2021reaching} reaches the level $10^\textsuperscript{-2} \rm{Hz}^\textsuperscript{2}/\rm{Hz}$ after 1 MHz. The integrated linewidth ($\tau_0 = $ 0.1 s) of 1.23 kHz is very similar to the SIL 250 MHz presented in the current work. However, this is important to keep the PIC size small to benefit from wafer-scale fabrication.
	
	It is important to mention that the presented SIL 1.5 GHz and SIL 250 MHz lasers are still not limited by TRN noise at almost all frequency offsets. Several factors critical for manufacturability and wafer-scale fabrication limit the FN performance but they allowed us to develop a robust laser architecture with high output power and frequency-agility. First, the top cladding is only 3 $\mu$m, in combination with metal microheaters and metal pads of piezo-actuators this introduces additional ohmic losses (at least 0.1 dB/m) and can reduce the Q-factors of cavities in comparison with >10 $\mu$m cladding. Second, the coupling rates to the bus waveguide and a drop-port are optimized to achieve high output power in combination with sufficient FN.
	
	\begin{figure*}[htb]
		\centering
		\includegraphics[width=\textwidth]{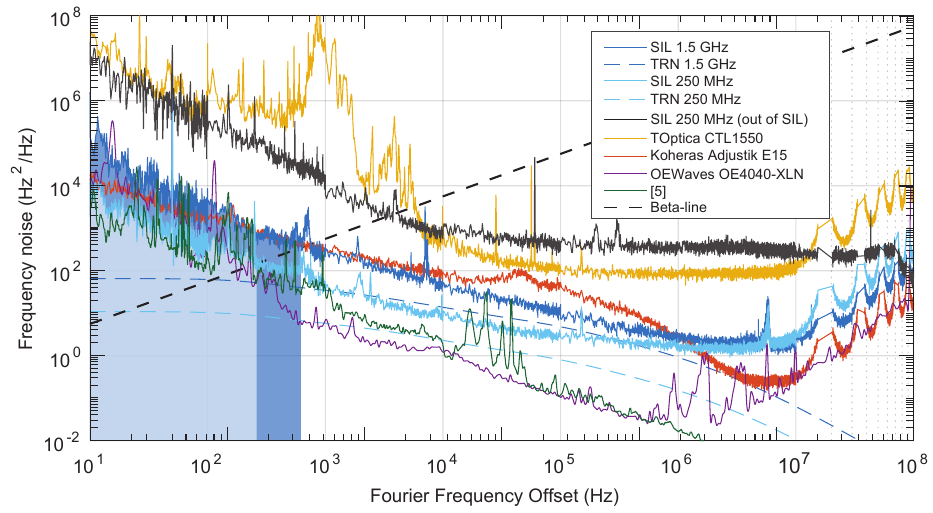}
		\caption{ 
			\footnotesize
			\textbf{Extended frequency noise spectra.} All lasers are measured using the OEWaves OE4000 laser frequency noise analyzer except two curves: the whispering gallery mode-based SIL laser OEWaves OE4040-XLN \cite{oewaves2023, liang2015ultralow} and the 125 MHz SIL laser \cite{bowers2021reaching}.
		}
		\label{Fig:FN_full}
	\end{figure*}
	
	\section{Analytical estimation of the SIL frequency noise}
	\noindent Finally, we compare the measured FN of SIL lasers with analytical estimations. Recently, different approaches have been proposed to quantitatively estimate the noise and linewidth reduction based on SIL  to PIC-based high-Q photonic microresonators \cite{voloshin2021dynamics, ulanov2024laser, jin2021hertz}. However, the verification of the models is complicated, as all these models predict noise levels for ultra-high-Q cavity-based SIL which are below the thermal refractive noise (TRN) of cavities.
	
	In the case of SIL of a fiber-coupled distributed feedback (DFB) laser to a high-Q fiber cavity, a simple model is employed to assess whether the observed noise performance could be accounted for by the rise of the resonator Q factor and to evaluate other possible noise contributions on the locked laser \cite{ousaid2024SIL}. This model is based on the theory describing the injection locking of microwave oscillators \cite{adler1946study}. It is then applied to the case of self-injection of microwave oscillators \cite{chang2003phase}. This model provides the frequency-dependent FN suppression coefficient, governing the noise reduction in a SIL laser with respect to that of the free-running DFB:
	
	\begin{equation}
		S_{\phi}\textsuperscript{locked}(f) = \frac{1 + 4 \left(\frac{f}{\nu_0}\right)^2 Q_r^2}{\left(1 + \rho \frac{Q_r}{Q_{\text{laser}}}\right)^2 + 4 \left(\frac{f}{\nu_0}\right)^2 Q_r^2} S_{\phi}\textsuperscript{free}(f)
		\label{eq:SIL_model}
	\end{equation}
	
	where $f$ is the Fourier frequency offset, $S_{\phi}\textsuperscript{free}$ the FN of a free-running DFB laser, $S_{\phi}\textsuperscript{locked}$ the FN of a SIL DFB laser, $Q_{\text{laser}}$ and $Q_{\text{r}}$ the Q-factors of a DBF laser cavity and of a high-Q spiral cavity respectively, and $\nu_0$ the optical frequency. Note that, $\rho$ is the absolute amplitude ratio of the feedback signal to the laser output signal, which includes all coupling losses. In our case, we assumed total $\rho$ = 0.05, $Q_{\text{r}} = 10^7$, $Q_{\text{laser}} = 10^4$.
	
	\section{Pull-back fabrication process of electrodes}
	\noindent The "Pull-back process" for fabrication of electrodes is developed for this work and the photonic packaging of hybrid integrated lasers based on PICs with piezo-actuators. It minimizes the capacitance of bond-pads and for the ease of wire-bonding to the bond-pads. The details of fabrication proces are discussed \cite{siddharth2023hertz}. We start with a layer stack of top Mo, AlN, and bottom Mo on top of the SiO$_2$. 
	We have used deep reactive-ion etching to pattern the Mo and AlN layers.
	Prior to the pullback process, the top-electrode bond pad had a layer of AlN and bottom Mo underneath it, creating a capacitor.
	Given the width of the bond pad is more than double the width of the device itself, the capacitance added is significant and should be avoided to reduce the RC time constant.
	We can eliminate the capacitance by etching through the top Mo, AlN, and bottom Mo layers and depositing a layer of Al to connect the top electrode bond pad.
	However, the deposited Al must be electrically isolated from the bottom Mo. 
	Thus, we add the “Pull-back” step for which the process is named to create an air gap between the top and bottom metals.
	XeF$_2$ etches Mo but has a near infinite selectivity with AlN. 
	Therefore, using an isotropic etching process, the Mo can be slightly etched away without harming the AlN. 
	
\begin{figure*}[htb]
		\centering
		\includegraphics[width=\textwidth]{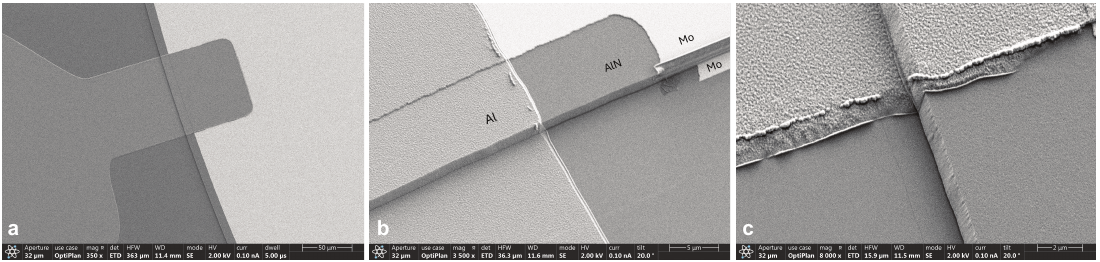}
		\caption{ 
			\footnotesize
			\textbf{Pull-back fabrication process.} (a, b, c) SEM images of the electrodes fabricated by this fabrication technique.
		}
		\label{Fig:Pull-back}
	\end{figure*}

	\noindent We “pull-back” the Mo from underneath the AlN between 1 and 2 $\mu$m, and then deposit a layer of Al with a lift-off process.
	A glancing angle deposition tool evaporates 200~nm of Al with a 15$^\circ$ angle on a rotating plate to make sure the Al covers the 1 $\mu$m AlN step height.
	The final result is a gap of air between the bottom Mo and top Al leaving them electrically isolated, but Al is electrically connected to the top Mo layer.
	Figure \ref{Fig:Pull-back} shows trials of this technique, and it can be seen that the Mo has been pulled back from underneath the AlN and the Al has been deposited on top and fully covers the step height of the AlN.
	In this trial, Al served as the entire top metal without top Mo, however the device described in this work has a layer of top Mo between the Al and AlN layers. 
	
	\section{Fabrication of the \SiN-AlN photonic chips}
	\noindent The \SiN photonic chips are fabricated with a custom subtractive waveguide manufacturing process based on 248~nm DUV stepper lithography with fluoride-based dry reactive ion etching chemistry. 
	\SiN thin films of 200~nm are procured externally and waveguide cores are etched using a fluoride-based dry reactive ion etching chemistry.
	After etching we grow a 10~nm thin layer of \SiN to conformally cover the waveguides using low-pressure chemical vapor deposition to reduce sidewall roughness \cite{torres2022ultralow} and absorption by surface located defect states induced in the etching \cite{puckett2021422}. 
	A high-temperature anneal (11~h, 1200$^{\circ}$) is conducted before a 3~$\mu$m top oxide cladding is deposited. 
	The AlN actuators are fabricated with the same process used in \cite{tian2020hybrid}. The electrodes are fabricated using a 'Pull-back' process which was developed for this work. The details are presented in \cite{siddharth2023hertz}.
	After chip release, a low temperature (30~min, 300°C)) annealing cycle is conducted to remove long-lived defect states induced by deep-UV exposure.
	
	\section{Photonic chip layout and passive characterization of optical cavities}
	\noindent The bus waveguide has a microheater to control the optical phase delay, which is an important parameter of optical SIL \cite{voloshin2021dynamics}. The bus waveguide is coupled to an optical waveguide cavity formed by two connected spiral structures with $\kappa_0/2\pi = 30\pm10$~MHz. The back-reflection of spiral cavities caused by the Rayleigh back-scattering is relatively low, usually  below 1\%. We enhanced the back-reflection using a loop reflector in a drop-port up to more than 10\%.
	
	We perform passive characterization of all photonic structures. The setup allows to measure transmission, reflection, dispersion and propagation loss of \SiN waveguide spirals with a custom frequency-comb calibrated scanning diode laser spectrometer. The setup is based on our earlier work \cite{Liu_char_setup} and is described in details in \cite{riemensberger2022TWPA}. One wideband mode-hop free tunable external-cavity diode laser (ECDL, Santec) covers the wavelength range 1500-1630 nm. Regular calibration markers are recorded by filtering a beat note between the lasers and a commercial optical frequency-comb (Menlo OFC-1500) using a balanced photoreceiver and logarithmic RF detector (LA, AD8307) for increased dynamic range. Furthermore, an imbalanced fiber-optical Mach-Zehnder Interferometer (MZI) and a molecular gas cell are used for further calibration and wavelength determination.
	
\begin{figure*}[htb]
		\centering
		\includegraphics[width=\textwidth]{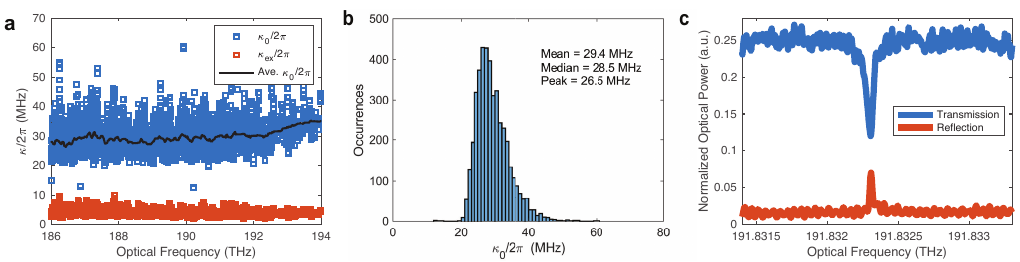}
		\caption{ 
			\footnotesize
			\textbf{Characterization of a PIC with 1.5 GHz spiral cavity.}
			(a) Intrinsic and external coupling rates at different optical frequencies. The intrinsic losses described by $\kappa_0/2\pi$ are in the range of 20-40 MHz in the region of target optical frequencies 188-193 THz. The $\kappa_{\rm{ex}}/2\pi$ describing the coupling rate to a bus waveguide is lower than 10 MHz making the cavity undercoupled.
			(b) The histogram of intrinsic losses $\kappa_0/2\pi$ with the mean value of 29.4 MHz.
			(c) The shape of a resonance in the transmitted light and reflected light.
		}
		\label{Fig:char}
	\end{figure*}

	Figure \ref{Fig:char} demonstrate the characterization data of an average sample of a PIC with 1.5 GHz spiral cavity. The histogram of intrinsic losses $\kappa_0/2\pi$ with the mean value of 29.4 MHz and the coupling losses of 5-10 MHz  in Fig. \ref{Fig:char}(a, b). Fig. \ref{Fig:char}(c) shows the reflection peaks which corresponds to the spiral cavity's resonance and is enhanced by a loop mirror up to more than 10\%.
	
	\section{Mechanical modes of mm-size piezoMEMS-PIC systems}
	
	\begin{figure*}[htb]
		\centering
		\includegraphics[width=\textwidth]{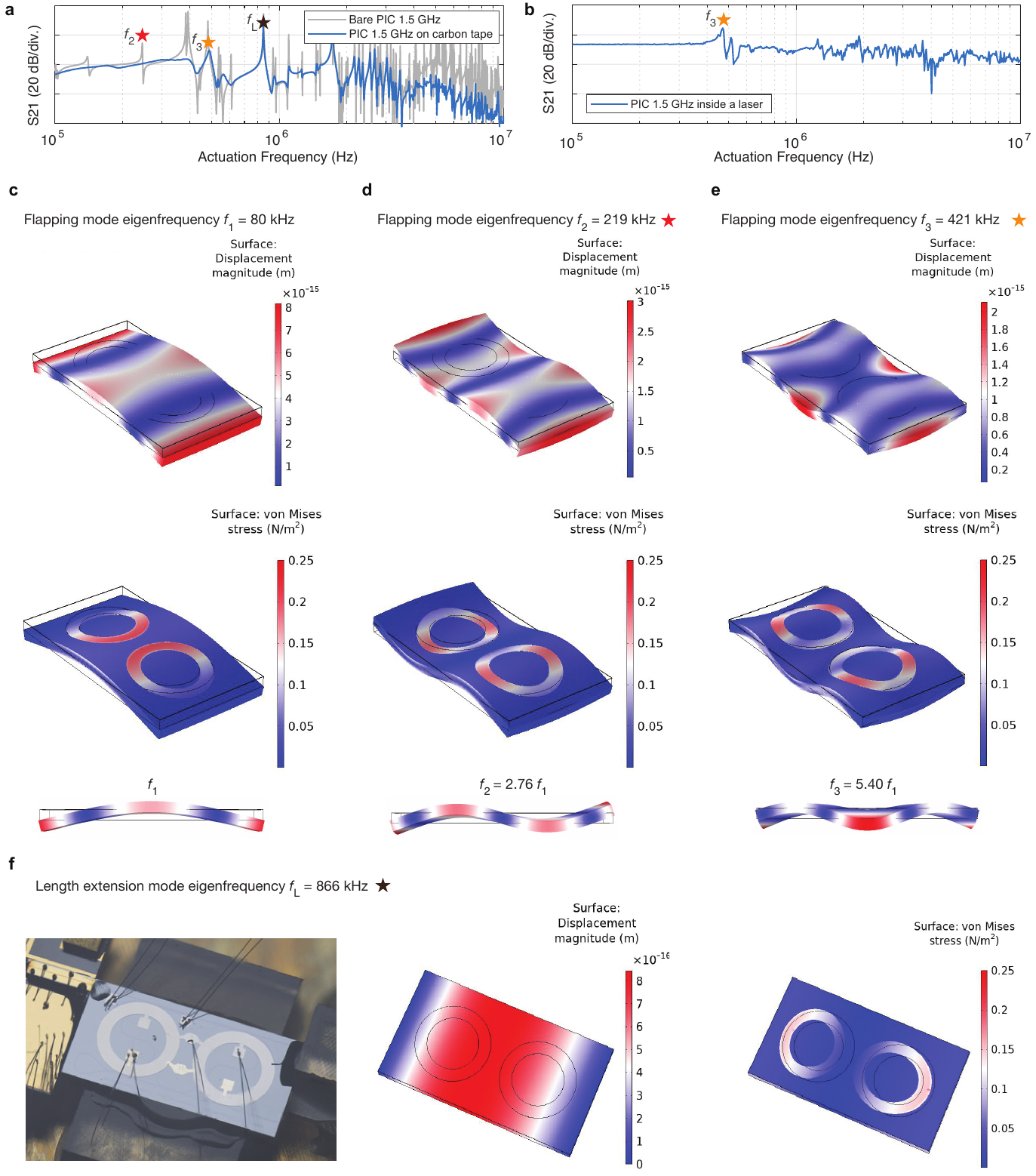}
		\caption{ 
			\footnotesize
			\textbf{Study of the mechanical modes of 1.5 GHz PIC.}
			(b) The $\rm{2}\textsuperscript{nd}$ flapping mode (the simulated eigenfrequency $f\textsubscript{2}$ = 220 kHz) and the $\rm{3}\textsuperscript{rd}$ flapping mode (the simulated eigenfrequency $f\textsubscript{3}$ = 421 kHz) are shown as well as the length extension mode (the simulated eigenfrequency $f\textsubscript{L}$ = 865 kHz). The flapping mode $f\textsubscript{2}$ is eliminated by a carbon tape, but not others.
			(b) $\rm{S}\textsubscript{21}$ of a 1.5 GHz PICs inside a packaged laser. A MEMS actuator inside a fully packaged laser supports flat actuation bandwidth up to 400 kHz. However, the $\rm{3}\textsuperscript{rd}$ flapping mode is still present according to the simulations and the experimental data. This mode is the limiting factor of the laser actuation response.
			(c, d, e) stress profile and vertical displacement of a PIC for the $\rm{1}\textsuperscript{st}$, $\rm{2}\textsuperscript{nd}$ and $\rm{3}\textsuperscript{rd}$ flapping mode. The approach to estimate eigenfrequencies analytically according to \cite{Demirci03IEEE} is shown in the bottom of panels.
			(f) stress and vertical displacement of the length extension mode $f\textsubscript{L}$ = 865 kHz.
		}
		\label{Fig:modes}
	\end{figure*}
	
	We simulated the mechanical modes of a PIC (220 $\mu$m thick \SiN) with 1.5 GHz spiral photonic cavity and a piezoactuator covering more than 20\% of the surface. In this case the actuation of the PIC is fully dominated by non-local mechanical stress caused by the first three flapping modes ($f_1$, $f_2$, $f_3$) and extension modes (the first mode $f_L$) \cite{Demirci03IEEE}. We experimentally observe these flapping mechanical modes in the $\rm{S}\textsubscript{21}$ response of bare and taped PICs (Fig. \ref{Fig:modes}(a)). The photonic packaging (compare Fig. \ref{Fig:modes}(a) and Fig. \ref{Fig:modes}(b)) helps to eliminate almost all of them, however, the $\rm{3}\textsuperscript{rd}$ flapping mode (the simulated eigenfrequency $f\textsubscript{3}$ = 421 kHz) cannot be suppressed in the currently used photonic packaging approach according to our simulations.
	
	We analytically estimated the eigenfrequencies of flapping modes using  the Euler-Bernoulli equations \cite{Demirci03IEEE}. The frequency of the \(n^{th}\) vibration flapping mode of a free-free PIC with the length \(L_r\) and the thickness \(h\) can be determined using the Euler-Bernoulli equation, given by \cite{Timoshenko1937}:
	
	\begin{equation}
		f_{nom} = \frac{-1}{2\pi \sqrt{12}} \left(\beta_n L_r \right)^2 \sqrt{\frac{E}{\rho}} \frac{h}{L_r^2}
	\end{equation}
	
	where \(E\) and \(\rho\) are the Young's modulus and density of the structural material, respectively, and \(\beta_n\) is the mode coefficient given by the \(n^{th}\) root of the equation \cite{Timoshenko1937}:
	
	\begin{equation}
		\cos \left( \beta_n L_r \right) \cosh \left( \beta_n L_r \right) = 1
	\end{equation}
	
	For the first three modes \(\beta_1 L_r\), \(\beta_2 L_r\), and \(\beta_3 L_r\) are 4.73, 7.853, and 10.996, respectively. It allows us to define the ratios between $f_1$ and the other modes: $f_2 = 2.76*f_1$ and $f_3 = 5.40*f_1$.
	
	In our case the analytical estimation gives: $f_1 = 83$ kHz, $f_2 = 229$ kHz, $f_3 = 448$ kHz. The finite element analysis gives: $f_1 = 80$ kHz, $f_2 = 219$ kHz, $f_3 = 421$ kHz. The experimental values are as follow: $f_2 = 245$ kHz, $f_3 = 480$ kHz. The experimental values could differ from eigenfrequencies since they are highly dependent on the mechanical environment, e.g., position of probes, orientation of a PIC on a mount, etc.
	
	It should be noted that the length extension mode of the same chip is analytically predicted to be 890 kHz, finite element analysis confirmed at 866 kHz. The experimental value is 851 kHz.
	
	The analytically estimated eigenfrequencies using the approach of \cite{Demirci03IEEE} (and confirmed with finite element analysis) and experimentally observed show that such a large piezoMEMS-PIC should be simulated and designed taking into account mechanical modes and stress distribution. After the photonic, mechanical and acoustic design we expect to extend the flat actuation bandwidth of a laser and potentially increase the actuation bandwidth.

	\section{Effect of laser linewidth and tuning linearity for homodyne detection}
	
	\begin{figure*}[htb]
		\centering
		\includegraphics[width=\textwidth]{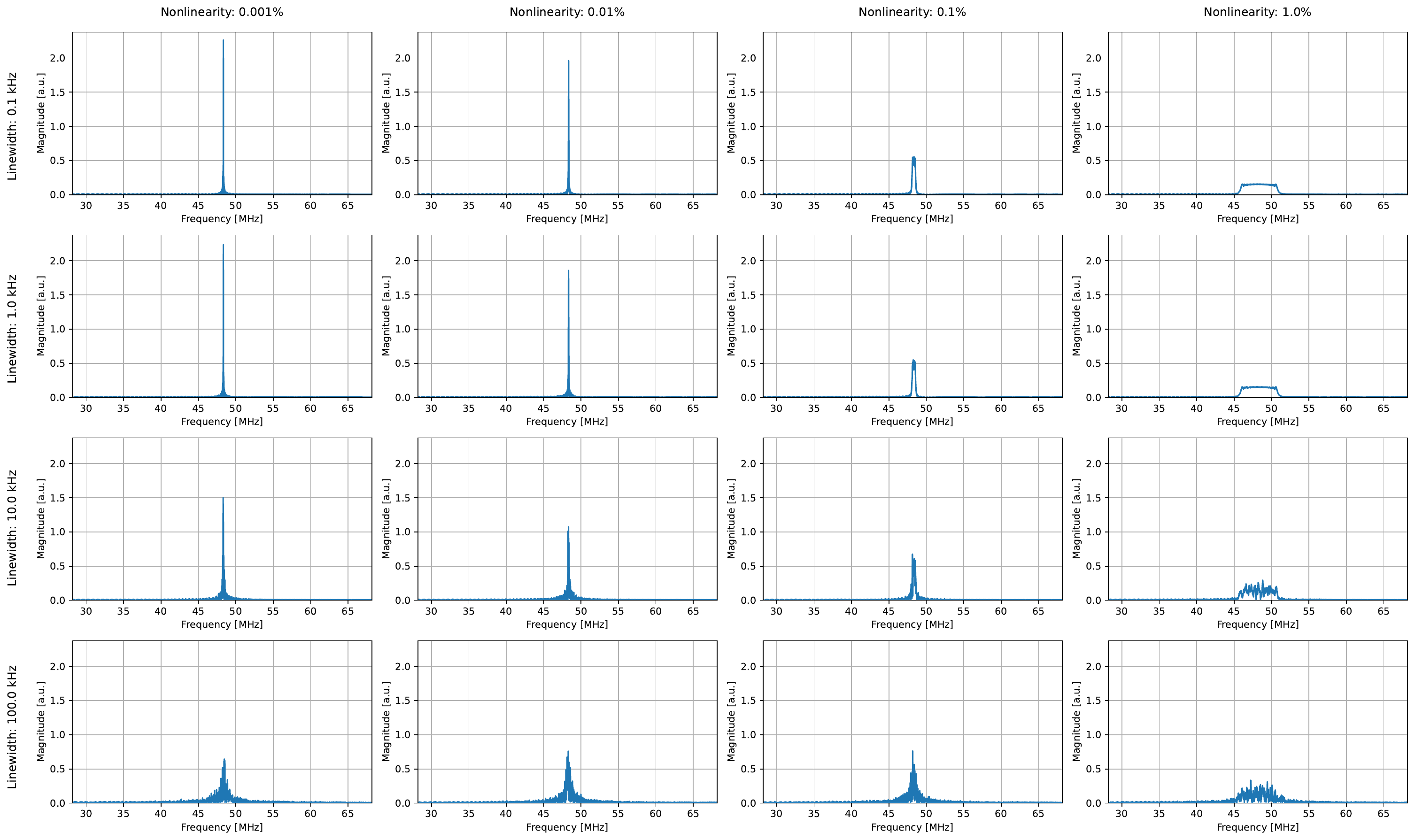}
		\caption{ 
			\footnotesize
			\textbf{Simulation of homodyne beatnote of FMCW LiDAR for various combinations of laser linewidth and chirp signal nonlinearity.}
			The figure presents a 4$\times$4 grid of plots. Each column corresponds to a different value of nonlinearity introduced in the chirp signal, increasing from left to right. Similarly, each row represents a different linewidth utilized introduced in the simulation, increasing from top to bottom. The plots within the grid display the resulting spectrum of the simulated homodyne beatnote of the frequency-modulated continuous wave signal with given nonlinearity and linewidth of the initial signal.
		}
		\label{Fig:linewidth_vs_linearity}
	\end{figure*}
	
	To estimate the effect of laser linewidth and nonlinearity on a frequency-modulated continuous wave (FMCW) signal for homodyne detection, we simulated the spectrum of the homodyne beatnote of an FMCW signal self-referenced over a delay line. The simulation used a chirp signal with a linearly changing instantaneous frequency ranging from 0 to 1 GHz over a duration of 50~$\mu$s. This corresponds to a periodic chirp signal with a base frequency of 10~kHz and an amplitude of 0.5~GHz. The signal was then self-referenced against a delayed copy, with a delay of 1.15~$\mu$s, which corresponds to a 500~m delay line at a wavelength of around 1550~nm. Before applying the delay, two types of deviations were introduced into the FMCW signal to simulate non-ideal conditions.
	
	The first deviation considered was linewidth broadening, now modeled using Lorentzian noise with zero mean and a variance corresponding to a linewidth level. Each sampling point of the signal was modified with noise from this distribution, altering the instantaneous frequency of the chirp and the cumulative phase of the signal. This aspect of the simulation is crucial, as linewidth broadening can significantly impact the resolution and clarity of frequency measurements, particularly in high-precision applications.
	
	The second deviation involved the introduction of quadratic nonlinearity into the chirp signal. This nonlinearity was quantified using a relative nonlinearity parameter, defined as the ratio of the estimated standard deviation from the ideal linear chirp to the entire chirp frequency range of 1~GHz. The calculated nonlinearity was then used to determine the parameters of a quadratic function. To maintain simplicity and the periodicity of the chirp signal, the quadratic function was set to zero at the beginning and end of the chirp. This parameter effectively measures the deviation of the chirp from a perfect linear sweep, with higher values indicating greater nonlinearity while preserving periodicity. Such nonlinearity can cause distortions in the frequency spectrum, leading to inaccuracies in signal interpretation.
	
	For accurate representation, the chirp signal was sampled at a rate of 10 GHz, ensuring that the frequency variations within the chirp were captured with high temporal resolution.
	
	The results of the simulation are presented in the figure \ref{Fig:linewidth_vs_linearity}. The figure is structured to demonstrate the effects of varying linewidth and nonlinearity on the simulated homodyne beatnote spectrum of FMCW signal. The figure is organized into a 4$\times$4 grid of plots. Each row corresponds to the simulation with the same linewidth, increasing from 0.1~kHz to 100~kHz as you move from top to bottom. Similarly, each column reflects a range of nonlinearity values, increasing from 0.001~\% to 1.0~\% of relative chirp nonlinearity from left to right. The plots within the grid display the resulting spectrum of the frequency-modulated continuous wave signal, providing a clear visualization of how different combinations of linewidth and nonlinearity influence the characteristics of the signal.
	
	This is evident that only lasers with the nonlinearity below 0.01\% and the linewidth below 10 kHz can provide excellent signal to noise ratio. However, using the noise reduction techniques lasers with higher linewidth (100 kHz) and poorer linearity (0.1\%) are used \cite{lihachev2023frequency}.
	
	\section{FMCW LiDAR demo}
	
	\begin{figure*}[htb]
		\centering
		\includegraphics[width=\textwidth]{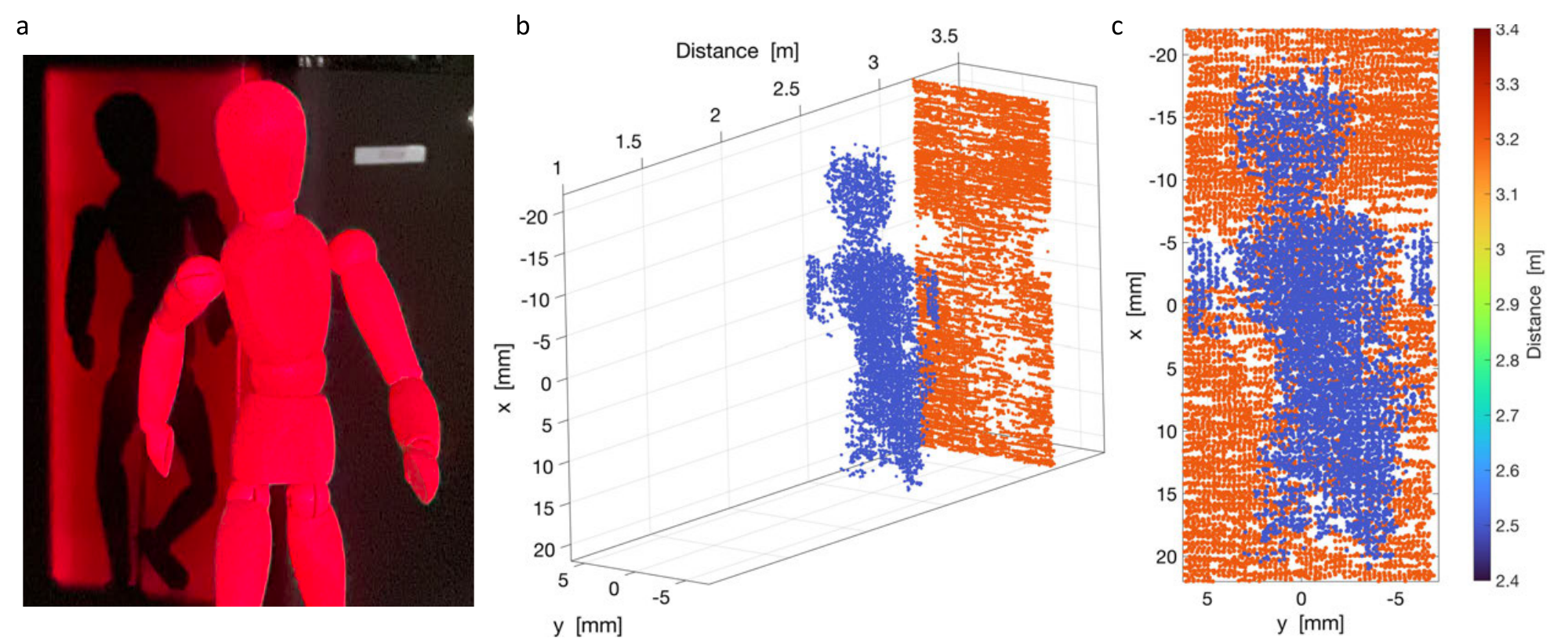}
		\caption{ 
			\footnotesize
			\textbf{FMCW LiDAR acquisition of a small wooden mannequin at 2.5 m distance using a frequency-agile PIC-based self injection-locked laser.} (a) Photograph of the target scanned using a red laser for visualization purposes. (b) Three-dimensional reconstruction of the ranged 3D target and scene. (c) Orthographic projection of the reconstructed 3D target and scene with color-coded ranging information.}
		\label{Fig:FMCW_demo}
	\end{figure*}
	
	To showcase the tuning bandwidth and chirp linearity of the photonic integrated circuit-based lasers, we conducted an optical coherent ranging demo in a lab environment using the FMCW LiDAR scheme.
	
	Here, the target range is inferred from the beat frequency of a laser source with triangular-shaped frequency modulation that is reflected from the target and detected by mixing it with the original signal on a fast photodiode. The FMCW range resolution is determined by the tuning range as \cite{behroozpour2017lidar}:
	
	\begin{equation}
		\Delta x = \dfrac{c}{2B},
	\end{equation}
	\label{eq:FMCW}
	where $B$ is the bandwidth of the linear frequency chirp. 
	
	The FMCW LiDAR acquisitions were performed using a packaged self PIC-based injection-locked laser linearly chirped by applying a triangular waveform of 100 V peak-to-peak to the on-chip actuators.
	The laser output is split into various channels, utilized for monitoring the chirp linearity, the output power (maintained around 2 mW in total), and as local oscillator. The transmitted output is passed through an fiber circulator for directional isolation and subsequently amplified to around 10 mW before being collimated onto a pair of galvo mirrors (Thorlabs GVS112) for scanning the scene. The return signal from the scene is coupled from the same optical path and isolated using the fiber circulator. This results in signal interference from any parasitic reflection at the circulator, collimator and fiber connectors, which produce beat notes frequencies significantly different from that of the target at 2.5 m, and hence can be easily ignored at the signal processing stage.
	
	The target used was a wooden mannequin placed at 2.5 m from the collimator output, with a scattering cardboard wall placed behind it at around 3.2 m from the collimator, as shown in \ref{Fig:FMCW_demo}(a). The fundamental depth resolution of the measurement is determined by the 200-MHz tuning range of the laser at this applied voltage, as given in eq. \ref{eq:FMCW}, resulting in around 75 cm.
	
	The acquired time-frequency spectrogram contains 20,000 time slices corresponding to each frequency ramp, with a typical SNR of 20 dB for the target. Each of these time slices is Fourier-transformed and filtered to determine the beat note from the target, which is converted into range information (i.e. radial coordinate) using calibrated tuning and ranging information. No signal linearization technique was used for this FMCW demo, which solely relied on the intrinsic chirp linearity of the laser.
	
	Figure \ref{Fig:FMCW_demo} shows point cloud representations of the reconstructed target scene with color-coded distance information: the mannequin profile is mostly visible in blue, and the wall behind in orange.
	
	This proof-of-concept coherent ranging demonstration showcases the PIC-based frequency-agile laser performance in FMCW LiDAR, where chirp linearity is a crucial requirement for the measurement accuracy. 
	
	%%%%%%%%%%%%%%%%%%%%%%%%%%%%%%%%%%%%%%%%%%%%%%%%%%%%%%%%%%%%%%%%
	
	\bibliographystyle{apsrev4-2}
	\bibliography{zotero_updated}